\documentclass{aa}  
\usepackage{chemmacros}

\newcommand{\ha}{H$\alpha$} 
\newcommand{\hb}{H$\beta$}
\newcommand{\hc}{H$\gamma$}
\newcommand{\hdelta}{H$\delta$}
\newcommand{\hydrogeni}{H\,{\sc i}}

\newcommand{\neoniii}{[Ne\,{\sc iii}]}
\newcommand{\neoniir}{Ne\,{\sc ii}}

\newcommand{\argoniv}{[Ar\,{\sc iv}]}

\newcommand{\argoniii}{[Ar\,{\sc iii}]}
\newcommand{\heliumi}{He\,{\sc i}}
\newcommand{\heliumii}{He\,{\sc ii}}

\newcommand{\nia}{[N\,{\sc i}]~5198~\AA}
\newcommand{\nib}{[N\,{\sc i}]~5200~\AA}

\newcommand{\nitrogenii}{[N\,{\sc ii}]}
\newcommand{\nitrogeniir}{N\,{\sc ii}}
\newcommand{\nitrogeniiir}{N\,{\sc iii}}
\newcommand{\nitrogeni}{[N\,{\sc i}]}

\newcommand{\cliv}{[Cl\,{\sc iv}]} 

\newcommand{\oxygeniii}{[O\,{\sc iii}]} 
\newcommand{\oxygeniiir}{O\,{\sc iii}}

\newcommand{\nitrogen}{[N\,{\sc ii}]}
\newcommand{\nitrogena}{[N\,{\sc i}]}

\newcommand{\oxygeni}{[O\,{\sc i}]}
\newcommand{\oxygenii}{[O\,{\sc ii}]}
\newcommand{\oxygeniir}{O\,{\sc ii}}

\newcommand{\sulfuriii}{[S\,{\sc iii}]}
\newcommand{\sulfurii}{[S\,{\sc ii}]}
\newcommand{\chloroiii}{[Cl\,{\sc iii}]}
\newcommand{\chloroii}{[Cl\,{\sc ii}]}

\newcommand{\ironii}{[Fe\,{\sc ii}]}
\newcommand{\ironiir}{Fe\,{\sc ii}}
\newcommand{\ironiii}{[Fe\,{\sc iii}]}
\newcommand{\ironi}{[Fe\,{\sc i}]}
\newcommand{\ironir}{Fe\,{\sc i}}

\newcommand{\carboni}{[C\,{\sc i}]}

\newcommand{\carboniir}{C\,{\sc ii}}

\newcommand{\carboniiir}{[C\,{\sc iii}]}

\newcommand{\nickelii}{{[Ni\,{\sc\,ii}]}}

\newcommand{\Phosii}{[P\,{\sc ii}]}

\def\vhel{\ifmmode{V_{{\rm HEL}}}\else{$V_{{\rm HEL}}$}\fi}
\def\vsys{\ifmmode{V_{\rm sys}}\else{$V_{\rm sys}$}\fi}
\def\kms{\ifmmode{~{\rm km\,s}^{-1}}\else{~km~s$^{-1}$}\fi}
\def\vlsr{\ifmmode{v_{\rm lsr}}\else{$v_{\rm lsr}$}\fi}

\newcommand{\Ne}{$n_{\rm e}$}
\newcommand{\Te}{$T_{\rm e}$}
\usepackage{textcomp}
\newcommand{\hmolec}{H\textsubscript{2}}

\newcommand{\nickeliii}{{[Ni\,{\sc\,iii}]}}

\usepackage[utf8]{inputenc}
\usepackage{longtable}
\usepackage{adjustbox}
\usepackage{caption}
\usepackage{subcaption}
\usepackage{soul}
\usepackage{natbib,twoopt}
\usepackage[colorlinks =true, urlcolor =blue, linkcolor =blue, citecolor =blue]{hyperref} 
\bibpunct{(}{)}{;}{a}{}{,}             
\makeatletter
  \newcommandtwoopt{\citeads}[3][][]{\href{http://adsabs.harvard.edu/abs/#3}%
    {\def\hyper@linkstart##1##2{}%
     \let\hyper@linkend\@empty\citealp[#1][#2]{#3}}}
  \newcommandtwoopt{\citepads}[3][][]{\href{http://adsabs.harvard.edu/abs/#3}%
    {\def\hyper@linkstart##1##2{}%
     \let\hyper@linkend\@empty\citep[#1][#2]{#3}}}
  \newcommandtwoopt{\citetads}[3][][]{\href{http://adsabs.harvard.edu/abs/#3}%
    {\def\hyper@linkstart##1##2{}%
     \let\hyper@linkend\@empty\citet[#1][#2]{#3}}}
  \newcommandtwoopt{\citeyearads}[3][][]%
    {\href{http://adsabs.harvard.edu/abs/#3}
    {\def\hyper@linkstart##1##2{}%
     \let\hyper@linkend\@empty\citeyear[#1][#2]{#3}}}
\makeatother

\usepackage{longtable}
\usepackage{graphicx}
\usepackage[utf8]{inputenc}
\usepackage{txfonts}
\usepackage{gensymb}

\definecolor{lime}{HTML}{A6CE39}
\DeclareRobustCommand{\orcidicon}{
        \begin{tikzpicture}
        \draw[lime, fill =lime] (0,0) 
        circle [radius =0.16] 
        node[white] {{\fontfamily{qag}\selectfont \tiny ID}};
        \draw[white, fill =white] (-0.0625,0.095) 
        circle [radius =0.007];
        \end{tikzpicture}
        \hspace{-2mm}
}

\foreach \x in {A, ..., Z}{\expandafter\xdef\csname orcid\x\endcsname{\noexpand\href{https://orcid.org/\csname orcidauthor\x\endcsname}
                        {\noexpand\orcidicon}}
}

\begin{document}

   \title{X-Shooter view of planetary nebulae microstructures}

   \subtitle{I. The case of NGC 7009 knots}

    \author{L. Konstantinou\inst{1,2}{\orcidD{}}, S. Akras\inst{1}{\orcidA{}},   K. Bouvis\inst{1,2}{\orcidC{}}, P. Boumis\inst{1}{\orcidG{}}, J. Garcia-Rojas\inst{3,4} {\orcidE{}},  D.R. Gon\c{c}alves\inst{5}{\orcidJ{}}  \and K.N. Gourgouliatos \inst{2}{\orcidK{}}}
         
    \authorrunning{L. Konstantinou et al.}
   \institute{Institute for Astronomy, Astrophysics, Space Applications and Remote Sensing, National Observatory of Athens,
              GR 15236 Penteli, Greece
              \email{lydiakonst@gmail.com} 
         \and
            Department of Physics, University of Patras, Patras, 26504 Rio, Greece    
         \and 
             Instituto de Astrof\'isica de Canarias, E-38205 La Laguna, Tenerife, Spain
         \and
             Departamento de Astrof\'isica, Universidad de La Laguna, E-38206 La Laguna, Tenerife, Spain
             \and
             Observat\'orio do Valongo, Universidade Federal do Rio de Janeiro, Ladeira Pedro Antonio 43, 20080-090, Rio de Janeiro, Brazil        
}

   \date{Received XXXXX, XXXX; accepted XXXXXX, XXXX}

 
 \abstract
   {The properties of NGC~7009  knots are investigated by employing UV-to-near-infrared data from the X-Shooter spectrograph. NGC~7009 is a widely studied elliptical planetary nebula (PN) that exhibits a pair of knots. These structures belong to the category of low-ionization structures (LISs), which are characterized by enhanced emission from low-ionization species and whose origin remains enigmatic.}
    { Our main goal is to conduct the first dedicated spectroscopic study of LISs in planetary nebulae that covers a wide spectral range from the UV to the near-IR, simultaneously. As a pilot study, we focus on NGC 7009 to investigate its ionized, atomic, and molecular gas, aiming to better characterize the physical properties of its LISs and explore their largely unexplored molecular content.}
   {We obtained spectroscopic data with the X-Shooter spectrograph, covering simultaneously and under the same conditions a spectral range from 300 nm to 2.5 $\mu$m. The spectrum was processed through the ESO Reflex pipeline, and the physical properties of LISs were derived with the latest version of {\sc PyNeb}. }
   {More than 250 emission lines were detected in the spectra of NGC~7009 LISs, including rare, forbidden transitions from low-ionization species such as Fe$^+$ and Ni$^+$, which have only very recently been identified in the spectra of LISs, as well as two transitions from P$^+$, identified here for the first time in LISs. The detection of multiple shock tracers, together with the shock excitation of \ironii~and \oxygeni, as revealed by emission-line diagnostics, supports the presence of shocks at LISs. The richness of the spectrum allowed us to derive electron temperatures and densities at the LISs using both collisionally excited and optical recombination lines. In the near-IR spectrum of the LISs, nine ro-vibrational \hmolec~emission lines were detected, seven of which were reported here for the first time in NGC~7009. These lines were used to construct the \hmolec~excitation diagram, providing new insights into the physical conditions of the nebular gas. The excitation of molecular hydrogen is most likely dominated by collisional processes through shocks. Finally, the estimated lifetime of \hmolec~at LISs is approximately 5.0~$\times$~10$^4$ years for K1a (East LIS) and 6.6~$\times$~10$^4$ years for K4 (West LIS).}
   {}


   \maketitle
%

\section{Introduction}

Planetary nebulae (PNe) represent the final evolutionary phase of low-to-intermediate mass stars. Among the most enigmatic features of PNe are small-scale structures, known as low-ionization structures \citep[LISs;][]{Balick1993a, Goncalves_2001,DeMarco2026}. These structures are characterized by enhanced emission in low-ionization lines, and their origin has been a
long-standing mystery. LISs exhibit a wide range of morphologies, such as knots, jets, jet-like structures, and even extended filaments. Kinematically, they are divided into two main categories: the fast low-ionization emission regions \citep[FLIERS,][]{Balick1993a} and the slow-moving low-ionization emitting regions \citep[SLOWERS;][]{Perinotto2000}. 
LISs have been the subject of various studies in the past, and they have attracted increasing attention in recent years \citep{Balick1993a, Corradi_1996, Concalves2001, Akras2017, Belen_2023}. Studies have mainly focused on the optical regime, where it is well known that LISs are prominent in low-ionization species such as \nitrogen, \sulfurii, and \oxygeni~\citep[e.g.,][]{Balick1993a, Corradi_1996,Concalves2001,Akras2015,Akras2022,Belen_2023}. 

Intriguingly, more recent studies have uncovered additional low-ionization lines at the LISs, including the \carboni~8727~\AA~\citep{Akras_2024,Konstantinou_2025}, \ironii~8617~\AA, and \nickelii~7378~\AA~\citep{Bouvis_2025,Konstantinou_2025}, as well as \ironii~1.644~$\mu$m~\citep{Akras_2024_Fe2}. Although these findings provide new insight into the nature of such structures, there is still no conclusive scenario for their origin.

Planetary nebulae LISs also exhibit electron temperatures, \Te, similar to the surrounding nebula, while this is not the case for electron densities, \Ne, which are often lower than the surrounding gas. This behavior has been attributed to the presence of a molecular component that is overlooked in the estimation of \Ne~\citep{Goncalves_2009}. Molecular hydrogen, \hmolec, has been detected in the ultraviolet (UV) and infrared (IR) spectra of several PNe \citep[e.g.,][]{Hora1999, Herald2004, McCandliss2007, Sterling2008}, and more recently
with JWST observations \citep[e.g.,][]{Wesson_2024} as well. However, in the case of LISs, this detection requires observations with high sensitivity, since in most cases \hmolec~emission is expected to be faint. Despite the difficulties, \hmolec~has been detected in the LISs of PNe through deep narrowband imaging \citep{Fang2015, Akras2017, Fang2018, Akras2020, mari2025}.

The optical spectra of LISs have been well studied compared to their IR characteristics, which remain largely unexplored, highlighting the need for targeted studies to comprehensively explore the nature and origin of LISs. Near-IR spectroscopic observations of PNe provide additional diagnostics for assessing the gas (either ionized or molecular) properties, revealing \hmolec~emission lines, among others. However, combining optical and near-IR observations is often challenging due to differences in epochs, telescopes, instruments, or configurations (e.g., slit positions). To overcome these limitations, X-shooter provides novel spectroscopic data covering a wavelength range from UV to visible and up to near-IR simultaneously. In this context, we present the first of a series of papers on multiwavelength long-slit studies focused on different morphological types of PN LISs. In this paper, we focus on the pair of LISs of the archetypal PN NGC\,7009. The remaining objects (nine LISs in seven PNe) will be presented in forthcoming publications.

NGC 7009, the Saturn nebula, is a bright ring-like PN with a size of approximately 50~$\times$~35 arcsec$^2$ at a distance of 1\,230~pc \citep{Bailer_2021}. NGC~7009 has been the subject of numerous investigations across the whole electromagnetic spectrum, spanning from X-ray to radio regimes \citep[e.g.,][] {Garay1989, Guerrero2002, Walsh_2016, Walsh_2018}. It consists of a bright inner core with a sharply defined elliptical rim, an outer elliptical shell, a nearly spherical halo that fades quickly in brightness, and two elongated LISs along its major axis. NGC 7009 has been classified as an oxygen-rich PN with a C/O ratio around 0.4 \citep{Kings_Barlow_1994}. Its central star is an H-rich O-type star with an effective temperature of $\sim$ 83\,000\,K \citep{Mendez_1988, Mendez_1992}. 

The LISs of NGC\,7009 are also known as ansae because they appear as knots at the tip of jet-like structures at almost the same distances from the central star. Numerous studies have focused on their physical and chemical properties, as well as their kinematics \citep[e.g.,][]{Goncalves_2003, Fernandez_2004, Sabbadin_2004, Akras2022, Hyung_2023}. Recently, \ironii~and \nickelii~emissions have been revealed emanating from these structures \citep{Bouvis_2025, Akras_2024_Fe2}. These findings reveal the presence of shock waves responsible for dust destruction and the subsequent release of refractory Fe and Ni into the gas phase \citep{Bouvis_2025}. Additionally, narrowband near-IR imaging has revealed \hmolec~2.12 and 2.24 $\mu$m emissions from NGC~7009 LISs \citep{Akras2020}. The presence of \hmolec~alongside \carboni~\citep{Akras_2024} indicates photodissociated or shock-heated gas within dense interstellar clumps. In this study, we present an extensive analysis of the spectrum of the NGC\,7009 LISs over the wavelength range of 300-2\,480~nm, using long-slit X-Shooter data. Given the large number of detected and identified emission lines, we are able to explore in detail the physical properties of these structures.

In Sect. \ref{Observations} we describe the observations and data reduction procedures. In Sect. \ref{Results} the spectroscopic results are presented, including the determination of the physical properties, chemical composition, and molecular hydrogen content. In Sect. \ref{Discussion}, we discuss the nature of the LISs along with the characteristics of the various gas components. Finally, the conclusions of this study are summarized in Sect. \ref{Conclusions}.

\section{Observations and data reduction}
\label{Observations}
The spectra presented in this work were acquired using the medium-resolution spectrograph X-Shooter \citep{Vernet_2011}, attached to the 8.2m UT3 of the Very Large Telescope (VLT) at the ESO Paranal Observatory in Chile (Program ID: 113.268E.001, PI: Akras S.). X-Shooter consists of three spectroscopic arms: UVB (300-559.5 nm), VIS (559.5-1\,024 nm), and NIR (1\,024-2\,480 nm). For this program, nine LISs embedded in seven PNe were observed under the same configuration.

\begin{figure}
    \centering
    \includegraphics[width=0.85\linewidth]{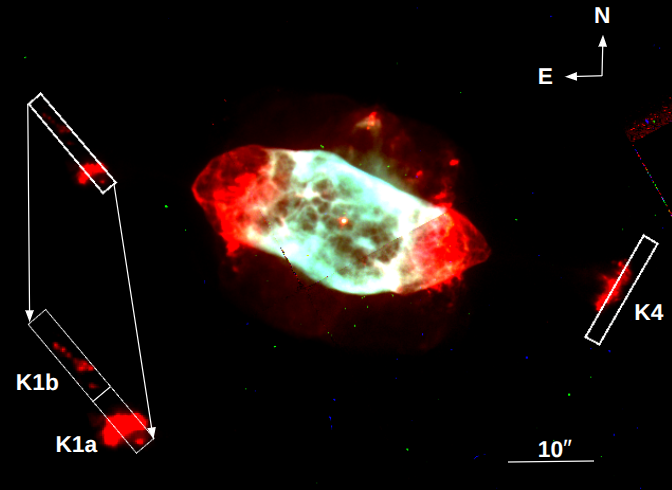}
    \caption{RGB image of NGC~7009 (R: \nitrogenii~ $\lambda$6584, G: \hydrogeni, B: \oxygenii~$\lambda$5007) from HST data. The white rectangles illustrate the position of the X-Shooter slits centered on the K1 and K4 LISs \citep{Goncalves_2003,Goncalves2004}. 
    The zoomed-in image of the eastern K1 LIS shows the individual regions that we chose to extract the spectra (K1a, K1b).}
    \label{NGC7009_slits}
\end{figure}

The LISs of NGC~7009 were observed with slit sizes of 1.6$^{\prime\prime}$~$\times$~11$^{\prime\prime}$, 1.5$^{\prime\prime}$~$\times$~11$^{\prime\prime}$, and 1.2$^{\prime\prime}$~$\times$~11\,$^{\prime\prime}$ and the spectral resolutions of R~=~3\,600, 5\,000, and 4\,300 for the UVB, VIS, and NIR arms, respectively, positioned as shown in Fig. \ref{NGC7009_slits}. To increase the signal-to-noise ratio (S/N), the data were acquired using 2~$\times$~2 binning in the UVB and VIS arms. No binning is offered for the NIR arm. Observations were conducted on the nights of 28-29 June (East LIS) and 6-7 July (West LIS) of 2024, under clear and stable conditions, with a seeing of 1.36$^{\prime\prime}$ and an airmass from 1.041 to 1.105.

We accounted for differences in slit width across wavelength ranges when estimating emission-line fluxes: We normalized the UVB and NIR spectra to the slit width of the VIS spectrum. For each arm, two sets of observations were obtained: one with a total exposure time of 250~s and shorter ones (40~s for UVB and VIS, 50~s for NIR) to avoid saturation of bright emission lines (e.g., \ha, \hb, \oxygeniii, \nitrogenii, etc.). For our observations, we used fixed sky offsets to achieve sufficient sky subtraction, with five cycles of an OSOS sequence.

The data were reduced using the \textit{ESOReflex} X-shooter pipeline \citep{Modigliani_2010, Esoreflex2013}, including wavelength and flux calibrations \citep{Vernet_2010, Hamuy_1994}. For the flux calibration, we used the spectrophotometric standard stars, Feige 110 (East LIS) and EG274 (West LIS) from the ESO database \citep{Moechler2014}, observed as part of the normal calibration plan. All data were reduced in offset mode. Both near-IR and VIS wavelengths are affected by molecular absorption in Earth's atmosphere, though the effect is less pronounced in the VIS arm. Telluric correction was developed using the \textit{Molecfit} tool, which fits a synthetic transmission spectrum to the provided science spectrum  \citep{Smette2015, Kaush2015}. It has to be noted, though, that \textit{Molecfit} is not always reliably applicable to low S/N data, leading to some residuals. Due to the low mean count level of our near-IR spectrum ($\approx$ 250 ADU), applying the telluric correction negatively impacted the spectrum quality.  In low S/N data, the fitting of the telluric features leads to overfitting of noise and the introduction of residuals. Therefore, we decided to proceed without applying the telluric correction in the near-IR spectrum, as recommended by \cite{Smette2015}.

To account for the effect of atmospheric differential refraction (ADR), we followed the approach described by \citet{Filippenko1982}. The effect is negligible across the three spectroscopic arms (losses of 1-2~\%). In the UVB arm, however, significant flux losses occur at the bluest wavelengths, reaching about $\approx$ 50~\%\ at 3\,000~\AA, dropping to about 15\%\ at 3\,500~\AA~and becoming negligible ($\sim$~1~\%) at 4\,000~\AA~ \citep[see also,][]{Mari2023b}. Therefore, we applied ADR correction to the highly affected part of the UVB arm (from 3\,000 to 3\,700 \AA).

In both slit positions, the emission from the LISs does not extend over the full length of the slit, so smaller extraction regions were chosen during the data reduction process. For the eastern LIS, the slit was placed at PA=~35$^{\circ}$ and encompasses the main K1a LIS and the fainter K1b component (Fig. \ref{NGC7009_slits}). Separate spectra were extracted for each of these regions, dividing the slit into subregions of length 5.2$^{\prime\prime}$. The western LIS (K4) was observed with a slit at PA~=~335$^{\circ}$, and the spectrum was extracted considering a length of 7.4$^{\prime\prime}$ for the extraction region. It should also be noted that the slit at the position of K4 was placed at the edge of the LIS, thus capturing only a part of it (see Fig. \ref{NGC7009_slits}).

\section{Results}
\label{Results}
\begin{figure*}[h!]
    \centering{
    \null\hfill
    \hfill
    \subfloat{\includegraphics[width =\textwidth]{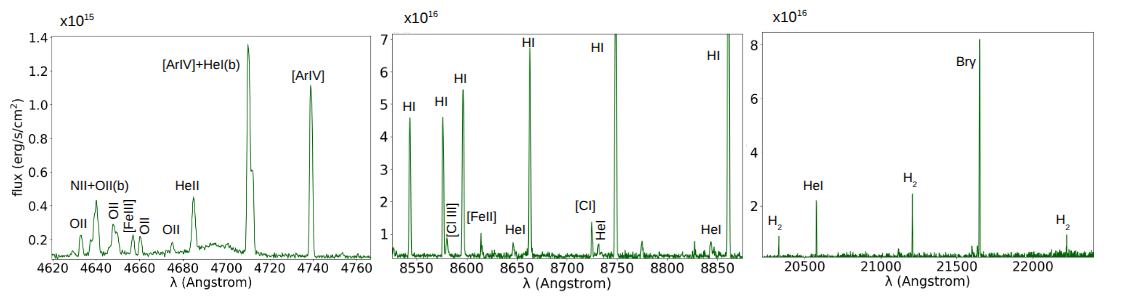}}
    \null\hfill
    }
\caption{Part of the X-Shooter spectrum at K1a LISs. From left to right: Spectrum from the UVB, VIS, and NIR arms. Emission lines noted with (b) are affected by blending: \nitrogeniiir~$\lambda$4640.64  with \oxygeniir~$\lambda$4641.81 and \nitrogeniiir~$\lambda$4641.85, \oxygeniir~$\lambda$4649.13 with \oxygeniir~$\lambda$4650.84, \argoniv~$\lambda$4711.37 with \heliumi~$\lambda$4713.17.}
\label{spectrum}
\end{figure*}

More than 250 emission lines are detected and identified in the spectra of the LISs in NGC\,7009 (see Table \ref{emission_lines}), utilizing the Atomic Line List database \citep{VanHoof2018}, the emission-line catalog of \cite{Fang_2011}, and line identifications provided by \cite{PyEmily_2026} from spectra analyzed with the {\sc PyEMILY} software \citep{PyEmily_2025}. The published line catalogs for molecular transitions by \cite{Hora_1999, Rudy2001, Nisini2002, Davis2011, Fang_2011} were used for molecular lines. We also recall the detection of $\approx$~700 emission lines from the observation of the PN by \cite{Fang_2011}. Flux uncertainties were estimated employing the new version (v2) of the Line Measuring library \citep[L{\sc i}M{\sc e},][]{LIME}. The faintest line detected with X-Shooter has a flux of 0.02 relative to \hb~=~100, with a signal-to-noise (S/N) of three. In the spectrum of NGC~7009 LISs, rare emission lines from low-ionization species are also detected, such as Fe$^+$, Ni$^+$, and P$^+$, with the latter being detected for the first time in LISs of PNe. In Figure \ref{spectrum} we present a part of the X-Shooter spectra, highlighting \ironii~$\lambda$8617, \carboni~$\lambda$8727, \oxygeniir~emission lines of the V1-multiplet (from $\lambda$4641 to $\lambda$4660) and finally three \hmolec~emission lines along with Br$\gamma$.

The emission line fluxes are generally in good agreement with previous studies, particularly with the most recent analysis of NGC~7009 LISs by \cite{Akras2020}. The only significant discrepancy concerns \nitrogeni. In our spectrum, the \nitrogeni~ doublet ($\lambda\lambda$5196, 5199) is clearly resolved, whereas in the previous study these lines are blended, resulting in an overestimation of the \nitrogeni~$\lambda$5199 flux. Minor differences in other lines may be attributed to variations in slit positioning, particularly for K4, where our slit samples only a small fraction of the structure. Moreover, contamination from the surrounding nebular gas is probably enhanced in K4, since it is fainter than K1a and not fully covered by the slit. This effect may lead to an overestimation of the measured emission-line fluxes in this region.

In the following subsections \ref{Physical_properties} and \ref{chemical_composition}, the physical properties such as interstellar extinction, electron temperature and density, and chemical composition of the K1a, K1b, and K4 LISs are derived. The molecular hydrogen content at the same structures is presented in Subsection \ref{H2}.

\subsection{Physical properties}
\label{Physical_properties}
In this section, we describe the physical properties derived from our spectra for K1a, K1b, and K4. For the estimations, we employed the {\sc PyNeb} tool \citep[version 1.1.18,][]{Luridiana_2015, Morisset_2020}. First, we estimated the extinction coefficients for the LISs by comparing the observed ratios of the Balmer lines (\ha, \hc, and \hdelta), as well as Pachen lines (Pa$\beta$, Pa$\gamma$, and Pa$\delta$) and Br$\gamma$, all relative to \hb, with their theoretical values in the low-density limit and for \Te=10\,000 K. For the estimations we employed the extinction law R$_v$ = 3.1 \citep{Seaton1979, Howarth1983, Cardelli1989}

The derived extinction coefficient, c(\hb), based on the \ha/\hb~ratio, is. 0.16 for K1a, 0.19 for K1b, and 0.16 for K4, in good agreement with the previous studies' estimations for the pair of LISs (0.16 for both K1a and K4) or the PN itself (0.16) \citep{Fang_2011, Goncalves_2003,Walsh_2018,Akras2022}. Extinction coefficients inferred from other hydrogen recombination lines relative to \hb\ (Table~\ref{table_extinction}) show no significant discrepancies, yielding mean values of 0.15, 0.19, and 0.15 for K1a, K1b, and K4, respectively. The consistency among coefficients derived from different spectral arms supports the reliability of the data reduction process across the entire X-Shooter spectra.
The measured fluxes were then corrected for the interstellar extinction.

\begin{table}[h!]
\centering
\caption{Interstellar extinction coefficients.}
\label{table_extinction}

    \begin{tabular}{|lcccc|}  
        \hline
       & & K1a & K1b & K4\\
        \hline
      Arm & line & &c(\hb)& \\
         \hline
    VIS   & \ha &0.16$\pm$0.04&0.19$\pm$0.09& 0.16$\pm$0.09 \\ \hline
     UVB  & \hc &0.17$\pm$0.10&0.20$\pm$0.05& 0.15$\pm$0.10\\
       & \hdelta &0.17$\pm$0.09&0.20$\pm$0.05& 0.15$\pm$0.10\\ \hline
    NIR  &   Pa$\delta$ &0.15$\pm$0.04&0.20$\pm$0.09& 0.12$\pm$0.08 \\
        &  Pa$\gamma$ &0.12$\pm$0.06&0.20$\pm$0.10& 0.12$\pm$0.08 \\
       & Pa$\beta$ &0.13$\pm$0.04&0.17$\pm$0.08& 0.17$\pm$0.02 \\
     &   Br$\gamma$ &0.14$\pm$0.02&0.18$\pm$0.08& 0.16$\pm$0.10 \\
    
        \hline

    \end{tabular}

\end{table}

To estimate \Te~and \Ne~of the LISs, we employed widely used diagnostics based on both collisionally excited lines (CELs) and optical recombination lines (ORLs). The results are presented in Table \ref{table_Te_Ne}, while the atomic data adopted for these calculations are listed in Table \ref{atomic_data}. To probe the physical conditions within the neutral atomic gas, the \oxygeni~$\lambda$5577/$\lambda$6300+$\lambda$6364 ratio is commonly used for estimating $T_{\rm e}$, while \nitrogena~$\lambda$5198/$\lambda$5200 serves as one of the few available density indicators for neutral species in the optical regime.
Additionally, the ratio \carboni~$\lambda$9824+$\lambda$9850/$\lambda$8727 can also be used as a temperature diagnostic for the atomic neutral gas \citep{Liu_1995}. The derived $T_{\rm e}$(\carboni)~is consistent with $T_{\rm e}$\oxygeni.

Electron temperatures derived from CELs are mutually consistent within the uncertainties, except for \Te (\oxygenii). The \oxygenii~$\lambda$7320+$\lambda$7330/$\lambda$3729 line ratio yields a higher electron temperature than the other diagnostics, with the behavior being more obvious for K1a and K1b. Similarly, elevated \Te (\oxygenii) ($\simeq$ 19\,000 K)\ is also reported in a long-slit study of the PN by \cite{Fang2018}, who assumed a density of 10\,000 cm$^{-3}$ to mitigate the otherwise abnormally high \Te(\oxygenii). Also, no significant discrepancy is reported for \Te (\carboni), \Te (\oxygeniii), and \Te (\nitrogenii), which may suggest that LISs do not follow the stratification expected in a transition zone between ionized and neutral gas. More precisely, in a photo-dissociation region (PDR), we would expect to observe \Te (\carboni)~lower than \Te (\oxygeniii)~and \Te (\nitrogenii)~\citep{Liu_1995}, which is not the case for the LISs of NGC~7009. 

Lower temperatures are estimated from the Paschen jump and \heliumi~diagnostics, with the former being higher than the latter, as it has been previously reported in several PNe, including NGC~7009 \citep[e.g.,][]{Zhang_2005a, Walsh_2018}. Particularly for \Te~(PJ), we computed 6\,060~K and 6\,000~K for K1a and K1b, respectively, which are in good agreement with the mean Te(PJ) of 6\,610~K found by \cite{Walsh_2018}, while a slightly higher value (7\,900~K) is observed for K4. The \Te~(PJ) map presented in \cite{Walsh_2016} shows elevated temperatures in localized regions, particularly in the LISs. This may account for the high temperatures obtained here compared to the mean value reported previously in the nebular main shell.

For the electron density in the ionized part of the gas, we employed \oxygenii, \sulfurii, \chloroiii, and \argoniv~diagnostic line ratios. For all the structures, \Ne~varies from $\approx$1\,000  to around 2\,500 cm$^{-3}$, with the higher values observed for \Ne (\sulfurii). Notably, \Ne~ measured from the atomic nitrogen lines (\nia, \nib) yields lower values, although they agree with the other diagnostics within the uncertainties. Our deep spectroscopic data also allow us to detect various optical iron lines, including \ironii~$\lambda\lambda$7155, 8617, whose ratio has been proposed as an electron density indicator \citep{Bautista_1998, Bautista_2015}. The density provided by this ratio is on the order of 10$^5$~cm$^{-3}$, two orders of magnitude higher than the rest of the density diagnostics. However, \cite{Bouvis_2025} demonstrate that this ratio is \Te~dependent in the \Ne~range between 10$^{3.5}$ to 10$^6$ cm$^{-3}$, and therefore it should be treated with caution. 

The electron density was also estimated from the ORL ratio \oxygeniir~4649/4661, yielding values slightly higher than those derived from CELs, with the exception of \Ne(\sulfurii), which is consistently higher than the other CEL diagnostics in all structures. Recently, \cite{Huang_2026} suggest that the trend for ORL diagnostics toward higher densities and lower temperatures compared to CELs may reflect their origin in physically distinct regions within the nebular gas. However, our ORL and CEL values do not differ significantly, so this scenario cannot explain the discrepancies observed in NGC~7009 LISs.

Overall, the \Te~and \Ne~values, estimated from CELs, are consistent with those reported in previous studies of NGC~7009 \citep[e.g.,][]{Akras2020, Goncalves_2003, Fang_2013, Walsh_2018}. However, some discrepancies may be attributed to differences in the datasets, as well as to the spatial positions selected for the measurements of \Te~and \Ne.  

It is evident that diagnostics of \Te~tracing the ionized and the atomic gas components do not reveal significant discrepancies within the associated uncertainties. This indicates that both the atomic and ionized components of the nebular gas reside at essentially the same temperature. On the contrary, \Te~estimations from ORLs (\Te~(\heliumi) and \Te~(PJ)) are systematically lower than those obtained from CELs. \cite{Peimbert_1967} suggested that, in the presence of temperature fluctuations within the gas, \Te (\oxygeniii) would be systematically higher than \Te (\hydrogeni), the latter being almost unaffected by such variations. The spatial mapping of temperature fluctuations in NGC~7009 reveals a temperature fluctuation parameter ($t^2$) of 0.03 in the LISs. This value is lower than those reported for other nebular structures, but still remains non-negligible \citep{Walsh_2018}. Therefore, temperature inhomogeneities within the LISs may play a role. \cite{Mendez-Delgado_2025} propose that this discrepancy may also arise from deviations from the case B approximations. We applied their toy model to the \heliumi~$\lambda$7281 emission line, which is the one most sensitive to this phenomenon. We found that approximately 70\% for K1a and K1b, and 76\% for K4 of the \heliumi~$\lambda$7281 emission follows case B conditions. 
Using these fractions and following the approach of \cite{Mendez-Delgado_2025}, together with the procedure described by \cite{Mendez-Delgado_2021} to estimate \Te~(\heliumi), we derive temperatures of (8\,700 $\pm$ 1\,500)~K, (8\,000 $\pm$ 1\,400)~K and (9\,500 $\pm$ 1\,500)~K, for K1a, K1b, and K4, respectively. Note that the previous values were $\approx$~5\,600~K, 5\,700 K, and 6,600~K. Thus, the application of the toy model results in an increase of about 3\,000 K in all regions. Despite this improvement, a discrepancy remains between \Te (\oxygeniii) and \Te (\heliumi), indicating that deviations from case B alone cannot fully explain the observed temperature difference. This suggests that additional physical processes may play a role, as proposed by \cite{Mendez-Delgado_2025}.

\begin{table}[h!]
\centering
\caption{Electron temperature and density.}
\label{table_Te_Ne}
\resizebox{.5\textwidth}{!}{
    \begin{tabular}{|lccc|}  
        \hline
        & K1a & K1b & K4 \\
        \hline
        \multicolumn{4}{c}{$T_{\rm e}$(K)}\\
        \hline
         {[O~\sc iii]} & 9\,140$\pm$380&  11\,300$\pm$1\,600 & 11\,140 $\pm$ 980  \\ $\lambda$4363/$\lambda$5007+ & & &  \\
        {[S~\sc iii]} & 9\,050$\pm$230&  9\,700$\pm$2\,000 & 10\,670$\pm$1\,200 \\ $\lambda$6312/$\lambda$9069 & &  &  \\
        {[N~\sc ii]}&  9\,820$\pm$1\,300& 10\,800$\pm$800& 10\,160$\pm$1\,610 \\$\lambda$5755/$\lambda$6584+&  & & \\
         {[S~\sc ii]} &  9\,200$\pm$1\,090& 10\,500$\pm$1\,600& 9\,200$\pm$1\,000 \\$\lambda$4069+/$\lambda$6717+&  & & \\
         {[O~\sc ii]} & 13\,200$\pm$1\,000 &13\,400$\pm$ 900 &11\,700$\pm$1\,500 \\ $\lambda$7320+/$\lambda$3729 & & & \\
         {[O~\sc i]}  & 8\,400$\pm$130&  - & 8\,640$\pm$240 \\ $\lambda$5577/$\lambda$6300 & &  & \\
         {[C~\sc i]} & 9\,900$\pm$960&  8\,740$\pm$850 & 9\,000$\pm$900 \\ $\lambda$9824+/$\lambda$8727 & &  &  \\
         \heliumi & 8\,700$\pm$1\,500&  8\,000$\pm$1\,000& 9\,500$\pm$1\,500\\
         $\lambda$7281/ $\lambda$6678 & &  &  \\
           Paschen Jump (PJ) & 6\,060$\pm$400&  6\,000$\pm$1\,000& 7\,900$\pm$1\,200\\
           \hline
           \multicolumn{4}{c}{\Ne~(cm$^{-3}$)}\\
           \hline
           \sulfurii & 2\,500$\pm$300 & 1\,800$\pm$285 & 2\,008$\pm$491 \\  $\lambda$6731/$\lambda$6716 &  & & \\ 
           {[O~\sc ii]}  & 1\,850$\pm$500& 1\,600$\pm$650 & 1\,220$\pm$575\\$\lambda$3726/$\lambda$3729 & & & \\
        {[Cl~\sc iii]}  & 1\,640$\pm$100& 1\,150$\pm$283 & 1\,460$\pm$190 \\$\lambda$5538/$\lambda$5518 & & &  \\
        {[Ar~\sc iv]} & 1\,200$\pm$180& 1\,000$\pm$370& 1\,000$\pm$600 \\$\lambda$4740/$\lambda$4711 & &  &  \\
        {[N~\sc i]} & 1\,040$\pm$500 & 1\,140$\pm$650& 800$\pm$400 \\
         $\lambda$5198/$\lambda$5200 &  & &  \\
         {O~\sc ii} & 2\,500$\pm$2\,000 &$-$& 1\,500$\pm$1\,000 \\
         $\lambda$4649/$\lambda$4661 &  &  & \\
        \hline
    \end{tabular}
}
\end{table}

\subsection{Chemical composition}
\label{chemical_composition}
The chemical characterization of NGC~7009 LISs was performed by estimating the elemental abundances utilizing both CELs and ORLs. In both cases, ionization correction factors (ICFs) were applied to account for unobserved ionization stages and derive the total elemental abundances.

\subsubsection{Elemental abundances from CELs}
\label{elem_abund_sect}
Electron temperatures and densities derived from CELs were used to determine the nebular chemical properties. The \Te(\oxygeniii)-\Ne(\chloroiii)~and \Te(\nitrogenii)-\Ne(\sulfurii)~pairs were adopted for the high- and low- ionization ions, respectively. When multiple emission lines were detected for the same ion, the ionic abundance was calculated independently from each line. The final ionic abundance was then determined by taking the average of the individual results (Table~\ref{abundances}).

\begin{table}
    \caption{Atomic data for CELs and ORLs}
    \label{atomic_data}
    \resizebox{.5\textwidth}{!}{

    \begin{centering}
        \begin{tabular}{|lcc|}
        \multicolumn{3}{c}{CELs}\\
    \hline
    Ion & Transition probabilities & Collision strengths \\\hline
    C$^0$ & \cite{Fischer1985}  & \cite{Johnson_1987}  \\
     O$^0$ &  \cite{Wiese1996} & \cite{Bhatia1995} \\
     O$^{+}$ &  \cite{Zeippen1982} & \cite{Kisielius2009}\\
     &  \cite{Wiese1996}&\\
     O$^{2+}$ &  \cite{Storey2000} & \cite{Storey2014}\\
        &  \cite{ FroeseFischer2004} & \\
        Ne$^{2+}$ & \cite{Galavis1997} & \cite{McLaughlin2000}\\
     N$^{0}$ &  \cite{Kaufman1986} & \cite{Peguignot1976}  \\
             & \cite{ Wiese1996}&\cite{Dopita1976} \\
     N$^{+}$ &  \cite{FroeseFischer2004} & \cite{Tayal2011}\\
    S$^{+}$ &  \cite{Rynkun2019} & \cite{Tayal2010}\\
    S$^{2+}$ &  \cite{FroeseFischer2006} & \cite{Tayal1999}\\
    Cl$^{+}$ & \cite{Mendoza1983} & \cite{Tayal2004}\\
    Cl$^{2+}$ & \cite{Rynkun2019} & \cite{Butler1989}\\
    Cl$^{3+}$ & \cite{Huang_1985}& \\
    & \cite{Kaufman_1986} & \\
    & \cite{Mendoza_1982}&  \cite{Galavis1995}\\
    & \cite{Ellis_1984}& \\
     Ar$^{2+}$ &  \cite{Burgos2009} & \cite{Burgos2009}\\ 
     Ar$^{3+}$ &  \cite{Rynkun2019} & \cite{Ramsbottom1997}\\ 
      Fe$^{+}$ & \cite{Bautista2015} & \cite{Bautista2015}\\
    Fe$^{2+}$ &\cite{BAndell2014}&\cite{BAndell2014}\\ 
    Ni$^{+}$ & \cite{Chianty_2021} & \cite{Chianty_2021}\\ \hline

     \multicolumn{3}{c}{ORLs}\\
      \hline
    \multicolumn{3}{|c|}{Effective recombination coefficients} \\
    \hline
    H$^{+}$ & \multicolumn{2}{c|}{\cite{Storey1995}}  \\
    He$^{+}$ &  \multicolumn{2}{c|}{\cite{Porter2012, Porter_2013}} \\
        He$^{2+}$ &   \multicolumn{2}{c|}{\cite{Storey1995}}\\
        O$^{2+}$ &   \multicolumn{2}{c|}{\cite{Storey2017}}\\ 
        N$^{2+}$ &   \multicolumn{2}{c|}{\cite{Davey2000}}\\ 
         C$^{2+}$ &   \multicolumn{2}{c|}{\cite{Fang2011_NII}}\\ 
         & \multicolumn{2}{c|}{\cite{Fang2013_CII}}\\ \hline

    \end{tabular}
    \end{centering}
    }
\end{table}

For the estimation of the total abundances, we used the ICFs derived from \cite{DIMS} and \cite{KB} to correct for the unobserved ionization stages. We found that there is no significant discrepancy between the two different approaches. Hence, in Table \ref{abundances}, we list only the ones estimated from \cite{DIMS}. Our results exhibit minor discrepancies relative to those reported in previous long-slit \citep{Goncalves_2003,Fang_2013} and integral field unit (IFU) data studies \citep{Walsh_2018,Akras2020}. These differences are primarily attributed to variations in the underlying datasets and to the specific regions selected for analysis.

Our deep spectroscopic observations of NGC~7009 LISs offer the opportunity to measure the fluxes of iron and nickel lines in these structures. In the case of Fe$^{+}$/H$^{+}$, we estimated the abundance using two different sets of atomic data \cite[TZ18, recommended by \citet{Mendoza_2023}]{Tayal_2018} and \cite[B15]{Bautista_2015}. The results may vary depending on the adopted dataset (see Table \ref{iron_abund}). These discrepancies may arise because, when TZ18 atomic data are adopted, \ironii~$\lambda$7155 emission line gives an Fe$^{+}$/H$^{+}$ abundance nearly one order of magnitude lower than \ironii~$\lambda$8617, thus decreasing the total Fe$^{+}$/H$^{+}$. To estimate the total iron abundance, we adopted two different approaches. In the first, we summed up Fe$^{+}$/H$^{+}$ (considering the B15 atomic data) with Fe$^{2+}$/H$^{+}$, leading to Fe/H=8.32~$\times$~10$^{-8}$ and 9.11~$\times$~10$^{-8}$. \cite{Bouvis_2025} report similar Fe/H values based on IFU MUSE data and assuming that Fe$^{+}$($\lambda$8617)/H$^{+}$ = Fe/H (Table \ref{iron_abund}). Yet, if we adopt the same assumption, we obtain values that are almost half those of \cite{Bouvis_2025}. The differences are attributed to the higher intensity of the \ironii~$\lambda$8617 line measured in MUSE data.

In the second approach, we derived the Fe abundance based on the detected \ironiii~emission lines, applying the ICF provided by \cite{Rodriguez_2005}. Two ICF formulae are provided by these authors. The first one (eq.~2) is based on photoionization models, while the second one (eq.~3) is empirically derived from observations. The two ICF schemes tend to give systematically different results, and it is suggested that the true iron abundance is likely between these two values. In the present study, the X-Shooter spectra reveal clear \ironiii~emission, including emission lines at 4881, 4658, and 5270 \AA. Using these transitions and considering both ICF schemes, we derive Fe abundances in the range of 1.90~$\times$~10$^{-7}$ $-$ 2.24~$\times$~10$^{-7}$ for K1a
and 6.41~$\times$~10$^{-8}$ $-$  7.30~$\times$~10$^{-8}$ for K4. For K1b, \ironii~emission lines were not detected, and only the \ironiii~$\lambda$4658 emission line was identified. Consequently, the elemental abundance of iron was estimated using only the second approach, yielding Fe/H values between 9.63~$\times$~10$^{-8}$ and 2.30 ~$\times$~10$^{-7}$.
In previous studies, only \ironii~lines were found to emanate from K1a, K1b, and K4 \citep{Bouvis_2025, Akras_2024_Fe2}, from MUSE emission-line maps.

For nickel abundance, we employed only the \nickelii~lines and atomic data from Chianty version 10.1 \citep{Chianty_2021}. Following the same approach as \citet{Bouvis_2025}, we assumed Ni$^{+}$/H$^{+}$=Ni/H~and we derive an elemental abundance of nickel (1.42~$\pm$~0.5)~$\times$~10$^{-9}$ and (4.20~$\pm$~2.10)~$\times$~10$^{-9}$ for K1a and K4, respectively. These values are in reasonable agreement, within the errors, with those from \cite{Bouvis_2025} (4.27~$\pm$~2.18)~$\times$~10$^{-9}$ and (6.41~$\pm$~3.27)~$\times$~10$^{-9}$ for K1a and K4, respectively. Since Fe$^{2+}$~and Ni$^{2+}$~share similar ionization potentials, one might expect \nickeliii~lines to be present in the optical spectrum (e.g., $\lambda\lambda$6755, 7890). However, no such lines are detected. In general, \nickeliii~emission lines are expected to be weaker than \ironiii~lines, which themselves are relatively faint, and their emission may therefore fall below our detection limit. Moreover, assuming that the ratios of Fe$^{2+}$/Fe$^{+}$ and Ni$^{2+}$/Ni$^{+}$ should be comparable, and using the values from Table \ref{abundances}, we obtain Fe$^{2+}$/Fe$^{+}$ around 1.4 for K1a and 1.7 for K4. Under that assumption, the Ni$^{2+}$/H$^{+}$ will not be negligible, and therefore the derived Ni/H abundances from our study should be treated as lower limits for the nickel abundance in the LISs of NGC~7009.

\subsubsection{Elemental abundances from ORLs}
Given the richness of our X-Shooter spectra in ORLs, we also derived ionic and total abundances from ORLs for oxygen, nitrogen, and carbon. The atomic data used for the following estimations are listed in Table \ref{atomic_data}. In a previous study, \citet{Fang_2013} (hereafter, FL13) carried out an ORL-based abundance analysis of NGC 7009 using long-slit optical spectroscopy. The present analysis, though, is focused on the LISs. 

For the determination of O/H, the only available ORLs are \oxygeniir. For the abundance estimation we used the \oxygeniir~from the V1-multiplet ($\lambda$4641, $\lambda$4649+50, $\lambda$4660), leading to O$^{2+}$/H$^{+}$ of (1.90 $\pm$ 0.90)~$\times$~10$^{-3}$ for K1a, (9.89 $\pm$ 2.30)~$\times$~10$^{-4}$ for K1b and (8.80 $\pm$ 3.22)~$\times$~10$^{-4}$ for K4. Following the approach of FL13, we assumed that the ionic ratio O$^{+}$/O$^{2+}$ for ORLs is the same as the one derived from CELs, in order to derive approximate values for O$^{2+}$/H$^{+}$. Employing this assumption in conjunction with the ICF(O) formula provided by DIMS, we derive O/H $\approx$ (2.3 $\pm$ 0.9)~$\times$~10$^{-3}$, (1.1 $\pm$ 0.5)~$\times$~10$^{-3}$, and (1.6 $\pm$ 0.8)~$\times$~10$^{-3}$ for K1a, K1b, and K4, respectively. FL13, performing the above analysis on long-slit data of NGC~7009 concluded to O$^{2+}$/H$^{+}$ = 1.40~$\times$~10$^{-3}$ and O/H~=~1.67~$\times$~10$^{-3}$. The difference in the total O/H arises from a discrepancy between the O$^{+}$/O$^{2+}$ ratio estimated from CELs in the LISs and the one reported by FL13. 

In the case of carbon, only \carboniir~lines were detected in our spectrum. For the abundance determination, we adopted the \carboniir~$\lambda$4267 line, as it is the brightest of the available transitions. This yields C$^{2+}$/H$^{+}$ abundances of $(3.21~\pm~0.50)\times10^{-4}$, $(3.80~\pm~0.70)\times10^{-4}$, and $(3.50~\pm~0.51)\times10^{-4}$ for K1, K1b, and K4, respectively. The C/H abundance was then derived using the ICF(C) prescription of \citep{DIMS}, yielding a value of $\sim 4.0 \times 10^{-4}$ for each LIS, which is nearly a factor of two lower than that reported by FL13. Unfortunately, optical and near- IR spectra do not include forbidden carbon lines of the same ionization stages detected in ORLs, preventing a direct comparison between CEL- and ORL- based abundances for carbon.

Finally, nitrogen emission lines from N$^{2+}$ were detected only in K1a, while N$^{3+}$ lines were detected in both K1a and K4. For N$^{2+}$, the \nitrogeniir\ $\lambda$5679 and $\lambda$4630 lines were used to determine the abundance. The N$^{3+}$ abundance was not estimated, since the \nitrogeniiir~$\lambda\lambda$4640, 4641 lines from multiplet V2 could not be properly deblended from the \oxygeniir~$\lambda$4641 line. This makes it difficult to estimate the total intensity of the multiplet, which is the most common way of estimating the N$^{3+}$ abundance from \nitrogeniiir\ ORLs. The resulting abundance N$^{2+}$/H$^{+}$ is (1.61 $\pm$ 0.20)~$\times$~10$^{-4}$, which is of the same order of the one estimated by FL13. The approach for estimating the total nitrogen abundance from ORLs, as suggested by FL13, as well as the more recent prescription of \cite{Huang_2026}, requires the N$^{3+}$/H$^{+}$. Since no N$^{3+}$ abundance is measured in the LISs of NGC~7009, neither approach can be applied to our data

\begin{table}[h!]
\centering
\caption{Ionic and elemental abundances.}
\label{abundances}
 \resizebox{.50\textwidth}{!}{
    \begin{tabular}{|lccc|}  
        \hline
        & K1a & K1b & K4 \\
        \hline
        \multicolumn{4}{|c|}{Ionic abundances}\\
        \hline
        He$^{+}$/H$^{+}$ & 0.10$\pm$0.03 & 0.11$\pm$0.08& 0.11$\pm$0.40 \\
        He$^{2+}$/H$^{+}$$\times$10$^{-4}$& 8.30$\pm$0.50 & 9.33$\pm$0.63 & 7.10$\pm$1.03 \\
        C$^{0}$/H$^{+}$$\times$10$^{-7}$ & 4.77$\pm$1.10 & 3.08$\pm$1.09& 10.40$\pm$4.77\\
         N$^{0}$/H$^{+}$$\times$10$^{-6}$ & 3.18$\pm$1.20 & 1.55$\pm$0.67 & 8.13$\pm$4.70\\
        N$^{+}$/H$^{+}$$\times$10$^{-5}$  & 6.30$\pm$1.50 & 3.55$\pm$1.00 & 9.20$\pm$4.20 \\
        O$^{0}$/H$^{+}$$\times$10$^{-5}$ & 3.30$\pm$1.00 & 2.06$\pm$0.95& 6.75$\pm$3.40\\
        O$^{+}$/H$^{+}$$\times$10$^{-4}$ & 1.60$\pm$0.90 & 0.59$\pm$0.20& 2.95$\pm$0.95\\
        O$^{2+}$/H$^{+}$$\times$10$^{-4}$& 7.40$\pm$1.50 & 5.07$\pm$1.60 & 3.40$\pm$0.90 \\
        Ne$^{2+}$/H$^{+}$$\times$10$^{-4}$ & 2.20$\pm$0.30 & 1.90$\pm$0.30 & 1.60$\pm$0.60 \\
        Ar$^{2+}$/H$^{+}$$\times$10$^{-6}$ & 3.88$\pm$0.28 & 2.18$\pm$0.90 & 2.55$\pm$0.80\\
        Ar$^{3+}$/H$^{+}$$\times$10$^{-7}$ & 6.06$\pm$0.85 & 8.70$\pm$2.30 & 2.19$\pm$1.10\\
        S$^{+}$/H$^{+}$$\times$10$^{-6}$ & 2.07$\pm$0.47 & 1.14$\pm$0.67 & 4.00$\pm$1.50\\
        S$^{2+}$/H$^{+}$$\times$10$^{-6}$ & 12.3$\pm$1.40 & 6.68$\pm$0.21 & 8.23$\pm$3.50\\
        Cl$^{+}$/H$^{+}$$\times$10$^{-8}$ & 2.30$\pm$0.24 & 1.19$\pm$0.50 & 4.37$\pm$1.30\\
        Cl$^{2+}$/H$^{+}$$\times$10$^{-7}$ & 1.40$\pm$0.24 & 1.13$\pm$0.54 & 0.84$\pm$0.32\\
        Cl$^{3+}$/H$^{+}$$\times$10$^{-8}$ & 2.64$\pm$0.22 & 3.66$\pm$1.01 & 1.85$\pm$0.50\\
        Fe$^{+}$/H$^{+}$$\times$10$^{-8}$ & 3.50$\pm$1.10 &- & 5.76$\pm$1.20\\
        Fe$^{2+}$/H$^{+}$$\times$10$^{-8}$ & 5.00$\pm$1.90 & 3.18$\pm$1.10 & 3.35$\pm$2.10\\
        Ni$^{+}$/H$^{+}$$\times$10$^{-9}$ & 1.42$\pm$0.50 & - & 4.20$\pm$2.10\\
        \hline
        \multicolumn{4}{|c|}{Total abundances (ICF)}\\
        \hline
        He/H & 0.11$\pm$0.01 & 0.11$\pm$0.08 & 0.11$\pm$0.09 \\
        N/H$\times$10$^{-4}$ & 5.05$\pm$1.50 (3.47) & 4.86$\pm$1.0 (13.8) & 2.29$\pm$0.50 (2.49) \\
        O/H$\times$10$^{-4}$ & 9.14$\pm$3.50 (1.0) & 6.0$\pm$1.60 (1.0)& 6.37$\pm$2.0 (0.9)\\
        Ne/H$\times$10$^{-4}$ & 2.91$\pm$0.35 (1.32) & 2.28$\pm$0.30 (1.22)& 3.11$\pm$0.60(2.0) \\
        Ar/H$\times$10$^{-6}$ & 5.05$\pm$0.40 (1.13) &3.10$\pm$1.10 (1.02) & 2.61 $\pm$0.90(1.03)\\
        S/H$\times$10$^{-6}$ & 8.64$\pm$2.20(1.12) & 11.7$\pm$0.90 (1.08)& 11.8$\pm$5.0(1.0)\\
        Cl/H$\times$ 10$^{-7}$ & 2.08$\pm$0.26 (1.25)& 1.85$\pm$0.64 (1.23) & 1.10$\pm$0.80(1.31) \\
        Fe/H$\times$ 10$^{-8}$& 8.32$\pm$3.00 & 23.00$\pm$9.50$^{\dag}$ & 9.11$\pm$3.30\\
        Ni/H$^{\dag\dag}$$\times$10$^{-9}$& 2.09$\pm$0.50  & - & 4.20$\pm$2.10\\
        \hline
        
    \end{tabular}

    }
 \tablefoot{
   \tablefoottext {${\dag}$}{ The value of the Fe/H in K1b is an upper limit and not an absolute value (see the analysis in Sect.\ref{elem_abund_sect}).} \\
   \tablefoottext {${\dag\dag}$}{ The Ni/H abundance is the lower limit, since Ni$^{2+}$/H$^{+}$, which was not measured here, is expected to be non-negligible.}
   
  }

\end{table}

\subsection{\hmolec~emission}
\label{H2}

We report the detection of nine \hmolec~emission lines in the K1a LIS of NGC7009 (Table \ref{table_H2}). These lines arise from the 1-0 and 2-1 vibrational transition states and span several rotational levels. Seven of them are new detections, while the 1-0 S(1) and 2-1 S(1) had been previously identified through narrowband imaging \citep{Akras2020}. In K1b and K4, five and seven \hmolec~lines are identified, respectively.

\begin{table}[h!]
\centering
\caption{\hmolec~emission-line fluxes}
\label{table_H2}

    \begin{tabular}{|lcccc|}  
        \hline
        & & K1a & K1b & K4 \\
        \hline
        $\lambda$ ($\mu$m)& Transition& \multicolumn{3}{c|}{F(\hmolec) ($\times$ 10$^{-16}$ergs\,s$^{-1}$\,cm$^{-2}$)} \\
        \hline
      1.96  &1-0 S(3) & 3.91$\pm$0.40& 0.71$\pm$0.11 & 0.68$\pm$0.15\\
        2.03&1-0 S(2) & 1.94$\pm$0.34& -& 4.02$\pm$0.27\\
       2.12& 1-0 S(1) & 7.41$\pm$0.54 & 2.27$\pm$0.25& 9.55$\pm$0.25\\
        2.22&1-0 S(0) & 1.85$\pm$0.57& $-$ & 2.30$\pm$0.43\\
        2.24&2-1 S(1) & 0.70$\pm$0.22& 0.29$\pm$0.08& 1.20$\pm$0.60 \\
        2.40&1-0 Q(1) & 6.85$\pm$1.40& -& 5.70$\pm$1.08\\
        2.41&1-0 Q(2) & 1.50$\pm$1.00& -& 3.20$\pm$2.00\\
        2.42&1-0 Q(3) & 7.05$\pm$1.22& -& 2.02$\pm$0.60\\
       2.45& 1-0 Q(5) & 3.44$\pm$0.80& 1.80$\pm$1.10& 2.70$\pm$0.65\\
        
        \hline
        
    \end{tabular}

\end{table}

\subsubsection{\hmolec~excitation plot and molecular mass}

\begin{figure}[h!]
     \centering
        \begin{subfigure}           {2.0\linewidth}
            \includegraphics[width =0.5\textwidth]{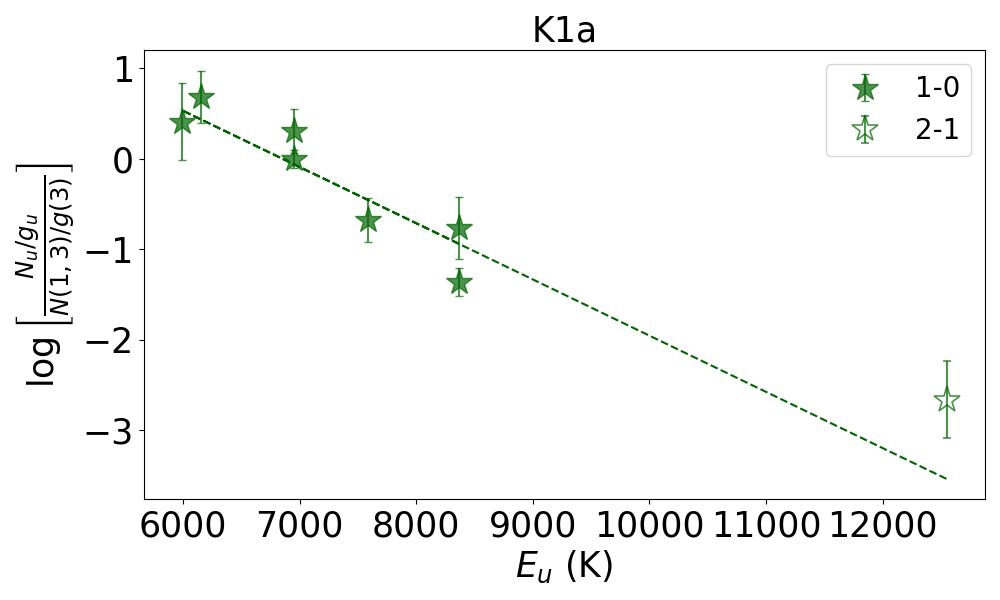}\\
            \includegraphics[width =0.5\textwidth]{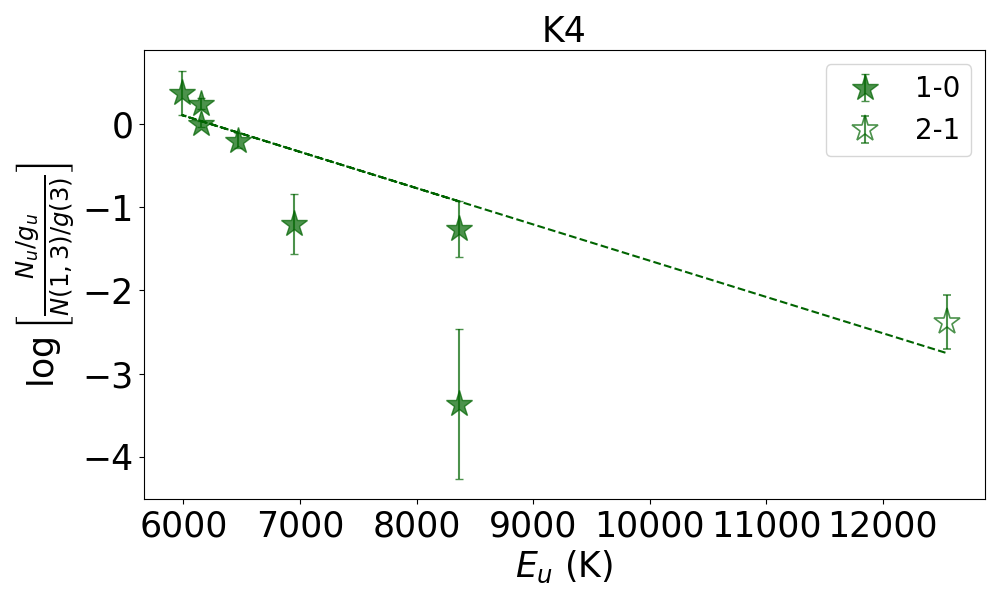}
        \end{subfigure}
     \caption{ Excitation diagram of \hmolec~for K1a (upper panel) and K4 (bottom panel). Filled stars indicate the 1-0 \hmolec~transitions, while the empty ones the indicate 2-1 \hmolec~transitions}
     \label{excitation_plot}
\end{figure}

An excitation diagram represents the column density (N$_u$(J)) divided by its statistical weight (g$_u$(J)), versus the energy of the upper level (E$_u$(K)). Given that most of the \hmolec~lines were identified in the pair of  LISs (K1a and K4), the following analysis focuses on these regions. The \hmolec~1-0 Q(2) lies significantly off the excitation diagram relative to the other \hmolec\ lines and was therefore excluded from the analysis. To estimate N$_u$(J), where J is the upper level, we employed the measured fluxes of \hmolec~(Table \ref{table_H2}) and the transition probabilities from \citet{Wolniewicz_1998}. The energy of the upper level E$_u$(K) was taken from \citet{Roueff2019}. The resulting excitation plots for K1a and K4 are presented in Figure \ref{excitation_plot}. Under local thermodynamic equilibrium (LTE) conditions, the level populations are fitted by a straight line following a Boltzmann distribution. The slope of this line is inversely proportional to the excitation temperature ($T_{\rm ex}$) of \hmolec~\citep[e.g.,][]{Likkel_2006}, while the intercept is a measure of the total column density of \hmolec~in the gas. The excitation temperature in K1a LIS is estimated $T_{\rm ex}$ = (2\,000 $\pm$ 300)~K, while for K4 $T_{\rm ex}$ = (2\,300 $\pm$ 250)~K. These relatively high excitation temperatures, as well as the LTE conditions, strongly support a scenario in which the gas has been collisionally excited \citep{Davis_2003, Hora_1999}. It has to be noted, though, that the nonthermal pattern, which is a characteristic of the UV excitation, emerges only when including levels at vibrational states higher than three \citep{Aleman_2011}, which is not the case here.

By summing up the apparent column densities of all the detected ro-vibrational and higher pure rotational lines, we obtain the first estimate for the total column density of \hmolec~in LISs, besides the cometary knots in the Helix \citep{Matsuura_2007}. The total column density of \hmolec~is N(\hmolec) = (5.90 $\pm$ 2.20)~$\times$~10$^{16}$ cm$^{-2}$ and (4.62 $\pm$ 1.20)~$\times$~10$^{16}$ cm$^{-2}$, for K1a and K4, respectively, and those are translated to \hmolec~mass of approximately (0.94~$\pm$ 0.50)~$\times$~10$^{-7}$ M$_{\sun}$ for K1a and (2.95~$\pm$~1.75)~$\times$~10$^{-7}$ M$_{\sun}$ for K4, considering an angular diameter of 2$^{\prime\prime}$ and 4$^{\prime\prime}$, for K1a and K4, respectively. The relation used for the estimation of \hmolec~mass is strongly weighted toward the warm molecular component of \hmolec, which is predominantly associated with PDRs \citep{BernardSalas2005}.

Recently, \citet{mari2025}, estimated the \hmolec~mass in the LISs of NGC~7009, using a method based on the 1-0 S(1) emission-line flux from the narrowband imaging study of \citet{Akras2020}, obtaining  three to four times higher values (2.9~$\times$~10$^{-7}$~M$_{\sun}$ and 6.2~$\times$~10$^{-7}$~M$_{\sun}$, for K1a and K4, respectively). The discrepancies between their estimates and ours, which are based on the \hmolec~column densities, can be attributed to the higher \hmolec~1-0 S(1) line fluxes adopted by the previous study. These higher fluxes naturally lead to higher \hmolec~mass estimates.
We also derived the mass of atomic hydrogen to be (8.00~$\pm$ 2.50)$\times$~10$^{-5}$ M$_{\sun}$ for K1a and (3.25~$\pm$ 2.00)$\times$~10$^{-5}$ M$_{\sun}$ for K4, consistent within the errors with \citet{mari2025} estimation. The molecular-to-atomic H mass ratio is 100 to 800 for K1a and K4, respectively.
This may suggest that \hmolec~occupies only a small fraction of the LISs volume, so its total mass is low, or that a significant reservoir of cold \hmolec~is present compared to the warm \hmolec~gas.

\section{Discussion}
\label{Discussion}
In this section, we further discuss the results of our analysis concerning the physico-chemical properties of NGC~7009. We investigate the dominant excitation mechanisms responsible for the observed \hmolec~emission, as well as those governing the emission from the LISs themselves. Finally, we examine whether \ hmolec~originated in situ within the LISs or if instead it represents material that has survived from earlier evolutionary stages of the progenitor star.

\subsection{LISs excitation traced by the atomic and ionic emission lines}
\label{carbon_H2}

The dominant excitation mechanism driving LISs in PNe remains one of the key unresolved questions in the field. Progress toward answering this problem can be achieved by employing diagnostic tools that probe the physical processes governing the excitation of the different ionic species observed within these structures.
Here, we focus on less-investigated emission lines of \carboni, \ironii~or \nickelii, which were very recently reported in nebular clumps \citep{Akras_2024, Akras_2024_Fe2, Bouvis_2025, Konstantinou_2025}. The detection of these transitions has provided a new perspective on the physical properties of the gas within LISs as well as their excitation processes.

At sufficiently high shock wave speeds, partial destruction of dust grains can occur, liberating refractory elements such as Fe and C into the gas phase. These freshly released atoms can then be collisionally excited in a hot gas \citep{Giannini2004}, giving rise to the observed forbidden lines. In a recent study by \citet{Akras_2024}, the far-red line of \carboni~at 8727\AA~was reported for the first time in the LISs of NGC~7009, using IFU MUSE data. In the present study, we confirm this detection spectroscopically.

The emission of \carboni~lines can arise from collisional or radiative recombination excitation processes \citep{Liu_1995}. Radiative recombination usually populates permitted transitions, but in low-temperature environments, it is possible that it populates metastable levels, leading to the emission of forbidden lines. The diagnostic ratio R(\carboni) = $\lambda$9824+$\lambda$9850/$\lambda$8727 can help distinguish among these mechanisms. Ratios around five are predicted from pure radiative recombination; values exceeding ten indicate collisional excitation, while intermediate values between five and ten suggest a combination of mechanisms. Here, we derive R(\carboni) = 16.6 $\pm$ 0.5, 21.5 $\pm$ 8.6, and 14.5 $\pm$ 2.0 for K1a, K1b, and K4, respectively, indicating that collisional excitation is the dominant mechanism. Similarly, in NGC 2440, where R(\carboni) has been found to be 9.5, the excitation was mainly attributed to collisions, but with a possible contribution of radiative recombination \citep{Liu_1995}.

Additionally, atomic carbon is often considered a reliable tracer of \hmolec, since photons with energies sufficient to dissociate \hmolec~can also ionize neutral carbon (C$^0$) and oxygen (O$^0$). Indeed, \carboni~and \hmolec, frequently coexist in PDRs associated with H{\sc ii} regions \citep[e.g.,][]{Henney_2024}. While such correlated emission is typically attributed to recombination-dominated processes in dense PDRs, the case of NGC~7009 LISs shows that \carboni\ and \hmolec\ may also coexist under shock conditions (see the following analysis and Sect. \ref{Diagnostic ratios_H2}). This finding supports the scenario in which \carboni~and \hmolec~are both excited by the same mechanism,  with \carboni\ potentially originating in shocks analogous to \hmolec. 

J-type, C-type shocks or J-type with a magnetic precursor are also suggested as viable mechanisms for generating \carboni~emission \citep[e.g., the case of HH99 from][]{McCoey_2004}. However, shock excitation is not restricted to purely dynamical scenarios. Even in the case of photoevaporation, a low-velocity shock may arise from the interaction between the photoevaporated flow and the stellar wind, potentially leading to the release of refractory elements such as Fe and C into the gas phase \citep{Mallema1998, Akras_2024}.

To investigate the \ironii~excitation properties, we compared the observed ratios of \ironii~$\lambda$7155/$\lambda$8617 (the $\lambda$7453 line was not detected) and \ironii~$\lambda$4815/$\lambda$8617, with theoretical predictions for UV and collisional excitation scenarios \citep[see][]{Bautista_1998, Giannini_2015}. For the first ratio, we found values of around 0.66~$\pm$~0.15 and 0.86~$\pm$~0.20 for K1a and K4, respectively, consistent with predictions for pure collisional excitation. The second ratio was estimated only for K1a due to the absence of the \ironii~$\lambda$4815 at K4. The \ironii~$\lambda$4815/$\lambda$8617 emission-line ratio was found to be $\approx$~0.40, lying in a regime indicative of a mixed excitation scenario, where fluorescence contributes alongside collisions. These results indicate that fluorescence plays a secondary role, enhancing the Fe emission, while collisions still dominate \citep{Kirwan_2023}. 

Photoionization and shock models are employed in a handful of studies to construct diagnostic diagrams aimed at distinguishing shock-excited from photoionized regions in planetary nebulae \citep[e.g.,][]{Belen_2023, Cala_2024,Cala_2025a, Cala_2025b, Bouvis_2025}. These studies considered {\sc cloudy} photoionization models of planetary nebulae \citep{DIMS} from the Mexican Million Models database \citep[3Mdb,][]{Morisset_2015}, and {\sc mappings} shock models \citep{Allen_2008, Dopita_2016, Sutherland_2017, alarie2019} available in the 3Mdb$^s$ database \citep{Alarie_2019}. The two model grids independently represent different excitation mechanisms, rather than models that simultaneously involve both shocks and photoionization. Of particular interest is the work of \cite{Bouvis_2025}, who applied this approach to diagnostics involving the ratios \ironii/\ha~and \oxygeni/\ha~(see their section 5.5 for further analysis on the chosen models). The emission-line ratios estimated here for NGC~7009 LISs place them in a region where slow shocks are responsible for exciting both \ironii~and \oxygeni. Our result is consistent with the findings of \cite{Bouvis_2025}, who employed MUSE data of the same object (see their fig. 16). So, the collisional excitation of \ironii~is mainly attributed to slow shocks propagating through the LISs of NGC~7009.

The emission of \nickelii~is also an important shock indicator. As was found recently, the \nickelii/\ha~versus \ironii/\ha~emission-line ratios efficiently separate shock-excited sources and photoionized nebulae \citep[see fig. 15 in][]{Bouvis_2025}. In NGC~7009 LISs, the observed \nickelii/\ha~and \ironii/\ha~line ratios are significantly below the upper limit of UV photoionization-dominated, as defined in \citet{Bouvis_2025}. However, this separation is based on observational data from predominantly shock-excited sources, such as supernova remnants, and purely photoionized nebulae. Therefore, slow shocks, such as those we suggest are present at NGC~7009 LISs (see  Section \ref{Comparison with PDR and shock models}), may overlap with the locus of photoionized regions. Without dedicated models, there is no robust way to separate the photoionized nebulae from the slow shocks based on \nickelii/\ha~and \ironii/\ha~emission-line ratios.

Overall, our findings suggest that the \ironii, \nickelii~and \oxygeni~lines are shock-excited. Given that these three species share comparable ionization potentials, we expect them to coexist at similar optical depths within the LISs. The presence of multiple shock tracers in the same region further supports a scenario in which shocks -likely slow, non-dissociative ones- play a significant role in the excitation and physical conditions of the LISs. 
\cite{Guerrero2013} suggested that the outer tips of the LISs in NGC~7009 are characterized by enhanced \oxygeniii/\ha\ emission, interpreted as the aftermath of bow shocks produced by fast collimated outflows. However, this enhancement is located in a different region of the LISs than the shock tracers identified in the present study.

\subsection{Chemical composition and excitation properties}

For the estimation of LISs' abundances, the prescription for photoionization was used to avoid assumptions regarding the relative contribution of shocks and photoionization. However, the presence of shocks may affect the abundance estimations following the standard nebular diagnostics. In this study, we propose, though, that shock contribution is important for emission lines such as \hmolec, \ironii~or \oxygeni, while, as shown by \citet{Guerrero2013}, shocks may also alter the \oxygeniii~emission at the tips of the LISs. 
Nevertheless, no significant increase in the \Te~(\oxygeniii)~ between the LISs and the PN has been found, as would be expected if shocks were dominating the \oxygeniii~emission. More specifically, \citet{Fang_2011} reported, from long-slit data, \Te~(\oxygeniii)~of 10\,900K, which generally is in good agreement, within the errors, with the values reported here for K1b and K4, but slightly higher compared to K1a~\Te. Moreover, the MUSE \Te~maps reported by \citet{Walsh_2018} reveal a slight increase in the  \Te~(\sulfuriii)~at the LISs location, which, however, is well within the errors again for K1b and K4. Therefore, we cannot exclude that shocks introduce some bias in CEL-based diagnostics through local temperature enhancements. Nevertheless, the comparable abundances derived for the LISs and those of the main nebular shell suggest that any such effect is not dominant at the level of our analysis.

The agreement between LIS abundances from CELs and those obtained from previous long-slit studies of this PN also suggests chemical homogeneity throughout the PN structures \citep{Goncalves_2006, Akras2015, Mari2023b}. This result provides additional evidence that the LISs of NGC~7009 were formed during the asymptotic giant branch (AGB) phase, similarly to the bulk of the nebula \citep[e.g.,][]{Concalves2001}.

A long-standing issue in the chemical analysis of ionized nebulae is that abundances from ORLs systematically exceed those obtained from CELs. This difference is quantified by the abundance discrepancy factor (ADF), defined as the ratio of the ionic abundance of a given element derived from ORLs to that derived from CELs. Several mechanisms are proposed to explain that discrepancy, but yet it is not completely understood  \citep{Peimbert_1980, Peimbert_1990,Liu_2000, Tsamis_2005, Nicholls_2012}.

Generally, the largest ADFs occur in nebulae with short-period binary central stars \citep{Corradi_2015, Jones_2016, Wesson_2018}. Most accurately, ADF can be determined for O$^{2+}$, since both ORLs and CELs of this ion are visible in the optical spectrum of nebulae. NGC~7009 is not considered a highly ADF PN, since it has an ADF between three and five \citep{Fang_2013, Liu_1995}. Here, focusing only on its LISs, we report ADF values for O/H of 2.5, 1.8, and 2.5 for K1a, K1b, and K4, respectively.

\subsection{The excitation of \hmolec~in the LISs}
\label{Diagnostic ratios_H2}
The excitation of \hmolec~can be attributed either to UV pumping or collisions in shocks or in illuminated dense gas. Another possibility involves X-ray excitation of \hmolec; however, this mechanism would require hard X-rays energetic enough to penetrate areas that are otherwise opaque to UV photons \citep{Tine_1997, Maloney1996}. In the case of LISs embedded within PNe, the X-ray contribution seems to be the least probable scenario, since in NGC~7009 the detected X-ray emission is restricted to the central regions of the nebula and is not associated with the locations of the LISs \citep{Guerrero2002}. Therefore, in the following discussion, we will focus on discriminating between the other two mechanisms.

A widely used diagnostic to examine the excitation mechanism of \hmolec~is the intensity ratio \hmolec~1–0 S(1)/2–1 S(1). In photoionized regions, the R(\hmolec) is expected to be low, ranging in most cases from 2-10, reflecting a cold gas that is not in LTE conditions. In the case of pure fluorescence UV excitation, the ratio is close to two \citep{Black1987}. On the other hand, if R(\hmolec) $>$ 10, then the excitation of \hmolec~is usually attributed to collisions through shocks \citep{Burton_1989}.

In the present study, we report R(\hmolec) = 10.60 $\pm$ 3.5, 7.83 $\pm$ 2.32 and 7.85 $\pm$ 2.43 for K1a, K1b, and K4 respectively.  This ratio
was previously estimated for the LISs of NGC~7009, through near-IR narrowband images \citep{Akras2020}, revealing that R(\hmolec) is higher at K1a compared to K4. In the same study, the \hmolec~excitation at K1a was attributed to a combination of UV pumping in a dense gas and collisions due to shocks, while for K4, UV excitation was considered the most probable mechanism. The ratios obtained from our work for K1a and K1b are in reasonable agreement with the ones estimated by \citet{Akras2020} (9.76 and 3.5 for K1a and K1b, respectively). However, for K4, R(\hmolec) is 2.3 times higher than the previous estimation (R(\hmolec)=3.5). Our value is closer to the value measured for K1a, as is expected assuming a common origin for the pair of LISs.
The high R(\hmolec) at K1a strongly supports the collisional excitation of \hmolec~in the gas. A similarly elevated ratio would be expected for K4 (R(\hmolec)$\approx$10), following the same trend as its counterpart K1a. It should be noted, though, that only a small fraction of K4 is encompassed by the slit, which may lead to an underestimation of the measured \hmolec~fluxes. 

Nevertheless, at high densities (up to 10$^5$ cm$^{-3}$), collisional de-excitation of FUV-pumped molecules can redistribute the populations of the lower-lying energy levels, leading to higher R(\hmolec) values \citep{Burton_1992, Sternberg_1989, Hollenbach_1995}. In such cases, it becomes challenging to determine whether the \hmolec~emission primarily originates from dense gas irradiated by UV photons or from thermal processes. Following the relation provided by \cite{Bohlin_1978} with an extinction coefficient of 0.16, corresponding to E(B-V)=0.11 mag and a resulting \hmolec~column density of 3.15~$\times$~10$^{20}$ cm$^{-2}$, we estimate a gas density of 3.5~$\times$~10${^4}$ cm$^{-3}$ for K1a and K4. This density is one order of magnitude smaller than the density needed for UV excitation to resemble the values of shock \hmolec~excitation.

Moreover, the  1-0 S(2)/1-0 S(0) \hmolec~ratio is another diagnostic to distinguish between the UV excitation in a dense gas (thermal UV excitation) and the shock excitation of \hmolec. We measure values of 1.05 $\pm$ 0.40 and 1.20 $\pm$ 0.24 for K1a and K4.  \citet{Mazzalay_2013} constructed a plot that includes the 2–1 S(1)/1–0 S(1) versus 1–0 S(2)/1–0 S(0) line-ratio diagnostics for selected regions within galaxies, along with predictions from different excitation models. When our values are placed on this diagram, the data points fall within the region where shock excitation dominates the emission.

The aforementioned ratios are also sensitive to the vibrational (1–0 S(1)/2–1 S(1)) and rotational (1-0 S(2)/1-0 S(0)) temperatures of \hmolec~\citep[e.g.,][]{Reunanen_2003}. Encompassing the equations (1) and (2) from \citet{Mazzalay_2013} we estimate T$_{vib}$(K1a) = (2\,200 $\pm$ 280)~K, T$_{rot}$(K1a) = (1\,030 $\pm$ 355)~K and T$_{vib}$(K4) = (2\,400 $\pm$ 300)~K, T$_{rot}$(K4) =(1\,160 $\pm$ 250)~K. Within the uncertainties, $T_{\rm rot}$ and $T_{\rm vib}$ are in good agreement in both K1a and K4. However, the higher $T_{\rm vib}$ compared to $T_{\rm rot}$ suggests a minor contribution from UV pumping \citep{Davis_2003}.

Finally, another diagnostic commonly used is the R(Br$\gamma$) versus R(\hmolec). In fig.6 of  \citet{Akras2020}, our points lie in the region where \hmolec~dominates the emission and shock excitation is the primary mechanism responsible for the observed \hmolec~emission lines.

\subsection{Comparison with PDR and shock models}
\label{Comparison with PDR and shock models}

To further investigate whether \hmolec~is excited inside a dense medium resembling a PDR or under shock conditions, we compared, for the pair of LISs (K1a and K4) the observed intensities for the \hmolec~lines at 2.12 and 2.24 $\mu$m with those found from \citep{Burton_1992} employing PDR, C-type, and J-type shock models. We adopted a luminosity of the central star L$_{*}$ = (3.8~$\pm$~0.9)$\times$~10$^3$L$_{\sun}$ at a distance of 1\,230~pc \citep{Bailer_2021} based on GAIA parallaxes \citep{GAIA_2016, GAIA_2021, GAIA_2023}. For comparison, \citet{Sabbadin_2004} estimated L$_{*}$ = 5.0~$\times$~10$^3$L$_{\sun}$ at a distance of 1\,400~pc and \citet{Mendez_1992} found L$_{*}$ = 9.3~$\times$~10$^3$L$_{\sun}$ considering a distance of 2\,100 pc. We then estimate a UV radiation field to be G$_0$ = 1\,400~$\pm$~600 for K1a and G$_0$ = 1\,100~$\pm$~340 for K4. Given the radiation field, together with the \hmolec~line intensities and intensity ratios, none of the PDR models seems to match our observations (see the upper left panel in Fig.\ref{Burton_models}).

To further explore this, we analyzed the constant-density grid of PDR models provided by version 7.1 of the Meudon PDR code \citep{Le_Petit_2006}, with model outputs accessed via the ISM Model Database (ISMDB). As observational constraint parameters, we used the emission-line intensities of the \hmolec~lines. We considered that the observer
sees the surface of the LISs perpendicularly, following the geometrical definition of the Meudon PDR Code. Under this configuration, the radiation field from the CS illuminates the surface of the LISs at 84 degrees, which is the inclination angle of the LISs relative to the plane of the sky.  However, such an inclination is not available in the current PDR model grid. Therefore, for the present analysis, we considered the highest available angle in the grid of models, which is 60 degrees. It has to be noted, though, that the reported results should be treated as upper limits rather than absolute estimates.  As the angle between the radiation source and the LIS surfaces increases, the effective incident radiation field is expected to decrease further. 
The best-fitting grid of models gives a radiation field, G$_0$, of 350 and 10 for K1a and K4, respectively. Both values are 2-3 orders of magnitude lower than the ones derived here. The n$_{H}$ is one order of magnitude lower for K1a, while for K4 there is better agreement. However, in a PDR with n$_H$ < 10$^5$ cm$^{-3}$, we would expect a value of R(\hmolec) close to 2. In contrast, to explain the observed ratio of R(\hmolec)$\approx$10 in a PDR framework, n$_H$ $>$ 10$^5$ cm$^{-3}$ and G$_0$ $>$ 10$^5$ are required \citep{Le_Petit_2006}, which are inconsistent with the conditions derived here.

Considering the relatively high $T_{\rm ex}$ and the reported \hmolec~line ratios (see Sect. \ref{Diagnostic ratios_H2}), together with the shock models of \citet{Burton_1992}, shocks seem to be the most plausible mechanism for the excitation of \hmolec. In particular, J-shock models with low velocities ($\leq$~20~km\,s$^{-1}$) provide the best match to our observed line intensities, as shown in Fig.\ref{Burton_models}. However, C-type shocks with velocities between 15 and 25~km\,s$^{-1}$ remain a good alternative.  In addition to this, \citet{Nesterenok_2019} demonstrate that such C-type shock models yield column densities in the higher energy states of  \hmolec~that are comparable to those estimated from our excitation diagram.  Moreover, the collisional excitation of \hmolec~through shocks agrees with what was found for \carboni, while the detection of iron and nickel lines further supports the presence of a shocked gas within LISs.

\subsection{Lifetime of \hmolec~}
\label{lifetime}
By estimating the lifetime of \hmolec~within the LISs, we can add constraints on the harshness of the local environment, the degree to which \hmolec~is protected against dissociation, and on its re-formation. Moreover, determining the timescale over which \hmolec~has persisted will allow us to better understand its evolutionary history, as a function of the PN evolution.

The main \hmolec~destruction mechanisms are: (1) UV photodissociation, and (2) dissociation through collisions. These two processes provide insights into the relative importance of radiative and dynamical processes in governing the molecular content in LISs. However, the dissociation rate of \hmolec~through collisions depends on several parameters, such as ionization rate, density, and shock velocity, which cannot be fully constrained for LISs. Existing shock models \citep[e.g.,][]{Hollenbach_1980, Flower_2003}, show that destruction of \hmolec~through shocks is significant at shock velocities higher than 25~km\,s$^{-1}$, or high pre-shock densities ($>$~10$^5$~cm$^{-3}$). In the present case, lower shock velocities are invoked, producing peak temperatures insufficient to trigger the dissociation of \hmolec. We therefore do not consider collisional dissociation as a significant destruction pathway of \hmolec~at the LISs.

For the photodissociation of \hmolec~we estimated the dissociation rate, considering the self-shielding
of \hmolec~\citep[see eq.~37 of][]{Draine_1996} and the shielding due to dust. The adopted \hmolec~density is n$_{\hmolec}$=1.76~$\times$~10$^4$ cm$^{-3}$, following the approach described in \citep{Bohlin_1978}, assuming that hydrogen inside the LISs is predominantly in the form of \hmolec. Under these assumptions, we estimate R$_{diss}$ of 6.30~$\times$~10$^{-13}$ s$^{-1}$ and 4.75~$\times$~10$^{-13}$ s$^{-1}$ for K1a and K4, respectively, which lead to an average lifetime of \hmolec~of around (6.0~$\pm$1.0)$\times$~10$^{4}$ years for both LISs.
The angular expansion of K1a was estimated to be 0.016 arcsec/year by \cite{Liller_1966}, 0.028 arcsec/year by \cite{Fernandez_2004} and 0.023 arcsec/year by \cite{Rodriguez2007}. The latter study also reported an angular expansion of 0.034 arcsec/year for K4. Adopting these angular expansion rates for K1a, we derive kinematic ages of 1\,440~$\pm$~600 years, 820~$\pm$~260 years, and 1\,000~$\pm$~260 years, respectively. For K4, we estimate a kinematical age of 790~$\pm$~200 years. Overall, these estimates suggest a characteristic kinematical age of the LISs of approximately 1\,000 years.
Under the current radiation field, \hmolec~is expected to undergo significant photodissociation in the absence of shielding. We propose that \hmolec~is still present, since LISs were initially much denser, providing stronger self-shielding and allowing the re-formation of \hmolec~on dust grains (requiring densities of order 10$^6$ cm$^{-3}$ \citep{VanHoof2010}, consistent with proto-planetary nebula jets~\citep{Riera_2011, Balick_2020}). As the density decreased over time, re-formation stopped and only a fraction of the original \hmolec~remained, which now undergoes photoevaporation. At the lower densities inferred here, \hmolec~ re-formation is negligible. Consequently, we suggest that \hmolec~observed in NGC 7009 is a remnant of a denser phase of the LISs and has survived through shielding within the LISs.

In comparison, \cite{Matsuura_2009} derived an \hmolec~lifetime in the cometary knots of the Helix nebula that is significantly longer (of the order of 10$^7$ years). This discrepancy is likely a consequence of the different physical conditions between the cometary knots of the Helix and the LISs of NGC~7009, with the former being denser than the latter, providing stronger shelf-shielding to \hmolec. Moreover, the Helix nebula is at an advanced evolutionary state, and thus its central star is currently on the cooling track and therefore significantly less luminous, resulting in a weaker radiation field, G$_0$~=~8.

\section{Conclusions}
\label{Conclusions}

The results presented in this work are the first ones of a project devoted to developing a homogeneous optical-to-near-IR spectroscopic analysis dedicated to LISs in PNe. Exploiting the capabilities of the X-Shooter spectrograph, we obtained simultaneous observations covering a broad spectral range, from 300 nm to 2.5$\mu$m. In this first paper of the series, we focused on the NGC~7009 LISs. We detect $\approx$300 emission lines from the substructures K1a, K1b, and K4 with fluxes down to 10$^{-17}~$erg\,s$^{-1}$\,cm$^{-2}$. 

Emission-line diagnostics tracing both ionized and atomic gas were employed for the estimation of the nebular physical properties (\Te~\Ne). We find no significant discrepancy between these values and the values for the main nebula. The mean \Te~estimated from collisionally excited lines is around 9\,000 K, while \Te~derived from optical recombination lines drops to 5\,000 to 6\,000 K. Electron densities derived from \nitrogeni, which trace the atomic gas at the LISs, tend to be lower than those determined from other diagnostics, although the differences remain within the measurement uncertainties. 

Regarding chemical abundances at the LISs, no significant differences are found compared to the PN itself, suggesting that both the LISs and the main nebular components likely formed under the same physical conditions. When comparing the abundances of the LISs with those of other nebular components, a potential caveat is the shock excitation of low-ionization elements in the LISs such as Fe$^+$, Ni$^+$, and C$^0$. However, the effect of shocks is not expected to significantly affect the abundance estimates.

In the near-IR spectrum of the LISs, nine \hmolec~lines are detected, seven of them for the first time. The observed \hmolec~ratios employed here, together with the high excitation temperature of the molecule ($T_{\rm ex}$~=~2\,050~K and 2\,300~K for K1a and K4, respectively), as well as comparisons with shock and PDR models, strongly suggest that molecular hydrogen is predominantly excited through collisions driven by shocks.  

Moreover, the \carboni~lines ratio R(\carboni) = $\lambda$9824+$\lambda$9850/$\lambda$8727 indicates collisional excitation of atomic carbon at the LISs. This result, in conjunction with the shock excitation of \hmolec~and the known correlation between these lines, suggests that shock-driven collisional processes can explain the excitation of both \hmolec~and possibly the \carboni\ as well. Shock tracers such as \ironii~and \nickelii~are confirmed spectroscopically to be emitted from the LISs. Furthermore, diagnostic line ratios reveal that the excitation of \ironii\ and \oxygeni\ at the LISs is driven by slow shocks. These detections further support the existence of a shock environment at the LISs, leading to dust destruction and the subsequent release of refractory elements into the gas phase.

Finally, the lifetime of \hmolec, under the current nebular conditions, is estimated to be 5.0~$\times$~10$^4$ years for K4 and 6.6~$\times$~10$^4$ years for K1a. We propose that \hmolec~in NGC 7009 LISs is likely a relic of its denser past, preserved by self-shielding within the LISs, and is currently under slow photoevaporation.

\section{Data availability}
The derived data products generated for this project are available at the ESO Phase 3 with DOI: \url{https://doi.org/10.18727/archive/116}.

\begin{acknowledgements}
     We would like to thank the anonymous referee for their careful review and constructive comments, which helped improve the content and the clarity of our work. Based on observations collected at the European Organisation for Astronomical Research in
the Southern Hemisphere under ESO programme 113.268E.001 (PI: Akras)
     The research project is implemented in the framework of H.F.R.I call “Basic research financing (Horizontal support of all Sciences)” under the National Recovery and Resilience Plan “Greece 2.0” funded by the European Union–NextGenerationEU (H.F.R.I. Project Number: 15665).  JGR acknowledges support from the Agencia Estatal de Investigaci\'on of the Ministerio de Ciencia e Innovaci\'on  y Universidades (AEI-MCIU) and from the European Regional Development Fund (ERDF) under grant ``Planetary nebulae as the key to understanding binary stellar evolution'' with reference PID2022-136653NA-I00 (DOI:10.13039/501100011033) and financial support through the Severo Ochoa Centre of Excellence accreditation awarded to the Instituto de Astrofísica de Canarias, grant CEX2025-001609-S, funded by MICIU/AEI/10.13039/501100011033. DRG acknowledges the grants of the Brazilian agencies FAPERJ (E-26/200.467/2026) and CNPq (315307/2023-4). This work has made use of data from the European Space Agency (ESA) mission {\it Gaia} (\url{https://www.cosmos.esa.int/gaia}), processed by the {\it Gaia} Data Processing and Analysis Consortium (DPAC, \url{https://www.cosmos.esa.int/web/gaia/dpac/consortium}). Funding for the DPAC has been provided by national institutions, in particular the institutions participating in the {\it Gaia} Multilateral Agreement. This work made use of Astropy: \footnote{https://www.astropy.org} a community-developed core Python package and an ecosystem of tools and resources for astronomy \citep{astropy_2022}, Matplotlib \citep{Hunter_2007}, NumPy \citep{harris2020}, Pandas \citep{reback2020pandas} and PyNeb \citep[version 1.1.18,][]{Luridiana_2015, Morisset_2020}
\end{acknowledgements}

\bibliographystyle{aa} 
\bibliography{biblio} 

@ARTICLE{Balick1993a,
       author = {{Balick}, Bruce and {Rugers}, M. and {Terzian}, Yervant and {Chengalur}, Jayaram N.},
        title = "{Fast, Low-Ionization Emission Regions and Other Microstructures in Planetary Nebulae}",
      journal = {\apj},
         year = 1993,
        month = jul,
       volume = {411},
        pages = {778},
          doi = {10.1086/172881},
       adsurl = {https://ui.adsabs.harvard.edu/abs/1993ApJ...411..778B}
}

@ARTICLE{Perinotto2000,
       author = {{Perinotto}, Mario},
        title = "{Gas Dynamics in Planetary Nebulae: From Macro-structures to FLIERs}",
      journal = {\apss},
         year = 2000,
        month = oct,
       volume = {274},
        pages = {205-219},
          doi = {10.1023/A:1026516527132},
       adsurl = {https://ui.adsabs.harvard.edu/abs/2000Ap&SS.274..205P}
}

@ARTICLE{Corradi_1996,
       author = {{Corradi}, R.~L.~M. and {Manso}, R. and {Mampaso}, A. and {Schwarz}, H.~E.},
        title = "{Unveiling low-ionization microstructures in planetary nebulae.}",
      journal = {\aap},
         year = 1996,
        month = sep,
       volume = {313},
        pages = {913-923},
       adsurl = {https://ui.adsabs.harvard.edu/abs/1996A\&A...313..913C}
}

@ARTICLE{Concalves2001,
       author = {{Gon{\c{c}}alves}, Denise R. and {Corradi}, Romano L.~M. and {Mampaso}, Antonio},
        title = "{Low-Ionization Structures in Planetary Nebulae: Confronting Models with Observations}",
      journal = {\apj},
         year = 2001,
        month = jan,
       volume = {547},
       number = {1},
        pages = {302-310},
          doi = {10.1086/318364},
archivePrefix = {arXiv},
       eprint = {astro-ph/0010282},
 primaryClass = {astro-ph},
       adsurl = {https://ui.adsabs.harvard.edu/abs/2001ApJ...547..302G}
}

@article{Akras2015,
    author = {Akras, Stavros and Gonçalves, Denise R},
    title = "{Low-ionization structures in planetary nebulae – I. Physical, kinematic and excitation properties }",
    journal = {\mnras},
    volume = {455},
    number = {1},
    pages = {930-961},
    year = {2016},
    month = {11},
    issn = {0035-8711},
    doi = {10.1093/mnras/stv2139},
    url = {https://doi.org/10.1093/mnras/stv2139},
    eprint = {https://academic.oup.com/mnras/article-pdf/455/1/930/3058746/stv2139.pdf},
}

@ARTICLE{Belen_2023,
       author = {{Mari}, M. Bel{\'e}n and {Akras}, Stavros and {Gon{\c{c}}alves}, Denise R.},
        title = "{Low-ionization structures in planetary nebulae - III. The statistical analysis of physico-chemical parameters and excitation mechanisms}",
      journal = {\mnras},
         year = 2023,
        month = oct,
       volume = {525},
       number = {2},
        pages = {1998-2014},
          doi = {10.1093/mnras/stad2256},
archivePrefix = {arXiv},
       eprint = {2308.04577},
 primaryClass = {astro-ph.GA},
       adsurl = {https://ui.adsabs.harvard.edu/abs/2023MNRAS.525.1998M}
}

@ARTICLE{Akras_2024,
       author = {{Akras}, S. and {Monteiro}, H. and {Walsh}, J.~R. and {Konstantinou}, L. and {Gon{\c{c}}alves}, D.~R. and {Garcia-Rojas}, J. and {Boumis}, P. and {Aleman}, I.},
        title = "{Detection of the [C I] {\ensuremath{\lambda}}8727 emission line. Low-ionization structures in NGC 7009}",
      journal = {\aap},
         year = 2024,
        month = sep,
       volume = {689},
          eid = {A14},
        pages = {A14},
          doi = {10.1051/0004-6361/202450325},
archivePrefix = {arXiv},
       eprint = {2409.00371},
 primaryClass = {astro-ph.SR},
       adsurl = {https://ui.adsabs.harvard.edu/abs/2024A&A...689A..14A}
}

@ARTICLE{Akras_2024_Fe2,
       author = {{Akras}, S. and {Aleman}, I. and {Gon{\c{c}}alves}, D.~R. and {Ramos-Larios}, G. and {Bouvis}, K.},
        title = "{[Fe II] 1.644 {\textmu}m imaging survey of planetary nebulae with low-ionisation structures}",
      journal = {\aap},
         year = 2024,
        month = sep,
       volume = {689},
          eid = {A70},
        pages = {A70},
          doi = {10.1051/0004-6361/202449580},
archivePrefix = {arXiv},
       eprint = {2409.00377},
 primaryClass = {astro-ph.SR},
       adsurl = {https://ui.adsabs.harvard.edu/abs/2024A&A...689A..70A}
}

@ARTICLE{Fang2015,
       author = {{Fang}, X. and {Guerrero}, M.~A. and {Miranda}, L.~F. and {Riera}, A. and {Vel{\'a}zquez}, P.~F. and {Raga}, A.~C.},
        title = "{Hu 1-2: a metal-poor bipolar planetary nebula with fast collimated outflows}",
      journal = {\mnras},
         year = 2015,
        month = sep,
       volume = {452},
       number = {3},
        pages = {2445-2462},
          doi = {10.1093/mnras/stv1477},
archivePrefix = {arXiv},
       eprint = {1507.00284},
 primaryClass = {astro-ph.SR},
       adsurl = {https://ui.adsabs.harvard.edu/abs/2015MNRAS.452.2445F}
}

@ARTICLE{Akras2017,
       author = {{Akras}, Stavros and {Gon{\c{c}}alves}, Denise R. and {Ramos-Larios}, Gerardo},
        title = "{H$_{2}$ in low-ionization structures of planetary nebulae}",
      journal = {\mnras},
         year = 2017,
        month = feb,
       volume = {465},
       number = {2},
        pages = {1289-1296},
          doi = {10.1093/mnras/stw2736},
archivePrefix = {arXiv},
       eprint = {1611.06578},
 primaryClass = {astro-ph.SR},
       adsurl = {https://ui.adsabs.harvard.edu/abs/2017MNRAS.465.1289A}
}

@ARTICLE{Fang2018,
       author = {{Fang}, Xuan and {Zhang}, Yong and {Kwok}, Sun and {Hsia}, Chih-Hao and {Chau}, Wayne and {Ramos-Larios}, Gerardo and {Guerrero}, Mart{\'\i}n A.},
        title = "{Extended Structures of Planetary Nebulae Detected in H$_{2}$ Emission}",
      journal = {\apj},
         year = 2018,
        month = jun,
       volume = {859},
       number = {2},
          eid = {92},
        pages = {92},
          doi = {10.3847/1538-4357/aac01e},
archivePrefix = {arXiv},
       eprint = {1804.08840},
 primaryClass = {astro-ph.SR},
       adsurl = {https://ui.adsabs.harvard.edu/abs/2018ApJ...859...92F}
}

@ARTICLE{Akras2020,
       author = {{Akras}, Stavros and {Gon{\c{c}}alves}, Denise R. and {Ramos-Larios}, Gerardo and {Aleman}, Isabel},
        title = "{H$_{2}$ emission in the low-ionization structures of the planetary nebulae NGC 7009 and NGC 6543}",
      journal = {\mnras},
         year = 2020,
        month = apr,
       volume = {493},
       number = {3},
        pages = {3800-3810},
          doi = {10.1093/mnras/staa515},
archivePrefix = {arXiv},
       eprint = {2003.01737},
 primaryClass = {astro-ph.SR},
       adsurl = {https://ui.adsabs.harvard.edu/abs/2020MNRAS.493.3800A}
}

@ARTICLE{Sterling2008,
       author = {{Sterling}, N.~C. and {Dinerstein}, Harriet L.},
        title = "{The Abundances of Light Neutron-Capture Elements in Planetary Nebulae. II. s-Process Enrichments and Interpretation}",
      journal = {\apjs},
         year = 2008,
        month = jan,
       volume = {174},
       number = {1},
        pages = {158-201},
          doi = {10.1086/520845},
       adsurl = {https://ui.adsabs.harvard.edu/abs/2008ApJS..174..158S}
}

@ARTICLE{Herald2004,
       author = {{Herald}, J.~E. and {Bianchi}, L.},
        title = "{Central Stars of Planetary Nebulae in the Large Magellanic Cloud: A Far-Ultraviolet Spectroscopic Analysis}",
      journal = {\apj},
         year = 2004,
        month = aug,
       volume = {611},
       number = {1},
        pages = {294-314},
          doi = {10.1086/422133},
archivePrefix = {arXiv},
       eprint = {astro-ph/0404367},
 primaryClass = {astro-ph},
       adsurl = {https://ui.adsabs.harvard.edu/abs/2004ApJ...611..294H}
}

@ARTICLE{McCandliss2007,
       author = {{McCandliss}, Stephan R. and {France}, Kevin and {Lupu}, Roxana E. and {Burgh}, Eric B. and {Sembach}, Kenneth and {Kruk}, Jeffrey and {Andersson}, B. -G. and {Feldman}, Paul D.},
        title = "{Molecular and Atomic Excitation Stratification in the Outflow of the Planetary Nebula M27}",
      journal = {\apj},
         year = 2007,
        month = apr,
       volume = {659},
       number = {2},
        pages = {1291-1316},
          doi = {10.1086/512803},
archivePrefix = {arXiv},
       eprint = {astro-ph/0701439},
 primaryClass = {astro-ph},
       adsurl = {https://ui.adsabs.harvard.edu/abs/2007ApJ...659.1291M}
}

@ARTICLE{Hora1999,
       author = {{Hora}, Joseph L. and {Latter}, William B. and {Deutsch}, Lynne K.},
        title = "{Investigating the Near-Infrared Properties of Planetary Nebulae. II. Medium-Resolution Spectra}",
      journal = {\apjs},
         year = 1999,
        month = sep,
       volume = {124},
       number = {1},
        pages = {195-240},
          doi = {10.1086/313256},
archivePrefix = {arXiv},
       eprint = {astro-ph/9904202},
 primaryClass = {astro-ph},
       adsurl = {https://ui.adsabs.harvard.edu/abs/1999ApJS..124..195H}
}

@ARTICLE{Kaush2015,
       author = {{Kausch}, W. and {Noll}, S. and {Smette}, A. and {Kimeswenger}, S. and {Barden}, M. and {Szyszka}, C. and {Jones}, A.~M. and {Sana}, H. and {Horst}, H. and {Kerber}, F.},
        title = "{Molecfit: A general tool for telluric absorption correction. II. Quantitative evaluation on ESO-VLT/X-Shooterspectra}",
      journal = {\aap},
         year = 2015,
        month = apr,
       volume = {576},
          eid = {A78},
        pages = {A78},
          doi = {10.1051/0004-6361/201423909},
archivePrefix = {arXiv},
       eprint = {1501.07265},
 primaryClass = {astro-ph.IM},
       adsurl = {https://ui.adsabs.harvard.edu/abs/2015A&A...576A..78K}
}

@ARTICLE{Smette2015,
       author = {{Smette}, A. and {Sana}, H. and {Noll}, S. and {Horst}, H. and {Kausch}, W. and {Kimeswenger}, S. and {Barden}, M. and {Szyszka}, C. and {Jones}, A.~M. and {Gallenne}, A. and {Vinther}, J. and {Ballester}, P. and {Taylor}, J.},
        title = "{Molecfit: A general tool for telluric absorption correction. I. Method and application to ESO instruments}",
      journal = {\aap},
         year = 2015,
        month = apr,
       volume = {576},
          eid = {A77},
        pages = {A77},
          doi = {10.1051/0004-6361/201423932},
archivePrefix = {arXiv},
       eprint = {1501.07239},
 primaryClass = {astro-ph.IM},
       adsurl = {https://ui.adsabs.harvard.edu/abs/2015A&A...576A..77S}
}

@article{Rodriguez_2005,
doi = {10.1086/429958},
url = {https://dx.doi.org/10.1086/429958},
year = {2005},
month = {jun},
publisher = {},
volume = {626},
number = {2},
pages = {900},
author = {Rodríguez, Mónica and Rubin, Robert H.},
title = {The [Fe IV] Discrepancy: Constraining the Iron Abundances in Nebulae},
journal = {\apj}
}

@article{Tayal_2018,
  title = {Electron-impact excitation of forbidden and allowed transitions in Fe ii},
  author = {Tayal, S. S. and Zatsarinny, O.},
  journal = {Phys. Rev. A},
  volume = {98},
  issue = {1},
  pages = {012706},
  numpages = {17},
  year = {2018},
  month = {Jul},
  publisher = {American Physical Society},
  doi = {10.1103/PhysRevA.98.012706},
  url = {https://link.aps.org/doi/10.1103/PhysRevA.98.012706}
}

@ARTICLE{Chianty_2021,
       author = {{Del Zanna}, G. and {Dere}, K.~P. and {Young}, P.~R. and {Landi}, E.},
        title = "{CHIANTI{\textemdash}An Atomic Database for Emission Lines. XVI. Version 10, Further Extensions}",
      journal = {\apj},
         year = 2021,
        month = mar,
       volume = {909},
       number = {1},
          eid = {38},
        pages = {38},
          doi = {10.3847/1538-4357/abd8ce},
archivePrefix = {arXiv},
       eprint = {2011.05211},
 primaryClass = {physics.atom-ph},
       adsurl = {https://ui.adsabs.harvard.edu/abs/2021ApJ...909...38D}
}

@article{DIMS,
    author = {Delgado-Inglada, Gloria and Morisset, Christophe and Stasińska, Grażyna},
    title = {Ionization correction factors for planetary nebulae – I. Using optical spectra},
    journal = {\mnras},
    volume = {440},
    number = {1},
    pages = {536-554},
    year = {2014},
    month = {03},
    issn = {0035-8711},
    doi = {10.1093/mnras/stu341},
    url = {https://doi.org/10.1093/mnras/stu341},
    eprint = {https://academic.oup.com/mnras/article-pdf/440/1/536/9389303/stu341.pdf},
}

@article{KB,
    author = {Kingsburgh, Robin L. and Barlow, M. J.},
    title = {Elemental abundances for a sample of southern galctic planetary nebulae},
    journal = {\mnras},
    volume = {271},
    number = {2},
    pages = {257-299},
    year = {1994},
    month = {11},
    issn = {0035-8711},
    doi = {10.1093/mnras/271.2.257},
    url = {https://doi.org/10.1093/mnras/271.2.257},
    eprint = {https://academic.oup.com/mnras/article-pdf/271/2/257/3156084/mnras271-0257.pdf},
}

@ARTICLE{Kings_Barlow_1994,
       author = {{Kingsburgh}, R.~L. and {Barlow}, M.~J.},
        title = "{Elemental abundances for a sample of southern galactic planetary nebulae.}",
      journal = {\mnras},
         year = 1994,
        month = nov,
       volume = {271},
        pages = {257-299},
          doi = {10.1093/mnras/271.2.257},
       adsurl = {https://ui.adsabs.harvard.edu/abs/1994MNRAS.271..257K}
}

@ARTICLE{Mendez_1988,
       author = {{Mendez}, R.~H. and {Kudritzki}, R.~P. and {Herrero}, A. and {Husfeld}, D. and {Groth}, H.~G.},
        title = "{High resolution spectroscopy of central stars of planetary nebulae. I. Basic atmospheric parameters and their interpretation.}",
      journal = {\aap},
         year = 1988,
        month = jan,
       volume = {190},
        pages = {113-136},
       adsurl = {https://ui.adsabs.harvard.edu/abs/1988A&A...190..113M}
}

@ARTICLE{Mendez_1992,
       author = {{Mendez}, R.~H. and {Kudritzki}, R.~P. and {Herrero}, A.},
        title = "{On central star luminosities and optical thicknesses in planetary nebulae.}",
      journal = {\aap},
         year = 1992,
        month = jul,
       volume = {260},
        pages = {329-340},
       adsurl = {https://ui.adsabs.harvard.edu/abs/1992A&A...260..329M}
}

@ARTICLE{Luridiana_2015,
       author = {{Luridiana}, V. and {Morisset}, C. and {Shaw}, R.~A.},
        title = "{PyNeb: a new tool for analyzing emission lines. I. Code description and validation of results}",
      journal = {\aap},
         year = 2015,
        month = jan,
       volume = {573},
          eid = {A42},
        pages = {A42},
          doi = {10.1051/0004-6361/201323152},
archivePrefix = {arXiv},
       eprint = {1410.6662},
 primaryClass = {astro-ph.IM},
       adsurl = {https://ui.adsabs.harvard.edu/abs/2015A&A...573A..42L}
}

@ARTICLE{Morisset_2020,
       author = {{Morisset}, Christophe and {Luridiana}, Valentina and {Garc{\'\i}a-Rojas}, Jorge and {G{\'o}mez-Llanos}, Ver{\'o}nica and {Bautista}, Manuel and {Mendoza}, Claudio},
        title = "{Atomic Data Assessment with PyNeb}",
      journal = {Atoms},
         year = 2020,
        month = oct,
       volume = {8},
       number = {4},
          eid = {66},
        pages = {66},
          doi = {10.3390/atoms8040066},
archivePrefix = {arXiv},
       eprint = {2009.10586},
 primaryClass = {astro-ph.GA},
       adsurl = {https://ui.adsabs.harvard.edu/abs/2020Atoms...8...66M}
}

@article{Akras2022,
    author = {Akras, S and Monteiro, H and Walsh, J R and García-Rojas, J and Aleman, I and Boffin, H and Boumis, P and Chiotellis, A and Corradi, R M L and Gonçalves, D R and Gutiérrez-Soto, L A and Jones, D and Morisset, C and Papanikolaou, X},
    title = {Spectroscopic analysis tool for intEgraL fieLd unIt daTacubEs (satellite): case studies of NGC 7009 and NGC 6778 with MUSE},
    journal = {\mnras},
    volume = {512},
    number = {2},
    pages = {2202-2221},
    year = {2022},
    month = {03},
    issn = {0035-8711},
    doi = {10.1093/mnras/stac632},
    url = {https://doi.org/10.1093/mnras/stac632},
    eprint = {https://academic.oup.com/mnras/article-pdf/512/2/2202/43057153/stac632.pdf},
}

@ARTICLE{Walsh_2018,
       author = {{Walsh}, J.~R. and {Monreal-Ibero}, A. and {Barlow}, M.~J. and {Ueta}, T. and {Wesson}, R. and {Zijlstra}, A.~A. and {Kimeswenger}, S. and {Leal-Ferreira}, M.~L. and {Otsuka}, M.},
        title = "{An imaging spectroscopic survey of the planetary nebula NGC 7009 with MUSE★}",
      journal = {\aap},
         year = 2018,
        month = dec,
       volume = {620},
          eid = {A169},
        pages = {A169},
          doi = {10.1051/0004-6361/201833445},
archivePrefix = {arXiv},
       eprint = {1810.03984},
 primaryClass = {astro-ph.GA},
       adsurl = {https://ui.adsabs.harvard.edu/abs/2018A&A...620A.169W}
}

@article{Goncalves_2003,
doi = {10.1086/378496},
url = {https://dx.doi.org/10.1086/378496},
year = {2003},
month = {nov},
publisher = {},
volume = {597},
number = {2},
pages = {975},
author = {Gonçalves, Denise R. and Corradi, Romano L. M. and Mampaso, Antonio and Perinotto, Mario},
title = {The Physical Parameters, Excitation, and Chemistry of the Rim, Jets, and Knots of the Planetary Nebula NGC 7009*},
journal = {\apj}
}

@article{Fang_2011,
    author = {Fang, X. and Liu, X.-W.},
    title = {Very deep spectroscopy of the bright Saturn nebula NGC 7009 – I. Observations and plasma diagnostics},
    journal = {\mnras},
    volume = {415},
    number = {1},
    pages = {181-198},
    year = {2011},
    month = {07},
    issn = {0035-8711},
    doi = {10.1111/j.1365-2966.2011.18681.x},
    url = {https://doi.org/10.1111/j.1365-2966.2011.18681.x},
    eprint = {https://academic.oup.com/mnras/article-pdf/415/1/181/17328705/mnras0415-0181.pdf},
}

@ARTICLE{Konstantinou_2025,
       author = {{Konstantinou}, L. and {Akras}, S. and {Garcia-Rojas}, J. and {Bouvis}, K. and {Gon{\c{c}}alves}, D.~R. and {Monteiro}, H. and {Boumis}, P. and {Mari}, M.~B. and {Aleman}, I. and {Monreal-Ibero}, A.},
        title = "{Integral field spectroscopy of the planetary nebula NGC 3242 and the puzzling nature of its low-ionization structures}",
      journal = {\aap},
         year = 2025,
        month = may,
       volume = {697},
          eid = {A227},
        pages = {A227},
          doi = {10.1051/0004-6361/202453635},
archivePrefix = {arXiv},
       eprint = {2503.16662},
 primaryClass = {astro-ph.SR},
       adsurl = {https://ui.adsabs.harvard.edu/abs/2025A&A...697A.227K}
}

@ARTICLE{Fang_2013,
       author = {{Fang}, X. and {Liu}, X. -W.},
        title = "{Very deep spectroscopy of the bright Saturn nebula NGC 7009 - II. Analysis of the rich optical recombination spectrum}",
      journal = {\mnras},
         year = 2013,
        month = mar,
       volume = {429},
       number = {4},
        pages = {2791-2851},
          doi = {10.1093/mnras/sts414},
archivePrefix = {arXiv},
       eprint = {1212.0005},
 primaryClass = {astro-ph.SR},
       adsurl = {https://ui.adsabs.harvard.edu/abs/2013MNRAS.429.2791F}
}

@article{Liu_1995,
    author = {Liu, X.-W. and Barlow, M. J. and Danziger, I. J. and Clegg, R. E. S.},
    title = {Neutral carbon far-red forbidden line emission from planetary nubulae},
    journal = {\mnras},
    volume = {273},
    number = {1},
    pages = {47-58},
    year = {1995},
    month = {03},
    issn = {0035-8711},
    doi = {10.1093/mnras/273.1.47},
    url = {https://doi.org/10.1093/mnras/273.1.47},
    eprint = {https://academic.oup.com/mnras/article-pdf/273/1/47/3090652/mnras273-0047.pdf},
}

@ARTICLE{Bailer_2021,
       author = {{Bailer-Jones}, C.~A.~L. and {Rybizki}, J. and {Fouesneau}, M. and {Demleitner}, M. and {Andrae}, R.},
        title = "{Estimating Distances from Parallaxes. V. Geometric and Photogeometric Distances to 1.47 Billion Stars in Gaia Early Data Release 3}",
      journal = {\aj},
         year = 2021,
        month = mar,
       volume = {161},
       number = {3},
          eid = {147},
        pages = {147},
          doi = {10.3847/1538-3881/abd806},
archivePrefix = {arXiv},
       eprint = {2012.05220},
 primaryClass = {astro-ph.SR},
       adsurl = {https://ui.adsabs.harvard.edu/abs/2021AJ....161..147B}
}

@ARTICLE{Sternberg_1989,
       author = {{Sternberg}, A. and {Dalgarno}, A.},
        title = "{The Infrared Response of Molecular Hydrogen Gas to Ultraviolet Radiation: High-Density Regions}",
      journal = {\apj},
         year = 1989,
        month = mar,
       volume = {338},
        pages = {197},
          doi = {10.1086/167193},
       adsurl = {https://ui.adsabs.harvard.edu/abs/1989ApJ...338..197S}
}

@article{Matsuura_2007,
    author = {Matsuura, M. and Speck, A. K. and Smith, M. D. and Zijlstra, A. A. and Viti, S. and Lowe, K. T. E. and Redman, M. and Wareing, C. J. and Lagadec, E.},
    title = {VLT/near-infrared integral field spectrometer observations of molecular hydrogen lines in the knots of the planetary nebula NGC 7293 (the Helix Nebula)*},
    journal = {\mnras},
    volume = {382},
    number = {4},
    pages = {1447-1459},
    year = {2007},
    month = {11},
    issn = {0035-8711},
    doi = {10.1111/j.1365-2966.2007.12496.x},
    url = {https://doi.org/10.1111/j.1365-2966.2007.12496.x},
    eprint = {https://academic.oup.com/mnras/article-pdf/382/4/1447/3940990/mnras0382-1447.pdf},
}

@article{Matsuura_2009,
doi = {10.1088/0004-637X/700/2/1067},
url = {https://dx.doi.org/10.1088/0004-637X/700/2/1067},
year = {2009},
month = {jul},
publisher = {The American Astronomical Society},
volume = {700},
number = {2},
pages = {1067},
author = {Matsuura, M. and Speck, A. K. and McHunu, B. M. and Tanaka, I. and Wright, N. J. and Smith, M. D. and Zijlstra, A. A. and Viti, S. and Wesson, R.},
title = {A “FIREWORK” OF H2 KNOTS IN THE PLANETARY NEBULA NGC 7293 (THE HELIX NEBULA)*},
journal = {\apj}
}

@ARTICLE{Black1987,
       author = {{Black}, John H. and {van Dishoeck}, Ewine F.},
        title = "{Fluorescent Excitation of Interstellar H 2}",
      journal = {\apj},
         year = 1987,
        month = nov,
       volume = {322},
        pages = {412},
          doi = {10.1086/165740},
       adsurl = {https://ui.adsabs.harvard.edu/abs/1987ApJ...322..412B}
}

@ARTICLE{Likkel_2006,
       author = {{Likkel}, Lauren and {Dinerstein}, Harriet L. and {Lester}, Dan F. and {Kindt}, Anna and {Bartig}, Kevin},
        title = "{A Spectrophotometric Survey of K-Band Emission Lines in Planetary Nebulae}",
      journal = {\aj},
         year = 2006,
        month = mar,
       volume = {131},
       number = {3},
        pages = {1515-1529},
          doi = {10.1086/500050},
       adsurl = {https://ui.adsabs.harvard.edu/abs/2006AJ....131.1515L}
}

@ARTICLE{Roueff2019,
       author = {{Roueff}, E. and {Abgrall}, H. and {Czachorowski}, P. and {Pachucki}, K. and {Puchalski}, M. and {Komasa}, J.},
        title = "{The full infrared spectrum of molecular hydrogen}",
      journal = {\aap},
         year = 2019,
        month = oct,
       volume = {630},
          eid = {A58},
        pages = {A58},
          doi = {10.1051/0004-6361/201936249},
archivePrefix = {arXiv},
       eprint = {1909.11585},
 primaryClass = {physics.atom-ph},
       adsurl = {https://ui.adsabs.harvard.edu/abs/2019A&A...630A..58R}
}

@ARTICLE{Henney_2024,
       author = {{Henney}, William J. and {Valerdi}, Mabel},
        title = "{One hundred optical emission lines of molecular hydrogen from a low-metallicity photodissociation region}",
      journal = {\mnras},
         year = 2024,
        month = apr,
       volume = {529},
       number = {2},
        pages = {1723-1751},
          doi = {10.1093/mnras/stae486},
archivePrefix = {arXiv},
       eprint = {2306.00894},
 primaryClass = {astro-ph.GA},
       adsurl = {https://ui.adsabs.harvard.edu/abs/2024MNRAS.529.1723H}
}

@ARTICLE{McCoey_2004,
       author = {{McCoey}, C. and {Giannini}, T. and {Flower}, D.~R. and {Caratti o Garatti}, A.},
        title = "{On the excitation of the infrared knots in the HH99 outflow}",
      journal = {\mnras},
         year = 2004,
        month = sep,
       volume = {353},
       number = {3},
        pages = {813-824},
          doi = {10.1111/j.1365-2966.2004.08109.x},
archivePrefix = {arXiv},
       eprint = {astro-ph/0406262},
 primaryClass = {astro-ph},
       adsurl = {https://ui.adsabs.harvard.edu/abs/2004MNRAS.353..813M}
}

@ARTICLE{Bouvis_2025,
       author = {{Bouvis}, K. and {Akras}, S. and {Monteiro}, H. and {Konstantinou}, L. and {Boumis}, P. and {Garc{\'\i}a-Rojas}, J. and {Gon{\c{c}}alves}, D.~R. and {Aleman}, I. and {Monreal-Ibero}, A. and {Cami}, J.},
        title = "{Nickel- and iron-rich clumps in planetary nebulae: New discoveries and emission-line diagnostics}",
      journal = {\aap},
         year = 2025,
        month = aug,
       volume = {700},
          eid = {A155},
        pages = {A155},
          doi = {10.1051/0004-6361/202554879},
archivePrefix = {arXiv},
       eprint = {2507.05357},
 primaryClass = {astro-ph.SR},
       adsurl = {https://ui.adsabs.harvard.edu/abs/2025A&A...700A.155B}
}

@ARTICLE{Tsamis_2005,
       author = {{Tsamis}, Yiannis G. and {P{\'e}quignot}, Daniel},
        title = "{A photoionization-modelling study of 30 Doradus: the case for small-scale chemical inhomogeneity}",
      journal = {\mnras},
         year = 2005,
        month = dec,
       volume = {364},
       number = {2},
        pages = {687-704},
          doi = {10.1111/j.1365-2966.2005.09595.x},
archivePrefix = {arXiv},
       eprint = {astro-ph/0509463},
 primaryClass = {astro-ph},
       adsurl = {https://ui.adsabs.harvard.edu/abs/2005MNRAS.364..687T}
}

@ARTICLE{Nicholls_2012,
       author = {{Nicholls}, David C. and {Dopita}, Michael A. and {Sutherland}, Ralph S.},
        title = "{Resolving the Electron Temperature Discrepancies in H II Regions and Planetary Nebulae: {\ensuremath{\kappa}}-distributed Electrons}",
      journal = {\apj},
         year = 2012,
        month = jun,
       volume = {752},
       number = {2},
          eid = {148},
        pages = {148},
          doi = {10.1088/0004-637X/752/2/148},
archivePrefix = {arXiv},
       eprint = {1204.3880},
 primaryClass = {astro-ph.GA},
       adsurl = {https://ui.adsabs.harvard.edu/abs/2012ApJ...752..148N}
}

@ARTICLE{Corradi_2015,
       author = {{Corradi}, Romano L.~M. and {Garc{\'\i}a-Rojas}, Jorge and {Jones}, David and {Rodr{\'\i}guez-Gil}, Pablo},
        title = "{Binarity and the Abundance Discrepancy Problem in Planetary Nebulae}",
      journal = {\apj},
         year = 2015,
        month = apr,
       volume = {803},
       number = {2},
          eid = {99},
        pages = {99},
          doi = {10.1088/0004-637X/803/2/99},
archivePrefix = {arXiv},
       eprint = {1502.05182},
 primaryClass = {astro-ph.SR},
       adsurl = {https://ui.adsabs.harvard.edu/abs/2015ApJ...803...99C}
}

@ARTICLE{Liu_2000,
       author = {{Liu}, X. -W. and {Storey}, P.~J. and {Barlow}, M.~J. and {Danziger}, I.~J. and {Cohen}, M. and {Bryce}, M.},
        title = "{NGC 6153: a super-metal-rich planetary nebula?}",
      journal = {\mnras},
         year = 2000,
        month = mar,
       volume = {312},
       number = {3},
        pages = {585-628},
          doi = {10.1046/j.1365-8711.2000.03167.x},
       adsurl = {https://ui.adsabs.harvard.edu/abs/2000MNRAS.312..585L}
}

@ARTICLE{Peimbert_1980,
       author = {{Torres-Peimbert}, S. and {Peimbert}, M. and {Daltabuit}, E.},
        title = "{IUE and visual observations of the Orion nebula and IC 418 : the carbon abundance.}",
      journal = {\apj},
         year = 1980,
        month = may,
       volume = {238},
        pages = {133-139},
          doi = {10.1086/157966},
       adsurl = {https://ui.adsabs.harvard.edu/abs/1980ApJ...238..133T}
}

@ARTICLE{Peimbert_1990,
       author = {{Torres-Peimbert}, S. and {Peimbert}, M. and {Pena}, M.},
        title = "{Planetary nebulae with a high degree of ionization : NGC 2242 and NGC 4361.}",
      journal = {\aap},
         year = 1990,
        month = jul,
       volume = {233},
        pages = {540},
       adsurl = {https://ui.adsabs.harvard.edu/abs/1990A&A...233..540T}
}

@ARTICLE{Jones_2016,
       author = {{Jones}, D. and {Wesson}, R. and {Garc{\'\i}a-Rojas}, J. and {Corradi}, R.~L.~M. and {Boffin}, H.~M.~J.},
        title = "{NGC 6778: strengthening the link between extreme abundance discrepancy factors and central star binarity in planetary nebulae}",
      journal = {\mnras},
         year = 2016,
        month = jan,
       volume = {455},
       number = {3},
        pages = {3263-3272},
          doi = {10.1093/mnras/stv2519},
archivePrefix = {arXiv},
       eprint = {1510.07516},
 primaryClass = {astro-ph.SR},
       adsurl = {https://ui.adsabs.harvard.edu/abs/2016MNRAS.455.3263J}
}

@article{Wesson_2018,
    author = {Wesson, R and Jones, D and García-Rojas, J and Boffin, H M J and Corradi, R L M},
    title = {Confirmation of the link between central star binarity and extreme abundance discrepancy factors in planetary nebulae},
    journal = {\mnras},
    volume = {480},
    number = {4},
    pages = {4589-4613},
    year = {2018},
    month = {07},
    issn = {0035-8711},
    doi = {10.1093/mnras/sty1871},
    url = {https://doi.org/10.1093/mnras/sty1871},
    eprint = {https://academic.oup.com/mnras/article-pdf/480/4/4589/25541972/sty1871.pdf},
}

@ARTICLE{Bohlin_1978,
       author = {{Bohlin}, R.~C. and {Savage}, B.~D. and {Drake}, J.~F.},
        title = "{A survey of interstellar H I from Lalpha absorption measurements. II.}",
      journal = {\apj},
         year = 1978,
        month = aug,
       volume = {224},
        pages = {132-142},
          doi = {10.1086/156357},
       adsurl = {https://ui.adsabs.harvard.edu/abs/1978ApJ...224..132B}
}

@ARTICLE{Draine_1996,
       author = {{Draine}, B.~T. and {Bertoldi}, Frank},
        title = "{Structure of Stationary Photodissociation Fronts}",
      journal = {\apj},
         year = 1996,
        month = sep,
       volume = {468},
        pages = {269},
          doi = {10.1086/177689},
archivePrefix = {arXiv},
       eprint = {astro-ph/9603032},
 primaryClass = {astro-ph},
       adsurl = {https://ui.adsabs.harvard.edu/abs/1996ApJ...468..269D}
}

@ARTICLE{Hora_1999,
       author = {{Hora}, Joseph L. and {Latter}, William B. and {Deutsch}, Lynne K.},
        title = "{Investigating the Near-Infrared Properties of Planetary Nebulae. II. Medium-Resolution Spectra}",
      journal = {\apjs},
         year = 1999,
        month = sep,
       volume = {124},
       number = {1},
        pages = {195-240},
          doi = {10.1086/313256},
archivePrefix = {arXiv},
       eprint = {astro-ph/9904202},
 primaryClass = {astro-ph},
       adsurl = {https://ui.adsabs.harvard.edu/abs/1999ApJS..124..195H}
}

@article{Davis_2003,
    author = {Davis, Christopher J. and Smith, Michael D. and Stern, Luke and Kerr, Thomas H. and Chiar, Jean E.},
    title = {Near-infrared spectroscopy of (proto)-planetary nebulae: molecular hydrogen excitation as an evolutionary tracer},
    journal = {\mnras},
    volume = {344},
    number = {1},
    pages = {262-282},
    year = {2003},
    month = {09},
    issn = {0035-8711},
    doi = {10.1046/j.1365-8711.2003.06820.x},
    url = {https://doi.org/10.1046/j.1365-8711.2003.06820.x},
    eprint = {https://academic.oup.com/mnras/article-pdf/344/1/262/18649231/344-1-262.pdf},
}

@ARTICLE{Filippenko1982,
       author = {{Filippenko}, A.~V.},
        title = "{The importance of atmospheric differential refraction in spectrophotometry.}",
      journal = {\pasp},
         year = 1982,
        month = aug,
       volume = {94},
        pages = {715-721},
          doi = {10.1086/131052},
       adsurl = {https://ui.adsabs.harvard.edu/abs/1982PASP...94..715F}
}

@ARTICLE{Esoreflex2013,
   author = {{Freudling}, W. and {Romaniello}, M. and {Bramich}, D.~M. and
{Ballester}, P. and {Forchi}, V. and {Garc{\'{\i}}a-Dabl{\'o}}, C.~E. and
{Moehler}, S. and {Neeser}, M.~J.},
    title = "{Automated data reduction workflows for astronomy. The ESO Reflex environment}",
  journal = {\aap},
archivePrefix = "arXiv",
   eprint = {1311.5411},
 primaryClass = "astro-ph.IM",
     year = 2013,
    month = nov,
   volume = 559,
      eid = {A96},
    pages = {A96},
      doi = {10.1051/0004-6361/201322494},
   adsurl = {http://adsabs.harvard.edu/abs/2013A%26A...559A..96F}
}

@ARTICLE{Burton_1989,
       author = {{Burton}, Michael G. and {Geballe}, T.~R. and {Brand}, P.~W.~J.~L.},
        title = "{Shocked molecular hydrogen in the bipolar outflow NGC 2071.}",
      journal = {\mnras},
         year = 1989,
        month = jun,
       volume = {238},
        pages = {1513-1527},
          doi = {10.1093/mnras/238.4.1513},
       adsurl = {https://ui.adsabs.harvard.edu/abs/1989MNRAS.238.1513B}
}

@article{Goncalves_2001,
doi = {10.1086/318364},
url = {https://dx.doi.org/10.1086/318364},
year = {2001},
month = {jan},
publisher = {},
volume = {547},
number = {1},
pages = {302},
author = {Gonçalves, Denise R. and Corradi, Romano L. M. and Mampaso, Antonio},
title = {Low-Ionization Structures in Planetary Nebulae: Confronting
Models with
Observations},
journal = {\apj}
}

@ARTICLE{Seaton1979,
       author = {{Seaton}, M.~J.},
        title = "{Interstellar extinction in the UV.}",
      journal = {\mnras},
         year = 1979,
        month = jan,
       volume = {187},
        pages = {73},
          doi = {10.1093/mnras/187.1.73P},
       adsurl = {https://ui.adsabs.harvard.edu/abs/1979MNRAS.187P..73S}
}

@ARTICLE{Howarth1983,
       author = {{Howarth}, I.~D. and {Wilson}, B.},
        title = "{HZ Her: the nature and origin of the emission lines.}",
      journal = {\mnras},
         year = 1983,
        month = sep,
       volume = {204},
        pages = {1091-1104},
          doi = {10.1093/mnras/204.4.1091},
       adsurl = {https://ui.adsabs.harvard.edu/abs/1983MNRAS.204.1091H}
}

@ARTICLE{Cardelli1989,
       author = {{Cardelli}, Jason A. and {Clayton}, Geoffrey C. and {Mathis}, John S.},
        title = "{The Relationship between Infrared, Optical, and Ultraviolet Extinction}",
      journal = {\apj},
         year = 1989,
        month = oct,
       volume = {345},
        pages = {245},
          doi = {10.1086/167900},
       adsurl = {https://ui.adsabs.harvard.edu/abs/1989ApJ...345..245C}
}

@article{Goncalves_2009,
    author = {Gonçalves, D. R. and Mampaso, A. and Corradi, R. L. M. and Quireza, C.},
    title = {Low-ionization pairs of knots in planetary nebulae: physical properties and excitation},
    journal = {\mnras},
    volume = {398},
    number = {4},
    pages = {2166-2176},
    year = {2009},
    month = {09},
    issn = {0035-8711},
    doi = {10.1111/j.1365-2966.2009.15266.x},
    url = {https://doi.org/10.1111/j.1365-2966.2009.15266.x},
    eprint = {https://academic.oup.com/mnras/article-pdf/398/4/2166/3061113/mnras0398-2166.pdf},
}

@article{Wesson_2024,
    author = {Wesson, R and Matsuura, Mikako and Zijlstra, Albert A and Volk, Kevin and Kavanagh, Patrick J and García-Segura, Guillermo and McDonald, I and Sahai, Raghvendra and Barlow, M J and Cox, Nick L J and Bernard-Salas, Jeronimo and Aleman, Isabel and Cami, Jan and Clark, Nicholas and Dinerstein, Harriet L and Justtanont, K and Kaplan, Kyle F and Manchado, A and Peeters, Els and Van de Steene, Griet C and van Hoof, Peter A M},
    title = {JWST observations of the Ring Nebula (NGC 6720): I. Imaging of the rings, globules, and arcs},
    journal = {\mnras},
    volume = {528},
    number = {2},
    pages = {3392-3416},
    year = {2023},
    month = {12},
    issn = {0035-8711},
    doi = {10.1093/mnras/stad3670},
    url = {https://doi.org/10.1093/mnras/stad3670},
    eprint = {https://academic.oup.com/mnras/article-pdf/528/2/3392/56617376/stad3670.pdf},
}

@article{Zhang_2005a,
    author = {Zhang, Y. and Liu, X.-W. and Liu, Y. and Rubin, R. H.},
    title = {Helium recombination spectra as temperature diagnostics for planetary nebulae},
    journal = {\mnras},
    volume = {358},
    number = {2},
    pages = {457-467},
    year = {2005},
    month = {04},
    issn = {0035-8711},
    doi = {10.1111/j.1365-2966.2005.08810.x},
    url = {https://doi.org/10.1111/j.1365-2966.2005.08810.x},
    eprint = {https://academic.oup.com/mnras/article-pdf/358/2/457/18424516/358-2-457.pdf},
}

@article{Bautista_1998,
doi = {10.1086/305061},
url = {https://dx.doi.org/10.1086/305061},
year = {1998},
month = {jan},
publisher = {},
volume = {492},
number = {2},
pages = {650},
author = {Bautista, Manuel A. and Pradhan, Anil K.},
title = {Ionization Structure and Spectra of Iron in Gaseous Nebulae},
journal = {\apj}
}

@article{Bautista_2015,
doi = {10.1088/0004-637X/808/2/174},
url = {https://dx.doi.org/10.1088/0004-637X/808/2/174},
year = {2015},
month = {jul},
publisher = {The American Astronomical Society},
volume = {808},
number = {2},
pages = {174},
author = {Bautista, Manuel A. and Fivet, Vanessa and Ballance, Connor and Quinet, Pascal and Ferland, Gary and Mendoza, Claudio and Kallman, Timothy R.},
title = {ATOMIC DATA AND SPECTRAL MODEL FOR Fe II},
journal = {\apj}
}

@ARTICLE{Peimbert_1967,
       author = {{Peimbert}, Manuel},
        title = "{Temperature Determinations of H II Regions}",
      journal = {\apj},
         year = 1967,
        month = dec,
       volume = {150},
        pages = {825},
          doi = {10.1086/149385},
       adsurl = {https://ui.adsabs.harvard.edu/abs/1967ApJ...150..825P}
}

@ARTICLE{Hollenbach_1995,
       author = {{Hollenbach}, David and {Natta}, Antonella},
        title = "{Time-dependent Photodissociation Regions}",
      journal = {\apj},
         year = 1995,
        month = dec,
       volume = {455},
        pages = {133},
          doi = {10.1086/176562},
       adsurl = {https://ui.adsabs.harvard.edu/abs/1995ApJ...455..133H}
}

@article{mari2025,
       author = {{Mari}, M.~B. and {Akras}, S. and {Gon{\c{c}}alves}, D.~R.},
        title = "{Low-ionization structures in planetary nebulae: IV. The molecular hydrogen counterpart}",
      journal = {\aap},
         year = 2025,
        month = dec,
       volume = {704},
          eid = {A59},
        pages = {A59},
          doi = {10.1051/0004-6361/202554757},
archivePrefix = {arXiv},
       eprint = {2508.20240},
 primaryClass = {astro-ph.GA},
       adsurl = {https://ui.adsabs.harvard.edu/abs/2025A&A...704A..59M}
}

@ARTICLE{Burton_1992,
       author = {{Burton}, Michael G. and {Hollenbach}, D.~J. and {Tielens}, A.~G.~G.},
        title = "{Mid-Infrared Rotational Line Emission from Interstellar Molecular Hydrogen}",
      journal = {\apj},
         year = 1992,
        month = nov,
       volume = {399},
        pages = {563},
          doi = {10.1086/171947},
       adsurl = {https://ui.adsabs.harvard.edu/abs/1992ApJ...399..563B}
}

@article{Nesterenok_2019,
    author = {Nesterenok, A V and Bossion, D and Scribano, Y and Lique, F},
    title = {C-type shock modelling – the effect of new H2–H collisional rate coefficients},
    journal = {\mnras},
    volume = {489},
    number = {4},
    pages = {4520-4529},
    year = {2019},
    month = {09},
    issn = {0035-8711},
    doi = {10.1093/mnras/stz2441},
    url = {https://doi.org/10.1093/mnras/stz2441},
    eprint = {https://academic.oup.com/mnras/article-pdf/489/4/4520/30040322/stz2441.pdf},
}

@INPROCEEDINGS{Aleman_2011,
       author = {{Aleman}, I. and {Gruenwald}, R.},
        title = "{Photoionization Models of the H\_2 Emission of the Narrow Line Region of AGNs}",
    booktitle = {The Molecular Universe},
         year = 2011,
       editor = {{Cernicharo}, Jos{\'e} and {Bachiller}, Rafael},
       series = {IAU Symposium},
       volume = {280},
        month = may,
          eid = {81},
        pages = {81},
       adsurl = {https://ui.adsabs.harvard.edu/abs/2011IAUS..280P..81A}
}

@article{Mendez-Delgado_2025,
doi = {10.3847/1538-4357/adc67a},
url = {https://doi.org/10.3847/1538-4357/adc67a},
year = {2025},
month = {jun},
publisher = {The American Astronomical Society},
volume = {986},
number = {1},
pages = {74},
author = {Méndez-Delgado, J. E. and Skillman, E. D. and Aver, E. and Morisset, C. and Esteban, C. and García-Rojas, J. and Kreckel, K. and Rogers, N. S. J. and Rosales-Ortega, F. F. and Arellano-Córdova, K. Z. and Flury, S. R. and Reyes-Rodríguez, E. and Orte-García, M. and Tan, S.},
title = {Generalized Te([O iii])–Te(He i) Discrepancies in Ionized Nebulae: Possible Evidence of Case B Deviations and Temperature Inhomogeneities},
journal = {\apj}
}

@ARTICLE{Le_Petit_2006,
       author = {{Le Petit}, Franck and {Nehm{\'e}}, Cyrine and {Le Bourlot}, Jacques and {Roueff}, Evelyne},
        title = "{A Model for Atomic and Molecular Interstellar Gas: The Meudon PDR Code}",
      journal = {\apjs},
         year = 2006,
        month = jun,
       volume = {164},
       number = {2},
        pages = {506-529},
          doi = {10.1086/503252},
archivePrefix = {arXiv},
       eprint = {astro-ph/0602150},
 primaryClass = {astro-ph},
       adsurl = {https://ui.adsabs.harvard.edu/abs/2006ApJS..164..506L}
}

@ARTICLE{Fernandez_2004,
       author = {{Fern{\'a}ndez}, Rodrigo and {Monteiro}, Hektor and {Schwarz}, Hugo E.},
        title = "{Proper Motion and Kinematics of the Ansae in NGC 7009}",
      journal = {\apj},
         year = 2004,
        month = mar,
       volume = {603},
       number = {2},
        pages = {595-598},
          doi = {10.1086/381523},
archivePrefix = {arXiv},
       eprint = {astro-ph/0311612},
 primaryClass = {astro-ph},
       adsurl = {https://ui.adsabs.harvard.edu/abs/2004ApJ...603..595F}
}

@ARTICLE{Maloney1996,
       author = {{Maloney}, Philip R. and {Hollenbach}, David J. and {Tielens}, A.~G.~G.~M.},
        title = "{X-Ray--irradiated Molecular Gas. I. Physical Processes and General Results}",
      journal = {\apj},
         year = 1996,
        month = jul,
       volume = {466},
        pages = {561},
          doi = {10.1086/177532},
       adsurl = {https://ui.adsabs.harvard.edu/abs/1996ApJ...466..561M}
}

@ARTICLE{Tine_1997,
       author = {{Tin{\'e}}, S. and {Lepp}, S. and {Gredel}, R. and {Dalgarno}, A.},
        title = "{Infrared Response of H$_{2}$ to X-Rays in Dense Clouds}",
      journal = {\apj},
         year = 1997,
        month = may,
       volume = {481},
       number = {1},
        pages = {282-295},
          doi = {10.1086/304048},
       adsurl = {https://ui.adsabs.harvard.edu/abs/1997ApJ...481..282T}
}

@ARTICLE{Mazzalay_2013,
       author = {{Mazzalay}, X. and {Saglia}, R.~P. and {Erwin}, Peter and {Fabricius}, M.~H. and {Rusli}, S.~P. and {Thomas}, J. and {Bender}, R. and {Opitsch}, M. and {Nowak}, N. and {Williams}, Michael J.},
        title = "{Molecular gas in the centre of nearby galaxies from VLT/SINFONI integral field spectroscopy - I. Morphology and mass inventory}",
      journal = {\mnras},
         year = 2013,
        month = jan,
       volume = {428},
       number = {3},
        pages = {2389-2406},
          doi = {10.1093/mnras/sts204},
archivePrefix = {arXiv},
       eprint = {1210.4171},
 primaryClass = {astro-ph.CO},
       adsurl = {https://ui.adsabs.harvard.edu/abs/2013MNRAS.428.2389M}
}

@ARTICLE{Reunanen_2003,
       author = {{Reunanen}, J. and {Kotilainen}, J.~K. and {Prieto}, M.~A.},
        title = "{Near-infrared spectroscopy of nearby Seyfert galaxies - II. Molecular content and coronal emission}",
      journal = {\mnras},
         year = 2003,
        month = jul,
       volume = {343},
       number = {1},
        pages = {192-208},
          doi = {10.1046/j.1365-8711.2003.06771.x},
archivePrefix = {arXiv},
       eprint = {astro-ph/0305100},
 primaryClass = {astro-ph},
       adsurl = {https://ui.adsabs.harvard.edu/abs/2003MNRAS.343..192R}
}

@ARTICLE{Mari2023b,
       author = {{Mari}, M. Bel{\'e}n and {Gon{\c{c}}alves}, Denise R. and {Akras}, Stavros},
        title = "{Low-ionization structures in planetary nebulae - II. Densities, temperatures, abundances, and excitation of six PNe}",
      journal = {\mnras},
         year = 2023,
        month = jan,
       volume = {518},
       number = {3},
        pages = {3908-3934},
          doi = {10.1093/mnras/stac2729},
archivePrefix = {arXiv},
       eprint = {2209.10988},
 primaryClass = {astro-ph.GA},
       adsurl = {https://ui.adsabs.harvard.edu/abs/2023MNRAS.518.3908M}
}

@ARTICLE{Goncalves2004,
       author = {{Gon{\c{c}}alves}, Denise R. and {Corradi}, Romano L.~M. and {Mampaso}, Antonio and {Perinotto}, Mario},
        title = "{Erratum: ``The Physical Parameters, Excitation, and Chemistry of the Rim, Jets, and Knots of the Planetary Nebula NGC 7009`` (<A href=''/abs/2003ApJ...597..975G''>ApJ, 597, 975 [2003]</A>)}",
      journal = {\apj},
         year = 2004,
        month = feb,
       volume = {601},
       number = {2},
        pages = {1161-1161},
          doi = {10.1086/381527},
       adsurl = {https://ui.adsabs.harvard.edu/abs/2004ApJ...601.1161G}
}

@ARTICLE{Giannini2004,
       author = {{Giannini}, T. and {McCoey}, C. and {Caratti o Garatti}, A. and {Nisini}, B. and {Lorenzetti}, D. and {Flower}, D.~R.},
        title = "{On the excitation of the infrared knots along protostellar jets}",
      journal = {\aap},
         year = 2004,
        month = jun,
       volume = {419},
        pages = {999-1014},
          doi = {10.1051/0004-6361:20040087},
archivePrefix = {arXiv},
       eprint = {astro-ph/0402395},
 primaryClass = {astro-ph},
       adsurl = {https://ui.adsabs.harvard.edu/abs/2004A&A...419..999G}
}

@ARTICLE{Bautista_1996,
       author = {{Bautista}, Manuel A. and {Peng}, Jianfang and {Pradhan}, Anil K.},
        title = "{Excitation of [Ni II] and [Fe II] Lines in Gaseous Nebulae}",
      journal = {\apj},
         year = 1996,
        month = mar,
       volume = {460},
        pages = {372},
          doi = {10.1086/176976},
       adsurl = {https://ui.adsabs.harvard.edu/abs/1996ApJ...460..372B}
}

@ARTICLE{Giannini_2015,
       author = {{Giannini}, T. and {Antoniucci}, S. and {Nisini}, B. and {Bacciotti}, F. and {Podio}, L.},
        title = "{Solving the Excitation and Chemical Abundances in Shocks: The Case of HH 1}",
      journal = {\apj},
         year = 2015,
        month = nov,
       volume = {814},
       number = {1},
          eid = {52},
        pages = {52},
          doi = {10.1088/0004-637X/814/1/52},
archivePrefix = {arXiv},
       eprint = {1510.06880},
 primaryClass = {astro-ph.SR},
       adsurl = {https://ui.adsabs.harvard.edu/abs/2015ApJ...814...52G}
}

@ARTICLE{Kirwan_2023,
       author = {{Kirwan}, A. and {Manara}, C.~F. and {Whelan}, E.~T. and {Robberto}, M. and {McLeod}, A.~F. and {Facchini}, S. and {Beccari}, G. and {Miotello}, A. and {Schneider}, P.~C. and {Murphy}, A. and {Vicente}, S.},
        title = "{A spectacular jet from the bright 244-440 Orion proplyd: The MUSE NFM view}",
      journal = {\aap},
         year = 2023,
        month = may,
       volume = {673},
          eid = {A166},
        pages = {A166},
          doi = {10.1051/0004-6361/202245428},
archivePrefix = {arXiv},
       eprint = {2303.13205},
 primaryClass = {astro-ph.GA},
       adsurl = {https://ui.adsabs.harvard.edu/abs/2023A&A...673A.166K}
}

@ARTICLE{Sabbadin_2004,
       author = {{Sabbadin}, F. and {Turatto}, M. and {Cappellaro}, E. and {Benetti}, S. and {Ragazzoni}, R.},
        title = "{The 3-D ionization structure and evolution of NGC 7009 (Saturn Nebula)}",
      journal = {\aap},
         year = 2004,
        month = mar,
       volume = {416},
        pages = {955-981},
          doi = {10.1051/0004-6361:20031729},
archivePrefix = {arXiv},
       eprint = {astro-ph/0312261},
 primaryClass = {astro-ph},
       adsurl = {https://ui.adsabs.harvard.edu/abs/2004A&A...416..955S}
}

@article{Hollenbach_1980,
  author       = {Hollenbach, D and McKee, C F},
  title        = {Molecule formation and infrared emission in fast interstellar shocks. II. Dissociation speeds for interstellar shock waves},
  doi          = {10.1086/183358},
  url          = {https://www.osti.gov/biblio/6678023},
  journal      = {Astrophys. J., Lett. Ed.; (United States)},
  issn         = {ISSN AJLEA},
  volume       = {241:1},
  place        = {United States},
  year         = {1980},
  month        = {10}}

@article{Flower_2003,
    author = {Flower, D. R. and Bourlot, J. Le and Des Forêts, G. Pineau and Cabrit, S.},
    title = {The contributions of J-type shocks to the H2 emission from molecular outflow sources},
    journal = {\mnras},
    volume = {341},
    number = {1},
    pages = {70-80},
    year = {2003},
    month = {05},
    issn = {0035-8711},
    doi = {10.1046/j.1365-8711.2003.06463.x},
    url = {https://doi.org/10.1046/j.1365-8711.2003.06463.x},
    eprint = {https://academic.oup.com/mnras/article-pdf/341/1/70/18651862/341-1-70.pdf},
}

@ARTICLE{LIME,
       author = {{Fern{\'a}ndez}, V. and {Amor{\'\i}n}, R. and {Firpo}, V. and {Morisset}, C.},
        title = "{LIME: A LIne MEasuring library for large and complex spectroscopic data sets. I. Implementation of a virtual observatory for JWST spectra}",
      journal = {\aap},
         year = 2024,
        month = aug,
       volume = {688},
          eid = {A69},
        pages = {A69},
          doi = {10.1051/0004-6361/202449224},
archivePrefix = {arXiv},
       eprint = {2405.15072},
 primaryClass = {astro-ph.IM},
       adsurl = {https://ui.adsabs.harvard.edu/abs/2024A&A...688A..69F}
}

@Article{VanHoof2018,
AUTHOR = {Van Hoof, Peter A. M.},
TITLE = {Recent Development of the Atomic Line List},
JOURNAL = {Galaxies},
VOLUME = {6},
YEAR = {2018},
NUMBER = {2},
ARTICLE-NUMBER = {63},
URL = {https://www.mdpi.com/2075-4434/6/2/63},
ISSN = {2075-4434},
DOI = {10.3390/galaxies6020063}
}

@ARTICLE{Davis2011,
       author = {{Davis}, C.~J. and {Cervantes}, B. and {Nisini}, B. and {Giannini}, T. and {Takami}, M. and {Whelan}, E. and {Smith}, M.~D. and {Ray}, T.~P. and {Chrysostomou}, A. and {Pyo}, T.~S.},
        title = "{VLT integral field spectroscopy of embedded protostars: using near-infrared emission lines as tracers of accretion and outflow}",
      journal = {\aap},
         year = 2011,
        month = apr,
       volume = {528},
          eid = {A3},
        pages = {A3},
          doi = {10.1051/0004-6361/201015897},
       adsurl = {https://ui.adsabs.harvard.edu/abs/2011A&A...528A...3D}
}

@ARTICLE{Nisini2002,
       author = {{Nisini}, B. and {Caratti o Garatti}, A. and {Giannini}, T. and {Lorenzetti}, D.},
        title = "{1-2.5 mu m spectra of jets from young stars: Strong Fe II emission in HH111, HH240-241 and HH120}",
      journal = {\aap},
         year = 2002,
        month = oct,
       volume = {393},
        pages = {1035-1051},
          doi = {10.1051/0004-6361:20021062},
       adsurl = {https://ui.adsabs.harvard.edu/abs/2002A&A...393.1035N}
}

@article{Rudy2001,
doi = {10.1086/318014},
url = {https://doi.org/10.1086/318014},
year = {2001},
month = {jan},
publisher = {},
volume = {121},
number = {1},
pages = {362},
author = {Rudy, Richard J. and Lynch, David K. and Mazuk, S. and Puetter, R. C. and Dearborn, David S. P.},
title = {The Near-Infrared Spectrum of the Planetary Nebula IC
5117},
journal = {The Astronomical Journal}
}

@article{Huang_2026,
doi = {10.3847/1538-4365/ae2bfd},
url = {https://doi.org/10.3847/1538-4365/ae2bfd},
year = {2026},
month = {feb},
publisher = {The American Astronomical Society},
volume = {282},
number = {2},
pages = {63},
author = {Huang, Haomiao and Fang, Xuan and García-Rojas, Jorge and Tu, Zhijun and Liu, Jifeng and Liu, Xiaowei},
title = {A Morphokinematic Study of Galactic High-abundance Discrepancy Factor Planetary Nebulae Based on the VLT/UVES Deep Spectroscopy*},
journal = {The Astrophysical Journal Supplement Series}
}

@ARTICLE{BernardSalas2005,
       author = {{Bernard-Salas}, J. and {Tielens}, A.~G.~G.~M.},
        title = "{Physical conditions in Photo-Dissociation Regions around Planetary Nebulae}",
      journal = {\aap},
         year = 2005,
        month = feb,
       volume = {431},
        pages = {523-538},
          doi = {10.1051/0004-6361:20035746},
       adsurl = {https://ui.adsabs.harvard.edu/abs/2005A&A...431..523B}
}

@ARTICLE{Garay1989,
       author = {{Garay}, G. and {Gathier}, R. and {Rodriguez}, L.~F.},
        title = "{Radio recombination line observations of compact planetary nebulae}",
      journal = {\aap},
         year = 1989,
        month = may,
       volume = {215},
       number = {1},
        pages = {101-112},
       adsurl = {https://ui.adsabs.harvard.edu/abs/1989A&A...215..101G}
}

@ARTICLE{Guerrero2002,
       author = {{Guerrero}, M.~A. and {Gruendl}, R.~A. and {Chu}, Y.-H.},
        title = "{Diffuse X-ray emission from the planetary nebula NGC 7009}",
      journal = {\aap},
         year = 2002,
        month = may,
       volume = {387},
        pages = {L1-L5},
          doi = {10.1051/0004-6361:20020330},
archivePrefix = {arXiv},
       eprint = {astro-ph/0203049},
 primaryClass = {astro-ph},
       adsurl = {https://ui.adsabs.harvard.edu/abs/2002A&A...387L...1G}
}

@ARTICLE{Walsh_2016,
       author = {{Walsh}, J.~R. and {Monreal-Ibero}, A. and {Barlow}, M.~J. and {Ueta}, T. and {Wesson}, R. and {Zijlstra}, A.~A.},
        title = "{The extinction and dust-to-gas structure of the planetary nebula NGC 7009 observed with MUSE}",
      journal = {\aap},
         year = 2016,
        month = apr,
       volume = {588},
          eid = {A106},
        pages = {A106},
          doi = {10.1051/0004-6361/201527988},
archivePrefix = {arXiv},
       eprint = {1602.06788},
 primaryClass = {astro-ph.GA},
       adsurl = {https://ui.adsabs.harvard.edu/abs/2016A&A...588A.106W}
}

@ARTICLE{Hyung_2023,
       author = {{Hyung}, Siek and {Lee}, Seong-Jae and {Otsuka}, Masaaki},
        title = "{The [S II] Spectral Images of the Planetary Nebula NGC 7009. II. Major Axis}",
      journal = {\apjs},
         year = 2023,
        month = mar,
       volume = {265},
       number = {1},
          eid = {2},
        pages = {2},
          doi = {10.3847/1538-4365/acad72},
archivePrefix = {arXiv},
       eprint = {2212.10056},
 primaryClass = {astro-ph.SR},
       adsurl = {https://ui.adsabs.harvard.edu/abs/2023ApJS..265....2H}
}

@misc{Wiese1996,
  author = {Wolfgang Wiese and Jeffrey Fuhr and T Deters},
  title = {Atomic Transition Probabilities for Carbon, Nitrogen and Oxygen},
  year = {1996},
  month = {1996-01-01},
  publisher = {J. Phys. & Chem. Ref. Data (JPCRD), National Institute of Standards and Technology, Gaithersburg, MD},
  language = {en},
}

@ARTICLE{Bhatia1995,
       author = {{Bhatia}, A.~K. and {Kastner}, S.~O.},
        title = "{The Neutral Oxygen Spectrum. I. Collisionally Excited Level Populations and Line Intensities under Optically Thin Conditions}",
      journal = {\apjs},
         year = 1995,
        month = jan,
       volume = {96},
        pages = {325},
          doi = {10.1086/192121},
       adsurl = {https://ui.adsabs.harvard.edu/abs/1995ApJS...96..325B}
}

@ARTICLE{Zeippen1982,
       author = {{Zeippen}, C.~J.},
        title = "{Transition probabilities for forbidden lines in the 2p3 configuration.}",
      journal = {\mnras},
         year = 1982,
        month = jan,
       volume = {198},
        pages = {111-125},
          doi = {10.1093/mnras/198.1.111},
       adsurl = {https://ui.adsabs.harvard.edu/abs/1982MNRAS.198..111Z}
}

@ARTICLE{Kisielius2009,
       author = {{Kisielius}, R. and {Storey}, P.~J. and {Ferland}, G.~J. and {Keenan}, F.~P.},
        title = "{Electron-impact excitation of OII fine-structure levels}",
      journal = {\mnras},
         year = 2009,
        month = aug,
       volume = {397},
       number = {2},
        pages = {903-912},
          doi = {10.1111/j.1365-2966.2009.14989.x},
archivePrefix = {arXiv},
       eprint = {0907.4209},
 primaryClass = {astro-ph.IM},
       adsurl = {https://ui.adsabs.harvard.edu/abs/2009MNRAS.397..903K}
}

@article{Storey2000,
    author = {Storey, P. J. and Zeippen, C. J.},
    title = "{Theoretical values for the [O iii] 5007/4959 line-intensity ratio and homologous cases}",
    journal = {MNRAS},
    volume = {312},
    number = {4},
    pages = {813-816},
    year = {2000},
    month = {03},
    issn = {0035-8711},
    doi = {10.1046/j.1365-8711.2000.03184.x},
    url = {https://doi.org/10.1046/j.1365-8711.2000.03184.x},
    eprint = {https://academic.oup.com/mnras/article-pdf/312/4/813/18634438/312-4-813.pdf},
}

@article{FroeseFischer2004,
title = {Breit–Pauli energy levels, lifetimes, and transition probabilities for the beryllium-like to neon-like sequences},
journal = {Atomic Data and Nuclear Data Tables},
volume = {87},
number = {1},
pages = {1-184},
year = {2004},
issn = {0092-640X},
doi = {https://doi.org/10.1016/j.adt.2004.02.001},
url = {https://www.sciencedirect.com/science/article/pii/S0092640X04000087},
author = {Charlotte {Froese Fischer} and Georgio Tachiev}
}

@ARTICLE{Storey2014,
       author = {{Storey}, P.~J. and {Sochi}, Taha and {Badnell}, N.~R.},
        title = "{Collision strengths for nebular [O III] optical and infrared lines}",
      journal = {\mnras},
         year = 2014,
        month = jul,
       volume = {441},
       number = {4},
        pages = {3028-3039},
          doi = {10.1093/mnras/stu777},
       adsurl = {https://ui.adsabs.harvard.edu/abs/2014MNRAS.441.3028S}
}

@article{Kaufman1986,
    author = {Kaufman, Victor and Sugar, Jack},
    title = "{Forbidden Lines in ns2npk Ground Configurations and nsnp Excited Configurations of Beryllium through Molybdenum Atoms and Ions}",
    journal = {JPCRD},
    volume = {15},
    number = {1},
    pages = {341},
    year = {1986},
    month = {01},
    issn = {0047-2689},
    doi = {10.1063/1.555775},
    url = {https://doi.org/10.1063/1.555775},
    eprint = {https://pubs.aip.org/aip/jpr/article-pdf/15/1/321/12129145/321\_1\_online.pdf},
}

@ARTICLE{Peguignot1976,
       author = {{Pequignot}, D. and {Aldrovandi}, S.~M.~V.},
        title = "{Effective collision strengths for the forbidden emission lines of neutral carbon, nitrogen and oxygen.}",
      journal = {\aap},
         year = 1976,
        month = jul,
       volume = {50},
       number = {1},
        pages = {141-144},
       adsurl = {https://ui.adsabs.harvard.edu/abs/1976A&A....50..141P}
}

@ARTICLE{Dopita1976,
       author = {{Dopita}, M.~A. and {Mason}, D.~J. and {Robb}, W.~D.},
        title = "{Atomic nitrogen as a probe of physical conditions in the interstellar medium.}",
      journal = {\apj},
         year = 1976,
        month = jul,
       volume = {207},
        pages = {102-109},
          doi = {10.1086/154472},
       adsurl = {https://ui.adsabs.harvard.edu/abs/1976ApJ...207..102D}
}

@article{Tayal2010,
doi = {10.1088/0067-0049/188/1/32},
url = {https://dx.doi.org/10.1088/0067-0049/188/1/32},
year = {2010},
month = {apr},
publisher = {AAS},
volume = {188},
number = {1},
pages = {32},
author = {S. S. Tayal and O. Zatsarinny},
title = {BREIT–PAULI TRANSITION PROBABILITIES AND ELECTRON EXCITATION COLLISION STRENGTHS FOR SINGLY IONIZED SULFUR},
journal = {\apjs}
}

@article{FroeseFischer2006,
title = {Relativistic energy levels, lifetimes, and transition probabilities for the sodium-like to argon-like sequences},
journal = {Atomic Data and Nuclear Data Tables},
volume = {92},
number = {5},
pages = {607-812},
year = {2006},
issn = {0092-640X},
doi = {https://doi.org/10.1016/j.adt.2006.03.001},
url = {https://www.sciencedirect.com/science/article/pii/S0092640X06000155},
author = {Charlotte {Froese Fischer} and Georgio Tachiev and Andrei Irimia}
}

@article{Tayal1999,
doi = {10.1086/307971},
url = {https://dx.doi.org/10.1086/307971},
year = {1999},
month = {nov},
publisher = {},
volume = {526},
number = {1},
pages = {544},
author = {S. S. Tayal and G. P. Gupta},
title = {Collision Strengths for Electron Collision Excitation of Fine-Structure Levels in S III},
journal = {APJ}
}

@article{Tayal2011,
doi = {10.1088/0067-0049/195/2/12},
url = {https://dx.doi.org/10.1088/0067-0049/195/2/12},
year = {2011},
month = {jul},
publisher = {AAS},
volume = {195},
number = {2},
pages = {12},
author = {S. S. Tayal},
title = {ELECTRON EXCITATION COLLISION STRENGTHS FOR SINGLY IONIZED NITROGEN},
journal = {\apjs}
}

@ARTICLE{Butler1989,
       author = {{Butler}, K. and {Zeippen}, C.~J.},
        title = "{Effective collision strengths for fine-structure forbidden transitions in the 3p3 configuration of Cl(III)}",
      journal = {\aap},
         year = 1989,
        month = jan,
       volume = {208},
       number = {1-2},
        pages = {337-344},
       adsurl = {https://ui.adsabs.harvard.edu/abs/1989A&A...208..337B}
}

@ARTICLE{Burgos2009,
       author = {{Munoz Burgos}, J.~M. and {Loch}, S.~D. and {Ballance}, C.~P. and {Boivin}, R.~F.},
        title = "{Electron-impact excitation of Ar$^{2+}$}",
      journal = {\aap},
         year = 2009,
        month = jun,
       volume = {500},
       number = {3},
        pages = {1253-1261},
          doi = {10.1051/0004-6361/200911743},
       adsurl = {https://ui.adsabs.harvard.edu/abs/2009A&A...500.1253M}
}

@article{Porter_2013,
    author = {Porter, R L and Ferland, G J and Storey, P J and Detisch, M J},
    title = "{Erratum: ‘Improved He i emissivities in the Case B approximation’}",
    journal = {\mnras: Letters},
    volume = {433},
    number = {1},
    pages = {L89-L90},
    year = {2013},
    month = {04},
    issn = {1745-3925},
    doi = {10.1093/mnrasl/slt049},
    url = {https://doi.org/10.1093/mnrasl/slt049},
    eprint = {https://academic.oup.com/mnrasl/article-pdf/433/1/L89/54658423/mnrasl\_433\_1\_l89.pdf},
}

@article{Storey1995,
    author = {Storey, P. J. and Hummer, D. G.},
    title = "{Recombination line intensities for hydrogenic ions-IV. Total recombination coefficients and machine-readable tables for Z=1 to 8}",
    journal = {MNRAS},
    volume = {272},
    number = {1},
    pages = {41-48},
    year = {1995},
    month = {01},
    issn = {0035-8711},
    doi = {10.1093/mnras/272.1.41},
    url = {https://doi.org/10.1093/mnras/272.1.41},
    eprint = {https://academic.oup.com/mnras/article-pdf/272/1/41/2897226/mnras272-0041.pdf},
}

@article{Porter2012,
    author = {Porter, R. L. and Ferland, G. J. and Storey, P. J. and Detisch, M. J.},
    title = "{Improved He I emissivities in the case B approximation}",
    journal = {MNRAS: Letters},
    volume = {425},
    number = {1},
    pages = {L28-L31},
    year = {2012},
    month = {09},
    issn = {1745-3925},
    doi = {10.1111/j.1745-3933.2012.01300.x},
    url = {https://doi.org/10.1111/j.1745-3933.2012.01300.x},
    eprint = {https://academic.oup.com/mnrasl/article-pdf/425/1/L28/56939875/mnrasl\_425\_1\_l28.pdf},
}

@ARTICLE{Bautista2015,
       author = {{Bautista}, Manuel A. and {Fivet}, Vanessa and {Ballance}, Connor and {Quinet}, Pascal and {Ferland}, Gary and {Mendoza}, Claudio and {Kallman}, Timothy R.},
        title = "{Atomic Data and Spectral Model for Fe II}",
      journal = {\apj},
         year = 2015,
        month = aug,
       volume = {808},
       number = {2},
          eid = {174},
        pages = {174},
          doi = {10.1088/0004-637X/808/2/174},
       adsurl = {https://ui.adsabs.harvard.edu/abs/2015ApJ...808..174B}
}

@ARTICLE{BAndell2014,
       author = {{Badnell}, N.~R. and {Ballance}, C.~P.},
        title = "{Electron-impact Excitation of Fe$^{2 +}$: A Comparison of Intermediate Coupling Frame Transformation, Breit-Pauli and Dirac R-matrix Calculations}",
      journal = {\apj},
         year = 2014,
        month = apr,
       volume = {785},
       number = {2},
          eid = {99},
        pages = {99},
          doi = {10.1088/0004-637X/785/2/99},
archivePrefix = {arXiv},
       eprint = {1402.4743},
 primaryClass = {astro-ph.GA},
       adsurl = {https://ui.adsabs.harvard.edu/abs/2014ApJ...785...99B}
}

@ARTICLE{Fischer1985,
       author = {{Froese Fischer}, Charlotte and {Saha}, H.~P.},
        title = "{Multiconfiguration Hartree-Fock Results with Breit-Pauli Corrections for Transitions in the Carbon Sequence}",
      journal = {\physscr},
         year = 1985,
        month = sep,
       volume = {32},
       number = {3},
        pages = {181-194},
          doi = {10.1088/0031-8949/32/3/004},
       adsurl = {https://ui.adsabs.harvard.edu/abs/1985PhyS...32..181F}
}

@article{Johnson_1987,
doi = {10.1088/0022-3700/20/11/022},
url = {https://doi.org/10.1088/0022-3700/20/11/022},
year = {1987},
month = {jun},
publisher = {},
volume = {20},
number = {11},
pages = {2553},
author = {C T Johnson and P G Burke and A E Kingston},
title = {Electron scattering from the fine structure levels within the 1s22s22p2 3Pe ground state of C I},
journal = {Journal of Physics B: Atomic and Molecular Physics}
}

@ARTICLE{Rynkun2019,
       author = {{Rynkun}, P. and {Gaigalas}, G. and {J{\"o}nsson}, P.},
        title = "{Theoretical investigation of energy levels and transition data for S II, Cl III, Ar IV}",
      journal = {\aap},
         year = 2019,
        month = mar,
       volume = {623},
          eid = {A155},
        pages = {A155},
          doi = {10.1051/0004-6361/201834931},
       adsurl = {https://ui.adsabs.harvard.edu/abs/2019A&A...623A.155R}
}

@ARTICLE{Ramsbottom1997,
       author = {{Ramsbottom}, C.~A. and {Bell}, K.~L.},
        title = "{Effective Collision Strengths for Electron-Impact Excitation of Triphy Ionized Argon}",
      journal = {Atomic Data and Nuclear Data Tables},
         year = 1997,
        month = jan,
       volume = {66},
        pages = {65},
          doi = {10.1006/adnd.1997.0741},
       adsurl = {https://ui.adsabs.harvard.edu/abs/1997ADNDT..66...65R}
}

@article{Mendoza1983,
    author = {Mendoza, C. and Zeippen, C. J.},
    title = {Transition probabilities for forbidden lines in the 3p4 configuration – III},
    journal = {\mnras},
    volume = {202},
    number = {4},
    pages = {981-986},
    year = {1983},
    month = {04},
    issn = {0035-8711},
    doi = {10.1093/mnras/202.4.981},
    url = {https://doi.org/10.1093/mnras/202.4.981},
    eprint = {https://academic.oup.com/mnras/article-pdf/202/4/981/18194432/mnras202-0981.pdf},
}

@ARTICLE{Tayal2004,
       author = {{Tayal}, S.~S.},
        title = "{Electron impact excitation of fine-structure levels in Cl II}",
      journal = {\aap},
         year = 2004,
        month = apr,
       volume = {418},
        pages = {363-369},
          doi = {10.1051/0004-6361:20034384},
       adsurl = {https://ui.adsabs.harvard.edu/abs/2004A&A...418..363T}
}

@ARTICLE{Galavis1997,
       author = {{Galavis}, M.~E. and {Mendoza}, C. and {Zeippen}, C.~J.},
        title = "{Atomic data from the IRON Project. XXII. Radiative rates for forbidden transitions within the ground configuration of ions in the carbon and oxygen isoelectronic sequences}",
      journal = {\aaps},
         year = 1997,
        month = may,
       volume = {123},
        pages = {159-171},
          doi = {10.1051/aas:1997344},
       adsurl = {https://ui.adsabs.harvard.edu/abs/1997A&AS..123..159G}
}

@ARTICLE{McLaughlin2000,
       author = {{McLaughlin}, Brendan M. and {Bell}, Kenneth L.},
        title = "{Electron collisional excitation of Ne III: (1s$^{2}$2s$^{2}$2p$^{4}$ $^{3}$P$_{2,1,0}$,$^{1}$D$_{2}$,$^{1}$S$_{0}$) fine-structure transitions}",
      journal = {Journal of Physics B Atomic Molecular Physics},
         year = 2000,
        month = feb,
       volume = {33},
       number = {4},
        pages = {597-613},
          doi = {10.1088/0953-4075/33/4/301},
       adsurl = {https://ui.adsabs.harvard.edu/abs/2000JPhB...33..597M}
}

@ARTICLE{Galavis1995,
       author = {{Galavis}, M.~E. and {Mendoza}, C. and {Zeippen}, C.~J.},
        title = "{Atomic data from the IRON Project. X. Effective collision strengths for infrared transitions in silicon- and sulphur-like ions.}",
      journal = {\aaps},
         year = 1995,
        month = jun,
       volume = {111},
        pages = {347},
       adsurl = {https://ui.adsabs.harvard.edu/abs/1995A&AS..111..347G}
}

@ARTICLE{Mendoza_2023,
       author = {{Mendoza}, Claudio and {M{\'e}ndez-Delgado}, Jos{\'e} E. and {Bautista}, Manuel and {Garc{\'\i}a-Rojas}, Jorge and {Morisset}, Christophe},
        title = "{Atomic Data Assessment with PyNeb: Radiative and Electron Impact Excitation Rates for [Fe II] and [Fe III]}",
      journal = {Atoms},
         year = 2023,
        month = apr,
       volume = {11},
       number = {4},
          eid = {63},
        pages = {63},
          doi = {10.3390/atoms11040063},
archivePrefix = {arXiv},
       eprint = {2304.01298},
 primaryClass = {astro-ph.GA},
       adsurl = {https://ui.adsabs.harvard.edu/abs/2023Atoms..11...63M}
}

@article{Huang_1985,
title = {Energy-level scheme and transition probabilities of Si-like ions},
journal = {Atomic Data and Nuclear Data Tables},
volume = {32},
number = {3},
pages = {503-566},
year = {1985},
issn = {0092-640X},
doi = {https://doi.org/10.1016/0092-640X(85)90022-1},
url = {https://www.sciencedirect.com/science/article/pii/0092640X85900221},
author = {Keh-Ning Huang}
}

@ARTICLE{Kaufman_1986,
       author = {{Kaufman}, Victor and {Sugar}, Jack},
        title = "{Forbidden Lines in ns$^{2}$np$^{k}$ Ground Configurations and nsnp Excited Configurations of Beryllium through Molybdenum Atoms and Ions}",
      journal = {Journal of Physical and Chemical Reference Data},
         year = 1986,
        month = jan,
       volume = {15},
       number = {1},
        pages = {321-426},
          doi = {10.1063/1.555775},
       adsurl = {https://ui.adsabs.harvard.edu/abs/1986JPCRD..15..321K}
}

@ARTICLE{Mendoza_1982,
       author = {{Mendoza}, C. and {Zeippen}, C.~J.},
        title = "{Transition probabilities for forbidden lines in the 3p$^{2}$ configuration -II.}",
      journal = {\mnras},
         year = 1982,
        month = jun,
       volume = {199},
        pages = {1025-1032},
          doi = {10.1093/mnras/199.4.1025},
       adsurl = {https://ui.adsabs.harvard.edu/abs/1982MNRAS.199.1025M}
}

@article{Ellis_1984,
doi = {10.1088/0031-8949/30/4/007},
url = {https://doi.org/10.1088/0031-8949/30/4/007},
year = {1984},
month = {oct},
publisher = {},
volume = {30},
number = {4},
pages = {255},
author = {D G Ellis and I Martinson},
title = {The 3s3p3 5So Level in the Silicon Isoelectronic Sequence},
journal = {Physica Scripta}
}

@ARTICLE{Moechler2014,
       author = {{Moehler}, S. and {Modigliani}, A. and {Freudling}, W. and {Giammichele}, N. and {Gianninas}, A. and {Gonneau}, A. and {Kausch}, W. and {Lan{\c{c}}on}, A. and {Noll}, S. and {Rauch}, T. and {Vinther}, J.},
        title = "{Flux calibration of medium-resolution spectra from 300 nm to 2500 nm: Model reference spectra and telluric correction}",
      journal = {\aap},
         year = 2014,
        month = aug,
       volume = {568},
          eid = {A9},
        pages = {A9},
          doi = {10.1051/0004-6361/201423790},
archivePrefix = {arXiv},
       eprint = {1408.1797},
 primaryClass = {astro-ph.IM},
       adsurl = {https://ui.adsabs.harvard.edu/abs/2014A&A...568A...9M}
}

@ARTICLE{Mallema1998,
       author = {{Mellema}, G. and {Raga}, A.~C. and {Canto}, J. and {Lundqvist}, P. and {Balick}, B. and {Steffen}, W. and {Noriega-Crespo}, A.},
        title = "{Photo-evaporation of clumps in planetary nebulae}",
      journal = {\aap},
         year = 1998,
        month = mar,
       volume = {331},
        pages = {335-346},
          doi = {10.48550/arXiv.astro-ph/9710205},
archivePrefix = {arXiv},
       eprint = {astro-ph/9710205},
 primaryClass = {astro-ph},
       adsurl = {https://ui.adsabs.harvard.edu/abs/1998A&A...331..335M}
}

@ARTICLE{PyEmily_2026,
       author = {{Tu}, Zhijun and {Fang}, Xuan and {Garc{\'\i}a-Rojas}, Jorge and {Williams}, Robert and {Liu}, Jifeng},
        title = "{PyEMILI: A New Generation Computer-aided Spectral-line Identifier. II. Emission-line Identification and Plasma Diagnostics of a Sample of Gaseous Nebulae}",
      journal = {\apjs},
         year = 2026,
        month = feb,
       volume = {282},
       number = {2},
          eid = {32},
        pages = {32},
          doi = {10.3847/1538-4365/ae2259},
archivePrefix = {arXiv},
       eprint = {2511.14451},
 primaryClass = {astro-ph.SR},
       adsurl = {https://ui.adsabs.harvard.edu/abs/2026ApJS..282...32T}
}

@ARTICLE{PyEmily_2025,
       author = {{Tu}, Zhijun and {Fang}, Xuan and {Williams}, Robert and {Liu}, Jifeng},
        title = "{PyEMILI: A New Generation Computer-aided Spectral Line Identifier}",
      journal = {\apjs},
         year = 2025,
        month = mar,
       volume = {277},
       number = {1},
          eid = {13},
        pages = {13},
          doi = {10.3847/1538-4365/adae00},
archivePrefix = {arXiv},
       eprint = {2501.11845},
 primaryClass = {astro-ph.SR},
       adsurl = {https://ui.adsabs.harvard.edu/abs/2025ApJS..277...13T}
}

@ARTICLE{Vernet_2011,
       author = {{Vernet}, J. and {Dekker}, H. and {D'Odorico}, S. and {Kaper}, L. and {Kjaergaard}, P. and {Hammer}, F. and {Randich}, S. and {Zerbi}, F. and {Groot}, P.~J. and {Hjorth}, J. and {Guinouard}, I. and {Navarro}, R. and {Adolfse}, T. and {Albers}, P.~W. and {Amans}, J.-P. and {Andersen}, J.~J. and {Andersen}, M.~I. and {Binetruy}, P. and {Bristow}, P. and {Castillo}, R. and {Chemla}, F. and {Christensen}, L. and {Conconi}, P. and {Conzelmann}, R. and {Dam}, J. and {de Caprio}, V. and {de Ugarte Postigo}, A. and {Delabre}, B. and {di Marcantonio}, P. and {Downing}, M. and {Elswijk}, E. and {Finger}, G. and {Fischer}, G. and {Flores}, H. and {Fran{\c{c}}ois}, P. and {Goldoni}, P. and {Guglielmi}, L. and {Haigron}, R. and {Hanenburg}, H. and {Hendriks}, I. and {Horrobin}, M. and {Horville}, D. and {Jessen}, N.~C. and {Kerber}, F. and {Kern}, L. and {Kiekebusch}, M. and {Kleszcz}, P. and {Klougart}, J. and {Kragt}, J. and {Larsen}, H.~H. and {Lizon}, J.-L. and {Lucuix}, C. and {Mainieri}, V. and {Manuputy}, R. and {Martayan}, C. and {Mason}, E. and {Mazzoleni}, R. and {Michaelsen}, N. and {Modigliani}, A. and {Moehler}, S. and {M{\o}ller}, P. and {Norup S{\o}rensen}, A. and {N{\o}rregaard}, P. and {P{\'e}roux}, C. and {Patat}, F. and {Pena}, E. and {Pragt}, J. and {Reinero}, C. and {Rigal}, F. and {Riva}, M. and {Roelfsema}, R. and {Royer}, F. and {Sacco}, G. and {Santin}, P. and {Schoenmaker}, T. and {Spano}, P. and {Sweers}, E. and {Ter Horst}, R. and {Tintori}, M. and {Tromp}, N. and {van Dael}, P. and {van der Vliet}, H. and {Venema}, L. and {Vidali}, M. and {Vinther}, J. and {Vola}, P. and {Winters}, R. and {Wistisen}, D. and {Wulterkens}, G. and {Zacchei}, A.},
        title = "{X-shooter, the new wide band intermediate resolution spectrograph at the ESO Very Large Telescope}",
      journal = {\aap},
         year = 2011,
        month = dec,
       volume = {536},
          eid = {A105},
        pages = {A105},
          doi = {10.1051/0004-6361/201117752},
archivePrefix = {arXiv},
       eprint = {1110.1944},
 primaryClass = {astro-ph.IM},
       adsurl = {https://ui.adsabs.harvard.edu/abs/2011A&A...536A.105V}
}

@ARTICLE{Mendez-Delgado_2021,
       author = {{M{\'e}ndez-Delgado}, J.~E. and {Esteban}, C. and {Garc{\'\i}a-Rojas}, J. and {Henney}, W.~J. and {Mesa-Delgado}, A. and {Arellano-C{\'o}rdova}, K.~Z.},
        title = "{Photoionized Herbig-Haro objects in the Orion Nebula through deep high-spectral resolution spectroscopy - I. HH 529 II and III}",
      journal = {\mnras},
         year = 2021,
        month = apr,
       volume = {502},
       number = {2},
        pages = {1703-1739},
          doi = {10.1093/mnras/stab068},
archivePrefix = {arXiv},
       eprint = {2101.02191},
 primaryClass = {astro-ph.GA},
       adsurl = {https://ui.adsabs.harvard.edu/abs/2021MNRAS.502.1703M}
}

@ARTICLE{Wolniewicz_1998,
       author = {{Wolniewicz}, L. and {Simbotin}, I. and {Dalgarno}, A.},
        title = "{Quadrupole Transition Probabilities for the Excited Rovibrational States of H$_{2}$}",
      journal = {\apjs},
         year = 1998,
        month = apr,
       volume = {115},
       number = {2},
        pages = {293-313},
          doi = {10.1086/313091},
       adsurl = {https://ui.adsabs.harvard.edu/abs/1998ApJS..115..293W}
}

@ARTICLE{Goncalves_2006,
       author = {{Gon{\c{c}}alves}, D.~R. and {Ercolano}, B. and {Carnero}, A. and {Mampaso}, A. and {Corradi}, R.~L.~M.},
        title = "{On the nitrogen abundance of fast, low-ionization emission regions: the outer knots of the planetary nebula NGC 7009}",
      journal = {\mnras},
         year = 2006,
        month = jan,
       volume = {365},
       number = {3},
        pages = {1039-1049},
          doi = {10.1111/j.1365-2966.2005.09794.x},
archivePrefix = {arXiv},
       eprint = {astro-ph/0511114},
 primaryClass = {astro-ph},
       adsurl = {https://ui.adsabs.harvard.edu/abs/2006MNRAS.365.1039G}
}

@ARTICLE{VanHoof2010,
       author = {{van Hoof}, P.~A.~M. and {van de Steene}, G.~C. and {Barlow}, M.~J. and {Exter}, K.~M. and {Sibthorpe}, B. and {Ueta}, T. and {Peris}, V. and {Groenewegen}, M.~A.~T. and {Blommaert}, J.~A.~D.~L. and {Cohen}, M. and {De Meester}, W. and {Ferland}, G.~J. and {Gear}, W.~K. and {Gomez}, H.~L. and {Hargrave}, P.~C. and {Huygen}, E. and {Ivison}, R.~J. and {Jean}, C. and {Leeks}, S.~J. and {Lim}, T.~L. and {Olofsson}, G. and {Polehampton}, E.~T. and {Regibo}, S. and {Royer}, P. and {Swinyard}, B.~M. and {Vandenbussche}, B. and {van Winckel}, H. and {Waelkens}, C. and {Walker}, H.~J. and {Wesson}, R.},
        title = "{Herschel images of NGC 6720: H$_{2}$ formation on dust grains}",
      journal = {\aap},
         year = 2010,
        month = jul,
       volume = {518},
          eid = {L137},
        pages = {L137},
          doi = {10.1051/0004-6361/201014590},
archivePrefix = {arXiv},
       eprint = {1005.1502},
 primaryClass = {astro-ph.SR},
       adsurl = {https://ui.adsabs.harvard.edu/abs/2010A&A...518L.137V}
}

@ARTICLE{Riera_2011,
       author = {{Riera}, A. and {Raga}, A.~C. and {Vel{\'a}zquez}, P.~F. and {Haro-Corzo}, S.~A.~R. and {Kajdic}, P.},
        title = "{STIS optical spectroscopy of the lobes of CRL 618}",
      journal = {\aap},
         year = 2011,
        month = sep,
       volume = {533},
          eid = {A118},
        pages = {A118},
          doi = {10.1051/0004-6361/201014891},
       adsurl = {https://ui.adsabs.harvard.edu/abs/2011A&A...533A.118R}
}

@ARTICLE{Hollenbach_1989,
       author = {{Hollenbach}, David and {McKee}, Christopher F.},
        title = "{Molecule Formation and Infrared Emission in Fast Interstellar Shocks. III. Results for J Shocks in Molecular Clouds}",
      journal = {\apj},
         year = 1989,
        month = jul,
       volume = {342},
        pages = {306},
          doi = {10.1086/167595},
       adsurl = {https://ui.adsabs.harvard.edu/abs/1989ApJ...342..306H}
}

@ARTICLE{Storey2017,
       author = {{Storey}, P.~J. and {Sochi}, Taha and {Bastin}, Robert},
        title = "{Recombination coefficients for O II lines in nebular conditions}",
      journal = {\mnras},
         year = 2017,
        month = sep,
       volume = {470},
       number = {1},
        pages = {379-389},
          doi = {10.1093/mnras/stx1189},
archivePrefix = {arXiv},
       eprint = {1703.09982},
 primaryClass = {physics.atom-ph},
       adsurl = {https://ui.adsabs.harvard.edu/abs/2017MNRAS.470..379S}
}

@ARTICLE{Davey2000,
       author = {{Davey}, A.~R. and {Storey}, P.~J. and {Kisielius}, R.},
        title = "{Recombination coefficients for C II lines}",
      journal = {\aaps},
         year = 2000,
        month = feb,
       volume = {142},
        pages = {85-94},
          doi = {10.1051/aas:2000139},
       adsurl = {https://ui.adsabs.harvard.edu/abs/2000A&AS..142...85D}
}

@ARTICLE{Fang2011_NII,
       author = {{Fang}, X. and {Storey}, P.~J. and {Liu}, X.-W.},
        title = "{New effective recombination coefficients for nebular N ii lines}",
      journal = {\aap},
         year = 2011,
        month = jun,
       volume = {530},
          eid = {A18},
        pages = {A18},
          doi = {10.1051/0004-6361/201116511},
archivePrefix = {arXiv},
       eprint = {1103.5558},
 primaryClass = {astro-ph.SR},
       adsurl = {https://ui.adsabs.harvard.edu/abs/2011A&A...530A..18F}
}

@MISC{Fang2013_CII,
       author = {{Fang}, X. and {Storey}, P.~J. and {Liu}, X.-W.},
        title = "{New effective recombination coefficients for nebular N II lines (Corrigendum)}",
         year = 2013,
        month = feb,
          eid = {C2},
        pages = {C2},
          doi = {10.1051/0004-6361/201116511e},
    publisher = {EDP},
       adsurl = {https://ui.adsabs.harvard.edu/abs/2013A&A...550C...2F}
}

@ARTICLE{Liller_1966,
       author = {{Liller}, Martha H. and {Welther}, Barbara L. and {Liller}, William},
        title = "{Angular Expansions of Planetary Nebulae}",
      journal = {\apj},
         year = 1966,
        month = apr,
       volume = {144},
        pages = {280},
          doi = {10.1086/148603},
       adsurl = {https://ui.adsabs.harvard.edu/abs/1966ApJ...144..280L}
}

@ARTICLE{Guerrero2013,
       author = {{Guerrero}, Mart{\'\i}n A. and {Toal{\'a}}, Jes{\'u}s A. and {Medina}, Juan J. and {Luridiana}, Valentina and {Miranda}, Luis F. and {Riera}, Angels and {Vel{\'a}zquez}, Pablo F.},
        title = "{Unveiling shocks in planetary nebulae}",
      journal = {\aap},
         year = 2013,
        month = sep,
       volume = {557},
          eid = {A121},
        pages = {A121},
          doi = {10.1051/0004-6361/201321786},
archivePrefix = {arXiv},
       eprint = {1307.8435},
 primaryClass = {astro-ph.SR},
       adsurl = {https://ui.adsabs.harvard.edu/abs/2013A&A...557A.121G}
}

@ARTICLE{GAIA_2021,
       author = {{Gaia Collaboration} and {Brown}, A.~G.~A. and {Vallenari}, A. and {Prusti}, T. and {de Bruijne}, J.~H.~J. and {Babusiaux}, C. and {Biermann}, M. and {Creevey}, O.~L. and {Evans}, D.~W. and {Eyer}, L. and {Hutton}, A. and {Jansen}, F. and {Jordi}, C. and {Klioner}, S.~A. and {Lammers}, U. and {Lindegren}, L. and {Luri}, X. and {Mignard}, F. and {Panem}, C. and {Pourbaix}, D. and {Randich}, S. and {Sartoretti}, P. and {Soubiran}, C. and {Walton}, N.~A. and {Arenou}, F. and {Bailer-Jones}, C.~A.~L. and {Bastian}, U. and {Cropper}, M. and {Drimmel}, R. and {Katz}, D. and {Lattanzi}, M.~G. and {van Leeuwen}, F. and {Bakker}, J. and {Cacciari}, C. and {Casta{\~n}eda}, J. and {De Angeli}, F. and {Ducourant}, C. and {Fabricius}, C. and {Fouesneau}, M. and {Fr{\'e}mat}, Y. and {Guerra}, R. and {Guerrier}, A. and {Guiraud}, J. and {Jean-Antoine Piccolo}, A. and {Masana}, E. and {Messineo}, R. and {Mowlavi}, N. and {Nicolas}, C. and {Nienartowicz}, K. and {Pailler}, F. and {Panuzzo}, P. and {Riclet}, F. and {Roux}, W. and {Seabroke}, G.~M. and {Sordo}, R. and {Tanga}, P. and {Th{\'e}venin}, F. and {Gracia-Abril}, G. and {Portell}, J. and {Teyssier}, D. and {Altmann}, M. and {Andrae}, R. and {Bellas-Velidis}, I. and {Benson}, K. and {Berthier}, J. and {Blomme}, R. and {Brugaletta}, E. and {Burgess}, P.~W. and {Busso}, G. and {Carry}, B. and {Cellino}, A. and {Cheek}, N. and {Clementini}, G. and {Damerdji}, Y. and {Davidson}, M. and {Delchambre}, L. and {Dell'Oro}, A. and {Fern{\'a}ndez-Hern{\'a}ndez}, J. and {Galluccio}, L. and {Garc{\'\i}a-Lario}, P. and {Garcia-Reinaldos}, M. and {Gonz{\'a}lez-N{\'u}{\~n}ez}, J. and {Gosset}, E. and {Haigron}, R. and {Halbwachs}, J.-L. and {Hambly}, N.~C. and {Harrison}, D.~L. and {Hatzidimitriou}, D. and {Heiter}, U. and {Hern{\'a}ndez}, J. and {Hestroffer}, D. and {Hodgkin}, S.~T. and {Holl}, B. and {Jan{\ss}en}, K. and {Jevardat de Fombelle}, G. and {Jordan}, S. and {Krone-Martins}, A. and {Lanzafame}, A.~C. and {L{\"o}ffler}, W. and {Lorca}, A. and {Manteiga}, M. and {Marchal}, O. and {Marrese}, P.~M. and {Moitinho}, A. and {Mora}, A. and {Muinonen}, K. and {Osborne}, P. and {Pancino}, E. and {Pauwels}, T. and {Petit}, J.-M. and {Recio-Blanco}, A. and {Richards}, P.~J. and {Riello}, M. and {Rimoldini}, L. and {Robin}, A.~C. and {Roegiers}, T. and {Rybizki}, J. and {Sarro}, L.~M. and {Siopis}, C. and {Smith}, M. and {Sozzetti}, A. and {Ulla}, A. and {Utrilla}, E. and {van Leeuwen}, M. and {van Reeven}, W. and {Abbas}, U. and {Abreu Aramburu}, A. and {Accart}, S. and {Aerts}, C. and {Aguado}, J.~J. and {Ajaj}, M. and {Altavilla}, G. and {{\'A}lvarez}, M.~A. and {{\'A}lvarez Cid-Fuentes}, J. and {Alves}, J. and {Anderson}, R.~I. and {Anglada Varela}, E. and {Antoja}, T. and {Audard}, M. and {Baines}, D. and {Baker}, S.~G. and {Balaguer-N{\'u}{\~n}ez}, L. and {Balbinot}, E. and {Balog}, Z. and {Barache}, C. and {Barbato}, D. and {Barros}, M. and {Barstow}, M.~A. and {Bartolom{\'e}}, S. and {Bassilana}, J.-L. and {Bauchet}, N. and {Baudesson-Stella}, A. and {Becciani}, U. and {Bellazzini}, M. and {Bernet}, M. and {Bertone}, S. and {Bianchi}, L. and {Blanco-Cuaresma}, S. and {Boch}, T. and {Bombrun}, A. and {Bossini}, D. and {Bouquillon}, S. and {Bragaglia}, A. and {Bramante}, L. and {Breedt}, E. and {Bressan}, A. and {Brouillet}, N. and {Bucciarelli}, B. and {Burlacu}, A. and {Busonero}, D. and {Butkevich}, A.~G. and {Buzzi}, R. and {Caffau}, E. and {Cancelliere}, R. and {C{\'a}novas}, H. and {Cantat-Gaudin}, T. and {Carballo}, R. and {Carlucci}, T. and {Carnerero}, M.~I. and {Carrasco}, J.~M. and {Casamiquela}, L. and {Castellani}, M. and {Castro-Ginard}, A. and {Castro Sampol}, P. and {Chaoul}, L. and {Charlot}, P. and {Chemin}, L. and {Chiavassa}, A. and {Cioni}, M.-R.~L. and {Comoretto}, G. and {Cooper}, W.~J. and {Cornez}, T. and {Cowell}, S. and {Crifo}, F. and {Crosta}, M. and {Crowley}, C. and {Dafonte}, C. and {Dapergolas}, A. and {David}, M. and {David}, P.},
        title = "{Gaia Early Data Release 3. Summary of the contents and survey properties}",
      journal = {\aap},
         year = 2021,
        month = may,
       volume = {649},
          eid = {A1},
        pages = {A1},
          doi = {10.1051/0004-6361/202039657},
archivePrefix = {arXiv},
       eprint = {2012.01533},
 primaryClass = {astro-ph.GA},
       adsurl = {https://ui.adsabs.harvard.edu/abs/2021A&A...649A...1G}
}

@INPROCEEDINGS{Modigliani_2010,
       author = {{Modigliani}, Andrea and {Goldoni}, Paolo and {Royer}, Fr{\'e}d{\'e}ric and {Haigron}, Regis and {Guglielmi}, Laurent and {Fran{\c{c}}ois}, Patrick and {Horrobin}, Matthew and {Bristow}, Paul and {Vernet}, Joel and {Moehler}, Sabine and {Kerber}, Florian and {Ballester}, Pascal and {Mason}, Elena and {Christensen}, Lise},
        title = "{The X-shooter pipeline}",
    booktitle = {Observatory Operations: Strategies, Processes, and Systems III},
         year = 2010,
       editor = {{Silva}, David R. and {Peck}, Alison B. and {Soifer}, B. Thomas},
       series = {Society of Photo-Optical Instrumentation Engineers (SPIE) Conference Series},
       volume = {7737},
        month = jul,
          eid = {773728},
        pages = {773728},
          doi = {10.1117/12.857211},
       adsurl = {https://ui.adsabs.harvard.edu/abs/2010SPIE.7737E..28M}
}

@ARTICLE{Vernet_2010,
       author = {{Vernet}, J. and {Kerber}, F. and {Mainieri}, V. and {Rauch}, T. and {Saitta}, F. and {D'Odorico}, S. and {Bohlin}, R. and {Ivanov}, V. and {Lidman}, C. and {Mason}, E. and {Smette}, A. and {Walsh}, J. and {Fosbury}, R. and {Goldoni}, P. and {Groot}, P. and {Hammer}, F. and {Kaper}, L. and {Horrobin}, M. and {Kjaergaard-Rasmussen}, P. and {Royer}, F.},
        title = "{Building-up a database of spectro-photometric standards from the UV to the NIR}",
      journal = {Highlights of Astronomy},
         year = 2010,
        month = nov,
       volume = {15},
        pages = {535-535},
          doi = {10.1017/S1743921310010574},
       adsurl = {https://ui.adsabs.harvard.edu/abs/2010HiA....15..535V}
}

@ARTICLE{Hamuy_1994,
       author = {{Hamuy}, Mario and {Suntzeff}, N.~B. and {Heathcote}, S.~R. and {Walker}, A.~R. and {Gigoux}, P. and {Phillips}, M.~M.},
        title = "{Southern Spectrophotometric Standards. II}",
      journal = {\pasp},
         year = 1994,
        month = jun,
       volume = {106},
        pages = {566},
          doi = {10.1086/133417},
       adsurl = {https://ui.adsabs.harvard.edu/abs/1994PASP..106..566H}
}

@ARTICLE{GAIA_2016,
       author = {{Gaia Collaboration} and {Prusti}, T. and {de Bruijne}, J.~H.~J. and {Brown}, A.~G.~A. and {Vallenari}, A. and {Babusiaux}, C. and {Bailer-Jones}, C.~A.~L. and {Bastian}, U. and {Biermann}, M. and {Evans}, D.~W. and {Eyer}, L. and {Jansen}, F. and {Jordi}, C. and {Klioner}, S.~A. and {Lammers}, U. and {Lindegren}, L. and {Luri}, X. and {Mignard}, F. and {Milligan}, D.~J. and {Panem}, C. and {Poinsignon}, V. and {Pourbaix}, D. and {Randich}, S. and {Sarri}, G. and {Sartoretti}, P. and {Siddiqui}, H.~I. and {Soubiran}, C. and {Valette}, V. and {van Leeuwen}, F. and {Walton}, N.~A. and {Aerts}, C. and {Arenou}, F. and {Cropper}, M. and {Drimmel}, R. and {H{\o}g}, E. and {Katz}, D. and {Lattanzi}, M.~G. and {O'Mullane}, W. and {Grebel}, E.~K. and {Holland}, A.~D. and {Huc}, C. and {Passot}, X. and {Bramante}, L. and {Cacciari}, C. and {Casta{\~n}eda}, J. and {Chaoul}, L. and {Cheek}, N. and {De Angeli}, F. and {Fabricius}, C. and {Guerra}, R. and {Hern{\'a}ndez}, J. and {Jean-Antoine-Piccolo}, A. and {Masana}, E. and {Messineo}, R. and {Mowlavi}, N. and {Nienartowicz}, K. and {Ord{\'o}{\~n}ez-Blanco}, D. and {Panuzzo}, P. and {Portell}, J. and {Richards}, P.~J. and {Riello}, M. and {Seabroke}, G.~M. and {Tanga}, P. and {Th{\'e}venin}, F. and {Torra}, J. and {Els}, S.~G. and {Gracia-Abril}, G. and {Comoretto}, G. and {Garcia-Reinaldos}, M. and {Lock}, T. and {Mercier}, E. and {Altmann}, M. and {Andrae}, R. and {Astraatmadja}, T.~L. and {Bellas-Velidis}, I. and {Benson}, K. and {Berthier}, J. and {Blomme}, R. and {Busso}, G. and {Carry}, B. and {Cellino}, A. and {Clementini}, G. and {Cowell}, S. and {Creevey}, O. and {Cuypers}, J. and {Davidson}, M. and {De Ridder}, J. and {de Torres}, A. and {Delchambre}, L. and {Dell'Oro}, A. and {Ducourant}, C. and {Fr{\'e}mat}, Y. and {Garc{\'\i}a-Torres}, M. and {Gosset}, E. and {Halbwachs}, J.-L. and {Hambly}, N.~C. and {Harrison}, D.~L. and {Hauser}, M. and {Hestroffer}, D. and {Hodgkin}, S.~T. and {Huckle}, H.~E. and {Hutton}, A. and {Jasniewicz}, G. and {Jordan}, S. and {Kontizas}, M. and {Korn}, A.~J. and {Lanzafame}, A.~C. and {Manteiga}, M. and {Moitinho}, A. and {Muinonen}, K. and {Osinde}, J. and {Pancino}, E. and {Pauwels}, T. and {Petit}, J.-M. and {Recio-Blanco}, A. and {Robin}, A.~C. and {Sarro}, L.~M. and {Siopis}, C. and {Smith}, M. and {Smith}, K.~W. and {Sozzetti}, A. and {Thuillot}, W. and {van Reeven}, W. and {Viala}, Y. and {Abbas}, U. and {Abreu Aramburu}, A. and {Accart}, S. and {Aguado}, J.~J. and {Allan}, P.~M. and {Allasia}, W. and {Altavilla}, G. and {{\'A}lvarez}, M.~A. and {Alves}, J. and {Anderson}, R.~I. and {Andrei}, A.~H. and {Anglada Varela}, E. and {Antiche}, E. and {Antoja}, T. and {Ant{\'o}n}, S. and {Arcay}, B. and {Atzei}, A. and {Ayache}, L. and {Bach}, N. and {Baker}, S.~G. and {Balaguer-N{\'u}{\~n}ez}, L. and {Barache}, C. and {Barata}, C. and {Barbier}, A. and {Barblan}, F. and {Baroni}, M. and {Barrado y Navascu{\'e}s}, D. and {Barros}, M. and {Barstow}, M.~A. and {Becciani}, U. and {Bellazzini}, M. and {Bellei}, G. and {Bello Garc{\'\i}a}, A. and {Belokurov}, V. and {Bendjoya}, P. and {Berihuete}, A. and {Bianchi}, L. and {Bienaym{\'e}}, O. and {Billebaud}, F. and {Blagorodnova}, N. and {Blanco-Cuaresma}, S. and {Boch}, T. and {Bombrun}, A. and {Borrachero}, R. and {Bouquillon}, S. and {Bourda}, G. and {Bouy}, H. and {Bragaglia}, A. and {Breddels}, M.~A. and {Brouillet}, N. and {Br{\"u}semeister}, T. and {Bucciarelli}, B. and {Budnik}, F. and {Burgess}, P. and {Burgon}, R. and {Burlacu}, A. and {Busonero}, D. and {Buzzi}, R. and {Caffau}, E. and {Cambras}, J. and {Campbell}, H. and {Cancelliere}, R. and {Cantat-Gaudin}, T. and {Carlucci}, T. and {Carrasco}, J.~M. and {Castellani}, M. and {Charlot}, P. and {Charnas}, J. and {Charvet}, P. and {Chassat}, F. and {Chiavassa}, A. and {Clotet}, M. and {Cocozza}, G. and {Collins}, R.~S. and {Collins}, P. and {Costigan}, G.},
        title = "{The Gaia mission}",
      journal = {\aap},
         year = 2016,
        month = nov,
       volume = {595},
          eid = {A1},
        pages = {A1},
          doi = {10.1051/0004-6361/201629272},
archivePrefix = {arXiv},
       eprint = {1609.04153},
 primaryClass = {astro-ph.IM},
       adsurl = {https://ui.adsabs.harvard.edu/abs/2016A&A...595A...1G}
}

@ARTICLE{GAIA_2023,
       author = {{Gaia Collaboration} and {Recio-Blanco}, A. and {Kordopatis}, G. and {de Laverny}, P. and {Palicio}, P.~A. and {Spagna}, A. and {Spina}, L. and {Katz}, D. and {Re Fiorentin}, P. and {Poggio}, E. and {McMillan}, P.~J. and {Vallenari}, A. and {Lattanzi}, M.~G. and {Seabroke}, G.~M. and {Casamiquela}, L. and {Bragaglia}, A. and {Antoja}, T. and {Bailer-Jones}, C.~A.~L. and {Schultheis}, M. and {Andrae}, R. and {Fouesneau}, M. and {Cropper}, M. and {Cantat-Gaudin}, T. and {Bijaoui}, A. and {Heiter}, U. and {Brown}, A.~G.~A. and {Prusti}, T. and {de Bruijne}, J.~H.~J. and {Arenou}, F. and {Babusiaux}, C. and {Biermann}, M. and {Creevey}, O.~L. and {Ducourant}, C. and {Evans}, D.~W. and {Eyer}, L. and {Guerra}, R. and {Hutton}, A. and {Jordi}, C. and {Klioner}, S.~A. and {Lammers}, U.~L. and {Lindegren}, L. and {Luri}, X. and {Mignard}, F. and {Panem}, C. and {Pourbaix}, D. and {Randich}, S. and {Sartoretti}, P. and {Soubiran}, C. and {Tanga}, P. and {Walton}, N.~A. and {Bastian}, U. and {Drimmel}, R. and {Jansen}, F. and {van Leeuwen}, F. and {Bakker}, J. and {Cacciari}, C. and {Casta{\~n}eda}, J. and {De Angeli}, F. and {Fabricius}, C. and {Fr{\'e}mat}, Y. and {Galluccio}, L. and {Guerrier}, A. and {Masana}, E. and {Messineo}, R. and {Mowlavi}, N. and {Nicolas}, C. and {Nienartowicz}, K. and {Pailler}, F. and {Panuzzo}, P. and {Riclet}, F. and {Roux}, W. and {Sordo}, R. and {Th{\'e}venin}, F. and {Gracia-Abril}, G. and {Portell}, J. and {Teyssier}, D. and {Altmann}, M. and {Audard}, M. and {Bellas-Velidis}, I. and {Benson}, K. and {Berthier}, J. and {Blomme}, R. and {Burgess}, P.~W. and {Busonero}, D. and {Busso}, G. and {C{\'a}novas}, H. and {Carry}, B. and {Cellino}, A. and {Cheek}, N. and {Clementini}, G. and {Damerdji}, Y. and {Davidson}, M. and {de Teodoro}, P. and {Nu{\~n}ez Campos}, M. and {Delchambre}, L. and {Dell'Oro}, A. and {Esquej}, P. and {Fern{\'a}ndez-Hern{\'a}ndez}, J. and {Fraile}, E. and {Garabato}, D. and {Garc{\'\i}a-Lario}, P. and {Gosset}, E. and {Haigron}, R. and {Halbwachs}, J.-L. and {Hambly}, N.~C. and {Harrison}, D.~L. and {Hern{\'a}ndez}, J. and {Hestroffer}, D. and {Hodgkin}, S.~T. and {Holl}, B. and {Jan{\ss}en}, K. and {Jevardat de Fombelle}, G. and {Jordan}, S. and {Krone-Martins}, A. and {Lanzafame}, A.~C. and {L{\"o}ffler}, W. and {Marchal}, O. and {Marrese}, P.~M. and {Moitinho}, A. and {Muinonen}, K. and {Osborne}, P. and {Pancino}, E. and {Pauwels}, T. and {Reyl{\'e}}, C. and {Riello}, M. and {Rimoldini}, L. and {Roegiers}, T. and {Rybizki}, J. and {Sarro}, L.~M. and {Siopis}, C. and {Smith}, M. and {Sozzetti}, A. and {Utrilla}, E. and {van Leeuwen}, M. and {Abbas}, U. and {{\'A}brah{\'a}m}, P. and {Abreu Aramburu}, A. and {Aerts}, C. and {Aguado}, J.~J. and {Ajaj}, M. and {Aldea-Montero}, F. and {Altavilla}, G. and {{\'A}lvarez}, M.~A. and {Alves}, J. and {Anders}, F. and {Anderson}, R.~I. and {Anglada Varela}, E. and {Baines}, D. and {Baker}, S.~G. and {Balaguer-N{\'u}{\~n}ez}, L. and {Balbinot}, E. and {Balog}, Z. and {Barache}, C. and {Barbato}, D. and {Barros}, M. and {Barstow}, M.~A. and {Bartolom{\'e}}, S. and {Bassilana}, J.-L. and {Bauchet}, N. and {Becciani}, U. and {Bellazzini}, M. and {Berihuete}, A. and {Bernet}, M. and {Bertone}, S. and {Bianchi}, L. and {Binnenfeld}, A. and {Blanco-Cuaresma}, S. and {Boch}, T. and {Bombrun}, A. and {Bossini}, D. and {Bouquillon}, S. and {Bramante}, L. and {Breedt}, E. and {Bressan}, A. and {Brouillet}, N. and {Brugaletta}, E. and {Bucciarelli}, B. and {Burlacu}, A. and {Butkevich}, A.~G. and {Buzzi}, R. and {Caffau}, E. and {Cancelliere}, R. and {Carballo}, R. and {Carlucci}, T. and {Carnerero}, M.~I. and {Carrasco}, J.~M. and {Castellani}, M. and {Castro-Ginard}, A. and {Chaoul}, L. and {Charlot}, P. and {Chemin}, L. and {Chiaramida}, V. and {Chiavassa}, A. and {Chornay}, N. and {Comoretto}, G. and {Contursi}, G. and {Cooper}, W.~J. and {Cornez}, T. and {Cowell}, S. and {Crifo}, F.},
        title = "{Gaia Data Release 3. Chemical cartography of the Milky Way}",
      journal = {\aap},
         year = 2023,
        month = jun,
       volume = {674},
          eid = {A38},
        pages = {A38},
          doi = {10.1051/0004-6361/202243511},
archivePrefix = {arXiv},
       eprint = {2206.05534},
 primaryClass = {astro-ph.GA},
       adsurl = {https://ui.adsabs.harvard.edu/abs/2023A&A...674A..38G}
}

@ARTICLE{Rodriguez2007,
       author = {{Rodr{\'\i}guez}, L.~F. and {G{\'o}mez}, Y.},
        title = "{Proper Motions of the Ansae in the Planetary Nebula NGC7009}",
      journal = {\rmxaa},
         year = 2007,
        month = apr,
       volume = {43},
        pages = {173-177},
          doi = {10.48550/arXiv.astro-ph/0611753},
archivePrefix = {arXiv},
       eprint = {astro-ph/0611753},
 primaryClass = {astro-ph},
       adsurl = {https://ui.adsabs.harvard.edu/abs/2007RMxAA..43..173R}
}

@INPROCEEDINGS{DeMarco2026,
       author = {{De Marco}, Orsola and {Aleman}, Isabel and {Akras}, Stavros},
        title = "{Planetary nebulae}",
    booktitle = {Encyclopedia of Astrophysics, Volume 2},
         year = 2026,
       volume = {2},
        month = jan,
        pages = {604-638},
          doi = {10.1016/B978-0-443-21439-4.00066-3},
       adsurl = {https://ui.adsabs.harvard.edu/abs/2026enap....2..604M}
}

@ARTICLE{Morisset_2015,
       author = {{Morisset}, C. and {Delgado-Inglada}, G. and {Flores-Fajardo}, N.},
        title = "{A virtual observatory for photoionized nebulae: the Mexican Million Models database (3MdB).}",
      journal = {\rmxaa},
         year = 2015,
        month = apr,
       volume = {51},
        pages = {103-120},
          doi = {10.48550/arXiv.1412.5349},
archivePrefix = {arXiv},
       eprint = {1412.5349},
 primaryClass = {astro-ph.GA},
       adsurl = {https://ui.adsabs.harvard.edu/abs/2015RMxAA..51..103M}
}

@ARTICLE{Alarie_2019,
       author = {{Alarie}, A. and {Morisset}, C.},
        title = "{Extensive Online Shock Model Database}",
      journal = {\rmxaa},
         year = 2019,
        month = oct,
       volume = {55},
        pages = {377-394},
          doi = {10.22201/ia.01851101p.2019.55.02.21},
archivePrefix = {arXiv},
       eprint = {1908.08579},
 primaryClass = {astro-ph.GA},
       adsurl = {https://ui.adsabs.harvard.edu/abs/2019RMxAA..55..377A}
}

@ARTICLE{Sutherland_2017,
       author = {{Sutherland}, Ralph S. and {Dopita}, Michael A.},
        title = "{Effects of Preionization in Radiative Shocks. I. Self-consistent Models}",
      journal = {\apjs},
         year = 2017,
        month = apr,
       volume = {229},
       number = {2},
          eid = {34},
        pages = {34},
          doi = {10.3847/1538-4365/aa6541},
archivePrefix = {arXiv},
       eprint = {1702.07453},
 primaryClass = {astro-ph.IM},
       adsurl = {https://ui.adsabs.harvard.edu/abs/2017ApJS..229...34S}
}

@ARTICLE{Allen_2008,
       author = {{Allen}, Mark G. and {Groves}, Brent A. and {Dopita}, Michael A. and {Sutherland}, Ralph S. and {Kewley}, Lisa J.},
        title = "{The MAPPINGS III Library of Fast Radiative Shock Models}",
      journal = {\apjs},
         year = 2008,
        month = sep,
       volume = {178},
       number = {1},
        pages = {20-55},
          doi = {10.1086/589652},
archivePrefix = {arXiv},
       eprint = {0805.0204},
 primaryClass = {astro-ph},
       adsurl = {https://ui.adsabs.harvard.edu/abs/2008ApJS..178...20A}
}

@ARTICLE{Dopita_2016,
       author = {{Dopita}, Michael A. and {Kewley}, Lisa J. and {Sutherland}, Ralph S. and {Nicholls}, David C.},
        title = "{Chemical abundances in high-redshift galaxies: a powerful new emission line diagnostic}",
      journal = {\apss},
         year = 2016,
        month = feb,
       volume = {361},
          eid = {61},
        pages = {61},
          doi = {10.1007/s10509-016-2657-8},
archivePrefix = {arXiv},
       eprint = {1601.01337},
 primaryClass = {astro-ph.GA},
       adsurl = {https://ui.adsabs.harvard.edu/abs/2016Ap&SS.361...61D}
}

@ARTICLE{alarie2019,
       author = {{Alarie}, Alexandre and {Drissen}, Laurent},
        title = "{A multispectral analysis of the northeastern shell of IC 443}",
      journal = {\mnras},
         year = 2019,
        month = nov,
       volume = {489},
       number = {3},
        pages = {3042-3058},
          doi = {10.1093/mnras/stz2279},
archivePrefix = {arXiv},
       eprint = {1908.02906},
 primaryClass = {astro-ph.GA},
       adsurl = {https://ui.adsabs.harvard.edu/abs/2019MNRAS.489.3042A}
}

@ARTICLE{Balick_2020,
       author = {{Balick}, Bruce and {Frank}, Adam and {Liu}, Baowei},
        title = "{Models of the Mass-ejection Histories of Pre-planetary Nebulae. IV. Magnetized Winds and the Origins of Jets, Bullets, and FLIERs}",
      journal = {\apj},
         year = 2020,
        month = jan,
       volume = {889},
       number = {1},
          eid = {13},
        pages = {13},
          doi = {10.3847/1538-4357/ab5651},
archivePrefix = {arXiv},
       eprint = {1911.12812},
 primaryClass = {astro-ph.SR},
       adsurl = {https://ui.adsabs.harvard.edu/abs/2020ApJ...889...13B}
}

@ARTICLE{Cala_2024,
       author = {{Cala}, Rold{\'a}n A. and {Miranda}, Luis F. and {G{\'o}mez}, Jos{\'e} F. and {Morisset}, Christophe and {Soto}, Federico and {Guill{\'e}n}, Pedro F. and {V{\'a}zquez}, Roberto},
        title = "{Long- and short-term variability of the possible nascent planetary nebula IRAS 22568+6141: A late thermal pulse?}",
      journal = {\aap},
         year = 2024,
        month = nov,
       volume = {691},
          eid = {A321},
        pages = {A321},
          doi = {10.1051/0004-6361/202451719},
archivePrefix = {arXiv},
       eprint = {2410.16786},
 primaryClass = {astro-ph.SR},
       adsurl = {https://ui.adsabs.harvard.edu/abs/2024A&A...691A.321C}
}

@ARTICLE{Cala_2025b,
       author = {{Cala}, Rold{\'a}n A. and {G{\'o}mez}, Jos{\'e} F. and {Miranda}, Luis F. and {Imai}, Hiroshi and {Anglada}, Guillem and {de Gregorio-Monsalvo}, Itziar and {Ohnaka}, Keiichi and {Su{\'a}rez}, Olga and {Tafoya}, Daniel and {Uscanga}, Lucero},
        title = "{Identification and characterization of nascent planetary nebulae with OH and H$_{2}$O masers}",
      journal = {\aap},
         year = 2025,
        month = nov,
       volume = {703},
          eid = {A268},
        pages = {A268},
          doi = {10.1051/0004-6361/202452213},
archivePrefix = {arXiv},
       eprint = {2509.15727},
 primaryClass = {astro-ph.SR},
       adsurl = {https://ui.adsabs.harvard.edu/abs/2025A&A...703A.268C}
}

@ARTICLE{Cala_2025a,
       author = {{Cala}, Rold{\'a}n A. and {Miranda}, Luis F. and {G{\'o}mez}, Jos{\'e} F. and {Ohnaka}, Keiichi},
        title = "{Low- and high-velocity ``water fountains'': Different evolutionary stages}",
      journal = {\aap},
         year = 2025,
        month = nov,
       volume = {703},
          eid = {L7},
        pages = {L7},
          doi = {10.1051/0004-6361/202557103},
archivePrefix = {arXiv},
       eprint = {2510.01856},
 primaryClass = {astro-ph.SR},
       adsurl = {https://ui.adsabs.harvard.edu/abs/2025A&A...703L...7C}
}

@Article{Hunter_2007,
  Author    = {Hunter, J. D.},
  Title     = {Matplotlib: A 2D graphics environment},
  Journal   = {Computing in Science \& Engineering},
  Volume    = {9},
  Number    = {3},
  Pages     = {90--95},
  publisher = {IEEE COMPUTER SOC},
  doi       = {10.1109/MCSE.2007.55},
  year      = 2007
}

@article{harris2020,
  title={Array programming with {NumPy}},
  author={Harris, Charles R. and Millman, K. Jarrod and van der Walt, Stéfan J. and Gommers, Ralf and Virtanen, Pauli and Cournapeau, David and Wieser, Eric and Taylor, Julian and Berg, Sebastian and Smith, Nathaniel J. and Kern, Robert and Picus, Matti and Hoyer, Stephan and van Kerkwijk, Marten H. and Brett, Matthew and Haldane, Allan and Fernández del Río, Jaime and Wiebe, Mark and Peterson, Pearu and Gérard-Marchant, Pierre and Sheppard, Kevin and Reddy, Tyler and Weckesser, Warren and Abbasi, Hameed and Gohlke, Christoph and Oliphant, Travis E.},
  journal={Nature},
  volume={585},
  number={7825},
  pages={357--362},
  year={2020},
  publisher={Springer Science and Business Media {LLC}},
  doi={10.1038/s41586-020-2649-2}
}

@software{reback2020pandas,
    author       = {The pandas development team},
    title        = {pandas-dev/pandas: Pandas},
    month        = feb,
    year         = 2020,
    publisher    = {Zenodo},
    version      = {latest},
    doi          = {10.5281/zenodo.3509134},
    url          = {https://doi.org/10.5281/zenodo.3509134}
}

\twocolumn
\begin{appendix}
\twocolumn

\onecolumn
\section{Emission-line fluxes}

In the spectrum of NGC~7009 LISs, more than 250 emission lines were detected in a spectral range from 300~nm to 2.5 $\mu$m. Line fluxes were corrected for extinction employing the mean extinction coefficient resulting from the values listed in Table \ref{table_extinction} for each one of the substructures (K1a, K1b, and K4). In Table \ref{emission_lines}, we list in the third, fourth, and fifth columns the extinction-corrected and normalized to \hb = 100 emission line fluxes for K1a, K1b, and K4, respectively. The identification of each emission line is given in the first column, while the observed and theoretical wavelengths are listed in the second and final columns, respectively. At the end of the table, the \hb~flux is also reported. For blended lines, the notes accompanying the table specify the exact wavelengths of the individual contributing components.

\setlength{\tabcolsep}{3pt}
\begin{longtable}{|cccccc|cccccc|}
\caption{De-reddened line fluxes relative to H$\beta$ = 100, for K1a, and K4.}\\
\hline
 ID & $\lambda_{obs}$ & I  & I  & I  & $\lambda_{lab}$ & ID & $\lambda_{obs}$ & I & I& I& $\lambda_{lab}$ \\
  &  & K1a & K1b& K4  &  &  & & K1a & K1b & K4 &  \\
\hline
\endfirsthead
\caption{continued.}\\
\hline
  ID & $\lambda_{obs}$ & I  & I  & I  & $\lambda_{lab}$ & ID & $\lambda_{obs}$ & I & I & I& $\lambda_{lab}$ \\
  &  & K1a & K1b& K4  &  &  & & K1a & K1b & K4 &  \\
\hline
\endhead

\label{emission_lines}
 \oxygeniiir    &  3046.29  & 0.74 & $-$& $-$& 3047.12&  \sulfurii & 4067.59& 6.10 & 3.86 &  11.40 & 4068.60\\
 \argoniii    &  3108.65  & 0.30 & $-$& 0.49 & 3109.17&  \oxygeniir       &  4071.15  & 0.19 &  0.20 &  0.14 & 4071.24   \\
\oxygeniiir      &  3121.12  & 0.05 & $-$&  $-$& 3121.64&$^b$~\oxygeniir  &  4075.36  & 2.04 &  1.53 &  3.70 & 4075.86  \\
 \oxygeniiir      &  3132.26  & 0.69 &  1.85& 1.11 & 3132.79& \hdelta       &  4100.80 & 25.44 & 26.30&  25.80 & 4101.74\\
 \heliumi       &  3186.95  & 2.85 & 5.41&  4.06 & 3187.74&  \oxygeniir       &  4096.23  & 0.47 & 0.80&  0.52 & 4097.26 \\
 \neoniir      &  3217.39  & 0.12 &  $-$& $-$& 3218.19&  \oxygeniir &  4118.22  & 0.04 & $-$&  $-$ & 4120.82 \\
 \oxygeniiir      &  3311.52  & 0.11 &  $-$& $-$& 3312.32&  \heliumi &  4119.28  & 0.23 &  0.27 & 0.29 &4120.82\\
 \neoniir      &  3326.35  & 0.03 & 0.30&  $-$& 3327.153& \heliumi  &  4142.75  & 0.37 &  0.41 & 0.40 & 4143.76 \\
 \neoniir      &  3334.04  & 0.23 & 0.48&  0.30 & 3334.84&   \heliumi&  4167.96  & 0.06 & $-$&  0.09 & 4168.97 \\
 \oxygeniiir      &  3339.95  & 0.34 &  $-$& 0.30 & 3340.76&  \carboniiir&  4185.88  & 0.05 & 0.09&  $-$& 4186.90 \\
 \neoniii     &  3341.86  & 0.15 & $-$&  0.38 & 3342.50 & \oxygeniir&  4188.77  & 0.03 & $-$&  0.06 & 4189.79  \\
 \neoniir      &  3345.07  & 0.05 &  $-$& $-$& 3345.88& \nitrogeniir &  4235.85  & 0.03 &  $-$& $-$&4236.91 \\
 \heliumi       &  3353.74  & 0.15 & 0.30&  $-$& 3354.55&\nitrogeniir &  4240.18  & 0.04 &  $-$& $-$& 4241.24  \\
 \neoniir      &  3357.01  & 0.05 & $-$&  $-$& 3357.82&  \carboniir  &  4266.08  & 0.36 & 0.40&  0.36 & 4267.15     \\
 \oxygeniiir      &  3443.45  & 0.26 &  0.54&  0.48 & 3444.07&\oxygeniir &  4272.04  & $-$ & 0.14 &  0.07 & 4273.10  \\
 \heliumi       &  3446.97  & 0.10 & $-$&  $-$& 3447.59& \oxygeniir   &  4275.10  & 0.08 & $-$&  0.05 & 4275.55 \\
 \heliumi       &  3465.32  & 0.12 & $-$&  0.21 & 3465.94& \oxygeniir &  4302.54  & 0.08 &  0.13 & 0.07 & 4303.61   \\
 \heliumi       &  3470.97  & 0.09 &  $-$& $-$& 3471.81&  \hc &  4340.92  & 43.98 &  46.30 & 47.14 & 4340.47 \\
 \heliumi       &  3478.13  & 0.09 & $-$&  $-$& 3478.97& \oxygeniir&  4348.76  & $-$&  0.08 & 0.06 & 4349.43 \\
 \heliumi       &  3486.88  & 0.11 &  $-$& $-$& 3487.72&  \oxygeniii &  4362.13  & 7.51 & 10.46 &  11.08 & 4363.21  \\
 \heliumi       &  3497.80  & 0.13 & $-$&  $-$& 3498.64&  \nitrogeniiir  &  4378.03  & 0.09 & 0.13 &  0.08 & 4379.11   \\
 \heliumi       &  3511.67  & 0.16 & 0.35&  0.20 & 3512.52&  \heliumi&  4386.85  & 0.65 & 0.70 &  0.68 & 4387.93\\
 \heliumi       &  3529.65  & 0.19 &  0.35& 0.25 & 3530.50&  \neoniir&  4408.21  & 0.02 & $-$&  $-$& 4409.30 \\
 \heliumi       &  3553.57  & 0.27 & 0.31&  0.39 & 3554.42& \oxygeniir&  4413.81  & 0.05 & $-$&  $-$& 4414.90 \\
 \heliumi       &  3586.43  & 0.35 & 0.53&  0.44 & 3587.28& \oxygeniir &  4415.88  & 0.05 & $-$&  0.06 & 4416.97  \\
 \heliumi       &  3612.78  & 0.24 &  0.30& 0.32 & 3613.64& \neoniir&  4427.78  & $-$&  $-$& 0.05 & 4428.54  \\
 \heliumi       &  3633.33  & 0.71 &  0.66& 0.74 & 3634.25&  \heliumi  &  4436.45  & 0.08 & 0.12 &  0.08 & 4437.55  \\
 \hydrogeni         &  3661.34  & 0.10 & $-$&  $-$& 3662.26&  \heliumi  &  4470.40  & 5.47 &  5.53&  5.90 & 4471.50 \\
 \hydrogeni         &  3663.76  & 0.20 &  $-$& 0.15 & 3664.07& \nitrogeniiir &  4509.79  & 0.08 &  0.08 & $-$& 4510.91 \\
 \hydrogeni         &  3665.18  & 0.12 &  0.08& 0.12 & 3666.10& \nitrogeniiir &  4522.46  & 0.04 & $-$&  $-$& 4523.58 \\
 \hydrogeni         &  3666.76  & 0.20 & 0.14&  $-$& 3667.68&  Mg~{\sc i}] &  4561.47  & 0.04 & 0.12 &  0.08 & 4562.60\\
 \hydrogeni         &  3667.12  & $-$& $-$&  0.25 & 3667.12& Mg~{\sc i}] &  4569.96  & 0.05 &  $-$& 0.10 & 4571.10  \\
 \hydrogeni         &  3668.54  & 0.33 &  $-$& 0.26 & 3669.46&  \oxygeniir  &  4589.78  & 0.03 &  0.09&  $-$& 4590.97 \\
 \hydrogeni         &  3670.55  & 0.33 & 0.31 &  0.31 & 3671.48&  \oxygeniir  &  4594.77  & 0.02 &  0.07 & $-$& 4595.96\\
 \hydrogeni         &  3672.81  & 0.36 &  0.32& 0.40 & 3673.74&  \oxygeniir &  4608.25  & 0.02 &  $-$& $-$& 4609.44  \\ 
 \hydrogeni         &  3675.43  & 0.33 & 0.39&  0.40 & 3676.36&  \nitrogeniir  &  4629.34  & 0.04 &  $-$& $-$& 4630.54 \\
 \hydrogeni         &  3678.43  & 0.43 &  0.57& 0.55 & 3679.36&\nitrogeniiir &  4632.94  & 0.16 &  0.26 & 0.18 & 4634.14\\
 \hydrogeni         &  3681.88  & 0.55 &  0.62 &  0.63 & 3682.81& \oxygeniir &  4637.66  & 0.11 & 0.12 &  0.14 & 4638.86 \\
 \hydrogeni         &  3685.9  & 0.64 & $-$&  0.72 & 3686.83& $^b$~\nitrogeniiir &  4640.61  & 0.69 &  0.56 & 0.70& 4640.64 \\
 \hydrogeni         &  3690.63  & 0.74 & 0.82 &  0.96 & 3691.56& \oxygeniir&  4648.04  & 0.28 & 0.32 &   0.24 & 4649.13\\
 \neoniir      &  3693.28  & 0.13 & $-$&  $-$& 3694.21&   \oxygeniir &  4649.87  & 0.18 &  0.34 & 0.17 & 4650.84\\
 \hydrogeni         &  3696.22  & 0.90 & 0.96&  1.15 & 3697.15&   \ironiii   &  4656.89  & 0.14 &  0.12 & 0.17 & 4658.10 \\  
 \hydrogeni         &  3702.93  & 1.01 &  $-$& 1.34 & 3703.86& \oxygeniir &  4660.42  & 0.14 &  0.16 & 0.13 & 4661.63 \\ 
 \heliumi       &  3704.09  & 0.77 & 0.89 &  0.86 & 3705.02&  \oxygeniir  &  4675.13  & 0.08 & 0.09 &  0.05 & 4676.24 \\ 
 \hydrogeni         &  3711.12  & 1.41 & 1.37 & 1.73 & 3711.97&\heliumii &  4684.57  & 0.42 &  0.86 & 0.71 & 4685.68\\ 
$^b$~\sulfuriii &  3720.78  & 3.69 &  3.66& 4.45 & 3721.69& \oxygeniir&  4695.24  & 0.11 & 0.40 &  0.13 & 4696.35 \\
 \oxygenii     &  3725.18  & 129.20 &  97.73 & 275.11 & 3726.03&  \oxygeniir &  4698.11  & 0.09 &  0.3 & 0.11 & 4699.22 \\
 \oxygenii     &  3727.97  & 83.80 &  62.88 & 199.70 & 3728.82& \oxygeniir       &  4704.24  & 0.05 &  $-$& $-$& 4705.35\\   
 \hydrogeni         &  3733.52  & 2.09 & 2.22 &  2.62 & 3734.37&\argoniv&  4710.26  & 1.62 & 2.90 &  1.77& 4711.37\\  
 \hydrogeni         &  3749.29  & 2.41 & 2.42 &  2.93 & 3750.15&    \heliumi       &  4712.05  & 0.60 &  0.64 & 0.68 & 4713.17\\  
 \oxygeniiir      &  3753.83  & 0.07 & 0.19&  0.20 & 3754.69&   \argoniv    &  4739.05  & 1.32 &  2.29 & 1.40 & 4740.17\\   
 \oxygeniiir      &  3759.01  & 0.12 & 0.22&  $-$& 3759.87&  \ironiii&  4753.43  & 0.03 & $-$&  $-$& 4754.69\\
 \hydrogeni         &  3769.77  & 3.44 &  3.68 & 4.00 & 3770.63&\ironii    &  4813.25  & 0.03 & $-$& $-$& 4814.53\\
\oxygeniiir &  3773.45  & $-$&  $-$& 0.05 & 3774.02&  \hb &  4860.12  & 100 & 100 & 100 & 4861.33\\
\hydrogeni &  3797.31  & 4.65& 4.92&  5.32 & 3797.90& \ironiii   &  4879.89  & 0.05 &  $-$& 0.06 & 4881.11\\
 \heliumi       &  3804.83  & 0.07 &  $-$& $-$& 3805.74&\heliumi       &  4920.71  & 1.46 &  $-$& 1.43 & 4921.93 \\
 \heliumi       &  3818.71  & 1.34 &  1.58& 1.58 & 3819.62& \oxygeniii &  4930.58  & 0.12 & 0.17 &  0.10 & 4931.23\\
 \hydrogeni         &  3834.48  & 6.52 & 7.02 &  7.34 & 3835.39&\oxygeniii &  4957.69  & 515.88 &  380.00 & 397.10 & 4958.91\\
\neoniii     &  3867.83  & 122.07 & 143.70&  179.05 & 3868.75& \oxygeniii    &  5005.39  & 1537.51 &  1056.40 & 1168.66 & 5006.80 \\
 \hydrogeni       &  3887.61  & 21.7 &  23.71& 24.9 & 3889.10&  \heliumi       &  5014.23  & 2.33 &  2.00 & 2.11 & 5015.68 \\
  \heliumi       &  3925.49  & 0.16 &  0.18 &  0.20 & 3926.54&  \heliumi       &  5046.28  & 0.18 &  0.15 & 0.18 & 5047.74\\ 
  \heliumi       &  3963.77  & 0.61 & 0.50 &  $-$& 3964.73&\argoniii    &  5190.69  & 0.14 &  0.10 & 0.21 & 5191.82\\
 \neoniii     &  3966.50  & 39.92 & 44.25&  56.63 & 3967.46&\nitrogeni      &  5196.77  & 1.46 &  1.30 & 4.80 & 5197.90 \\
 \hydrogeni         &  3969.11  & 14.93 & 15.54&  16.35 & 3970.07&\nitrogeni      &  5199.13  & 1.02 &  0.89 & 3.69 & 5200.26 \\
  \heliumi    &  4008.29  & 0.24 &  0.30 & 0.26 & 4009.26&  \ironiii  &  5269.10  & 0.03 &  $-$& 0.05 & 5270.40\\
  \heliumi       &  4025.11  & 2.61 & 2.70&  2.92 & 4026.08 &  \heliumii      &  5410.48  & 0.04 & 0.11 &  0.06 & 5411.52 \\
    \chloroiii   &  5516.33  & 0.77 &  0.83 & 0.78 & 5517.66 &   \chloroii    & 8576.61 & 0.45 & 0.27 & 0.93& 8578.69 \\
  \chloroiii   &  5536.26  & 0.71 & 0.71 &  0.70 & 5537.60 & \heliumi       & 8579.79 & 0.09 & 0.11 & 0.10 & 8581.87\\
   \oxygeni      & 5576.11 & 0.23 &  0.58 & 0.66 &5577.34 &  \hydrogeni & 8596.30 & 0.64 &  0.62 &  0.66& 8598.39\\
\nitrogeniir        & 5678.27 & 0.04&  $-$& $-$ & 5679.56 &    \ironii    & 8614.86 & 0.05 &  $-$& 0.07& 8616.95 \\ 
\nitrogenii     & 5753.05 & 4.33 & 3.13 &  9.20 &5754.606 &    \heliumi       & 8646.20 & 0.10 &  $-$& $-$ & 8648.25   \\
 \heliumi& 5874.25 & 16.46 & 16.20 & 17.6 & 5875.66 &   \hydrogeni         & 8662.97 & 0.80 & 0.80 &  0.82& 8665.02 \\
 \oxygeni & 6298.54 & 18.18 & 14.022 &  45.13 & 6300.34  & \carboni & 8724.76 & 0.10 &  0.07 & 0.24& 8727.13  \\
 \sulfuriii & 6310.31 & 2.86 &  2.58 & 3.7 & 6312.10 & \heliumi       &  8731.38 & 0.04 & $-$&  $-$ &8733.43   \\
 \oxygeni & 6361.98 & 6.04 & 5.24 &  15.75& 6363.78 &  \hydrogeni         & 8748.42 & 1.02 & $-$& 1.02& 8750.47 \\
\carboniir & 6460.36 & 0.04 & $-$&  $-$& 6462.05  & \heliumi       & 8774.91 & 0.04 & $-$&  $-$ &8776.97 \\
\oxygeniir& 6499.75 & 0.03 & $-$&  $-$& 6501.41&  \heliumi       & 8843.40 & 0.06 & $-$& $-$ & 8845.37 \\
 \heliumii      & 6525.51 & 0.06 & $-$&  0.1& 6527.11& \hydrogeni & 8860.75 & 1.32 & 1.30 & 1.33& 8862.78 \\
 \nitrogenii     & 6546.54 & 94.83 &  54.60 & 185.40 &6548.10 & \heliumi       & 8994.94 & 0.12 & $-$& 0.08 & 8997.02\\
\ha & 6561.21 & 286.00 & 286.00 & 285.50& 6562.77 & \hydrogeni         & 9012.83 & 1.71 &  1.77 & 1.78& 9014.91  \\
  \nitrogenii     & 6581.93 & 287.49 & 178.63&  574& 6583.50&  \sulfuriii    & 9066.78 & 53.93 & 41.27 &  47.5& 9068.60 \\
\heliumi       & 6676.58 & 4.79 & 4.38 & 4.70& 6678.16 & \neoniir      & 9082.78 & 0.03 & $-$&  $-$ &9084.60\\
 \sulfurii     & 6714.67 & 27.54 & $-$& 60.29& 6716.44 &\neoniir?        & 9093.63 & 0.03 & $-$& $-$ & 9095.95   \\
 \sulfurii     & 6729.05 & 39.38 & 23.57 & 78.90 & 6730.82 &\chloroii    & 9121.21 & 0.12 &  $-$& 0.24& 9123.6   \\
\heliumi       & 7063.53 & 3.70 & 3.37 &  4.30 &7065.25 & \heliumi       & 9208.19 & 0.12 & 0.10 & 0.11& 9210.33\\
 \argoniii    & 7134.04 & 26.33 &  25.01 & 30.50 & 7135.80  &  \hydrogeni         & 9226.86 & 2.43 &  2.45 &  2.43 & 9229.01\\
 \ironii& 7153.65 & 0.02 & $-$&  0.05 & 7155.17 & \heliumi       & 9514.82 & 0.11 & 0.12 & 0.12& 9516.63  \\
\heliumi &7158.80 & 0.02 & $-$&  - & 7160.56 &  \sulfuriii    & 9528.78 & 83.65 & 81.40 & 99.94& 9530.60 \\
\ironiir& 7169.20 & 0.05 & $-$&  0.03 & 7172.00 &  \hydrogeni         & 9544.15 & 2.30 & 2.07 & 1.68&  9545.97  \\
\carboniir        & 7229.75 & 0.03 & $-$& $-$ & 7231.34 &  ?      & 9694.88 & 0.03 & $-$& $-$ & ?  \\
\carboniir        & 7234.61 & 0.06 & 0.22 & $-$ & 7235.00 &   \neoniir      & 9717.50 & 0.02 & $-$& $-$ & 9720.13 \\
 \heliumi & 7261.45 & 0.03 & $-$& $-$ & 7262.76&   \oxygeniir        & 9737.56 & 0.02 & $-$& $-$ & 9741.50 \\
 \heliumi       & 7280.04 & 0.68 & 0.61 & 0.77 & 7281.35&  \carboni      & 9821.59 & 0.41 &  0.33 &  0.88& 9824.13\\
 \oxygenii     & 7318.67$^b$ & 6.80 &  4.05 & 12.37& 7319.99  &  \carboni       & 9847.41 & 1.22 & 1.12 & 2.66 & 9850.26\\
 \oxygenii     & 7328.35 & 4.98 &  3.64 & 5.43& 7330.73& ? & 9914.38 & 0.05 & $-$& $-$ & ?  \\
 \nickelii& 7375.84 & 0.03 &  $-$& 0.06 & 7377.87 & \ironiir     & 9961.34 & 0.03 & $-$& $-$ & 9963.77 \\
\heliumi       & 7497.83 & 0.04 & $-$& 0.05& 7499.84   &  \neoniir      & 10007.90 & 0.09 & $-$& $-$ & 10010.2  \\
 \cliv& 7528.81 & 0.13 & 0.23 &  0.11 &7530.80  & \heliumi       & 10025.4 & 0.20 & $-$& 4.9& 10027.7   \\
 \argoniii    & 7749.21 & 6.32 &  5.70 & 7.23& 7751.06&  Pa$\delta$ & 10047.05 & 6.15 & 5.19 &   5.13 & 10049.4    \\
  \heliumi       & 7814.30 & 0.05 & $-$&  $-$ &7816.16 & \sulfurii& 10284.17 & 1.23 & $-$& 2.03& 10286.73 \\
 \Phosii & 7873.79& 0.03&   $-$& $-$ & 7875.99 & \heliumi & 10308.36 & 0.25 & $-$& $-$ & 10311.23    \\
    \cliv & 8044.01 & 0.29 &  0.45 & 0.30 & 8046.30 &\sulfurii& 10317.00& 1.72& $-$&   2.43 &  10320.32\\
 \ironir        & 8194.67 & 0.05 &  $-$& $-$ & 8196.61 & \sulfurii &10333.78 & 1.26 & $-$& 1.94 & 10336.41\\
 \ironiir     & 8202.37 & 0.03 &  $-$& $-$  & 8204.42 & [S~{\sc ii}]  & 10367.93 & 0.54 & $-$& 0.96& 10370.49\\
\heliumii      & 8234.82 & 0.04 &  $-$& $-$ &8236.77   &  \nitrogeni &10395.17 & 0.30 & $-$ &0.60& 10397.74\\
\hydrogeni         & 8253.07 & 0.03 & $-$&  $-$ &8255.02 &  \nitrogeni & 10404.62 & 0.21 & $-$ &0.38& 10407.17 \\
\hydrogeni         & 8258.98 & 0.05 &  $-$& $-$ &8260.93& \heliumi & 10827.60 & 62.00 & 77.00&67.00 & 10830.00\\
\hydrogeni         & 8262.33 & 0.08 &  $-$& $-$ &8264.28& \heliumi& 10910.38 & 0.49 & 0.77 &0.38& 10913.05\\
  \hydrogeni         & 8265.99 & 0.06 &  $-$& $-$ &8267.94 &  \heliumi & 10914.56 & 0.18 & $-$ &0.15 & 10917.10\\
  \hydrogeni         & 8269.98 & 0.06 &  $-$& $-$ & 8271.93  &  Pa$\gamma$& 10935.38& 8.45 & 9.63 &8.20  &10938.00\\
\hydrogeni         & 8274.35 & 0.06 & $-$& $-$ & 8276.31&   [P{ \sc ii}] & 11465.50& 0.15& $-$&0.38& 11468.16\\
\hydrogeni         & 8279.16 & 0.07 & $-$& $-$ & 8281.12  &   ? & 11616.40 & 0.08 & $-$ &0.25& ?\\
 \hydrogeni         & 8284.47 & 0.08 & $-$& $-$ &  8286.43   &$^b$ [P{ \sc ii}]& 11880.09 & 0.72& 0.25 & 0.83& 11882.82\\
\hydrogeni         & 8290.35 & 0.07 & $-$& $-$ & 8292.31  &  \heliumi & 11966.12 & 0.25 & 0.15&0.26&  11969.00 \\
 \hydrogeni         & 8296.88 & 0.10 &  $-$& $-$ & 8298.83  &  \heliumi& 12524.13 & 0.13 &0.25& 0.20& 12527.52 \\
\hydrogeni         & 8304.16 & 0.08 &  $-$& 0.08 & 8306.11 & \ironii & 12563.67 & 0.23 & $-$&0.4& 12566.80\\
\hydrogeni         & 8312.31 & 0.12 &  0.06 & 0.14& 8314.26 & \hydrogeni & 12781.72 & 0.87 & 0.90 &0.66& 12784.92\\
\hydrogeni         & 8321.46 & 0.11 & 0.07 & 0.10& 8323.42  &   \heliumi& 12787.39 & 0.29 &  $-$ &0.20& 12790.50  \\
\hydrogeni         & 8331.82 & 0.15 &  0.09 & 0.17& 8333.78   &  Pa$\beta$ & 12814.85 & 15.56 &  15.60& 16.01& 12818.00   \\
  \heliumi & 8340.30 & 0.03 & $-$&$-$ &  8342.32  & \heliumi & 12965.08 & 0.08 & $-$&$-$ & 12968.43  \\
 \hydrogeni         & 8343.51 & 0.15 &  0.11 &0.15& 8345.47   &\ironii & 13202.34 & 0.09 & $-$& 0.13& 13205.50 \\
 \hydrogeni         & 8357.04 & 0.10 &  0.23 & 0.16& 8359.00  &  \nickelii & 13348.00 & 0.04 & $-$& $-$ &  13352.55\\
 \heliumi & 8359.00 & 0.05 & $-$& 0.08& 8360.99 &  \hydrogeni& 14997.30  & 0.09 & $-$& $-$ & 15000.86   \\
\hydrogeni    & 8372.51 & 0.21 & 0.15 &  0.22& 8374.48&  \heliumi & 15050.60  & 0.08 &  $-$&1.5 & 15054.00 \\
\hydrogeni    & 8390.46 & 0.22 & 0.19 & 0.23&  8392.40  & \hydrogeni& 15079.15 & 0.09 & $-$&$-$ & 15082.77  \\
\hydrogeni         & 8411.38 & 0.27 & 0.28 & 0.27& 8413.32  &\hydrogeni& 15129.68 & 0.11 & $-$&$-$ & 15133.22 \\
\hydrogeni         & 8436.00 & 0.31 &  0.28 & 0.32& 8437.95  &\hydrogeni & 15188.03 & 0.09 & $-$&0.10& 15191.84 \\
\heliumi       & 8443.94 & 0.10 & $-$&0.24& 8444.69&\hydrogeni & 15255.78 & 0.06 & $-$&0.09&  15260.54\\
\hydrogeni         & 8465.29 & 0.38 & 0.36 & 0.37& 8467.25&   \ironii & 15327.60 & 0.14 & $-$&0.76& 15330.66   \\
 \hydrogeni         & 8500.51 & 0.43 & 0.42 & 0.42& 8502.48&   \hydrogeni & 15337.86 & 0.19 & $-$&0.19& 15341.79  \\
\heliumi       & 8527.02 & 0.05 & $-$&$-$ & 8528.99& \hydrogeni & 15435.06 & 0.18 & 0.19 &0.15& 15438.90   \\
 \hydrogeni         & 8543.31 & 0.53 & 0.52 & 0.54& 8545.38& \hydrogeni & 15552.51 & 0.22 & 0.322&0.18& 15556.45\\
   \hydrogeni& 15696.43 & 0.23 &  $-$&0.11& 15700.66& Pa$\alpha$& 18745.40 & 0.98 & 0.71 & 1.08& 18751.1 \\
 \hydrogeni& 15876.56 & 0.32 &  0.33 & 0.30& 15880.54&  $^b$~Br$\delta$ & 19440.38 & 0.77 & 0.65 & 0.66& 19445.56\\
\ironii & 15989.60 & 0.03 & $-$&$-$& 15994.72  &   \hmolec~1-0S(3) & 19566.47 & 0.31 & 0.25 & 0.14& 19575.60 \\
    \hydrogeni & 16105.25 & 0.35 & 0.39 & 0.36& 16109.31   & \hmolec~1-0S(2) & 20328.15 & 0.15 &  $-$&0.80& 20337.60 \\
 \hydrogeni& 16403.12 & 0.49 & 0.46 & 0.47& 16407.19 &\heliumi& 20576.27 & 0.64 & 0.44& 0.73& 20581.28 \\
  \ironii & 16434.17 & 0.22 &  $-$&0.54& 16435.49&   \hmolec~1-0S(1) & 21208.55 & 0.56 & 0.79 &  1.95& 21218.00\\
  \hydrogeni& 16802.35 & 0.64 &  $-$&0.51 & 16806.50&   \heliumi & 21635.96 & 0.18 & 0.34 &$-$ & 21641.48\\
   \heliumi & 16998.16 & 0.40 & 0.41 &  0.80& 17002.00&  Br$\gamma$ & 21649.91 & 2.75 & 3.45 & 2.73& 21655.29 \\
    \hydrogeni& 17357.88 & 0.85 &  0.66 & 0.76 & 17362.11&  \hmolec~1-0S(0) & 22222.37 & 0.14 &  0.14 &0.47&22232.90 \\
      \heliumi& 17471.58 & 0.12 & 0.13& 0.40& 17476.89 &    \hmolec~2-1S(1) & 22465.00 & 0.05 &  0.10 &0.10& 22477.20    \\
        \ironi &  17872.00& 0.03 &  $-$&$-$ & 17878.48 &  \hmolec~1-0Q(1) & 24054.48 & 0.52 & $-$&1.17& 24065.90   \\
      \hydrogeni& 18135.78 & 0.16 &  $-$&$-$ & 18139.03& \hmolec~1-0Q(2) & 24123.00 & 0.12 &  $-$&0.66& 24134.00 \\
      \hydrogeni& 18169.57 & 1.12 & 1.10 &0.90 & 18172.33  &\hmolec~1-0Q(3) & 24225.48 & 0.53 &  0.15 &0.42& 24237.30\\
      \heliumi & 18680.00 & 0.20 &  0.33 & 0.7 & 18685.45 & \hmolec~1-0Q(5) & 24536.13 & 0.26 & 0.63 &0.55 & 24547.50 \\
      \ironii& 18738.51 & $-$& $-$&0.39& 18741.70& f(\hb) & & 1.32 $\times$10$^{-13}$&  2.85$\times$10$^{-14}$&4.6$\times$10$^{-14}$&\\
     \hline
\end{longtable}
\tablefoot{$\lambda_{obs}$ and $\lambda_{lab}$ represent the observed and the theoretical wavelength for each emission line, respectively. f(\hb) is in units ergs~s$^{-1}$ cm$^{-2}$. Flux uncertainties were estimated using the error spectrum produced during the X-Shooter data reduction. Final uncertainties were then derived by a Monte Carlo approach implemented in the L{\sc i}M{\sc e} algorithm. In the NIR arm, flux uncertainties range from 40\% for the faintest lines (flux<5\% \hb), to 20\% for brighter lines (5\% \hb<flux<20\% \hb) and less than 10\% for the brightest ones. In the UVB and VIS arms, the uncertainties are typically lower in every case (10\% and 20\%). The lines noted with $^b$ are blended: \sulfuriii~$\lambda$3721.69 with \hydrogeni~$\lambda$3721.93, \hydrogeni~$\lambda$ 3889.1  with \heliumi~$\lambda$3888.65 , \oxygeniir~$\lambda$4075.36  with \oxygeniir~$\lambda$4078.84, \nitrogeniiir~$\lambda$4640.64  with \oxygeniir~$\lambda$4641.81 and \nitrogeniiir~$\lambda$4641.85, [P{ \sc ii}]$^b$ $\lambda$11880.09 with \ironii $\lambda$11881.57 and Br$\delta$ 19445.56  with \hmolec~$\lambda$19448.}

\section{Iron abundances}
In Table \ref{iron_abund} we summarize the derived ionic and total iron abundances. For \ironii~two sets of atomic data are available in {\sc PyNeb}, namely TZ18 (\citealt{Tayal_2018}) and B15 (\citealt{Bautista_1996}), so the  Fe$^{+}$/H$^{+}$ abundance was computed using both of them. Total iron abundances were estimated following three different approaches: (i) assuming Fe/H = Fe$^{+}$($\lambda8617$)/H$^{+}$, using the same approach adopted by \citet{Bouvis_2025} (fourth row), (ii) summing the observed ionic abundances, Fe$^{+}$/H$^{+}$ and Fe$^{2+}$/H$^{+}$ (fifth row), and finally (iii) applying the two ionization correction factor (ICF) schemes proposed by \citet{Rodriguez_2005} (their equations 2 and 3), which are based on photoionization models and empirical calibrations, respectively (sixth and seventh columns).

\begin{table}[ht!]
\centering
\caption{\ironii, \ironiii, and total Fe abundances.}
\label{iron_abund}

    \begin{tabular}{|lcccccc|}  
        \hline
        
        & \multicolumn{3}{c}{This work} &
   \multicolumn{3}{c|}{Bouvis et al. (2025)} \\
 & K1a & K1b & K4 & K1a & K1b & K4 \\
        \hline
      Fe$^{+}$/H$^{+}$ $\times$10$^{-8}$ (TZ18) &2.09$\pm$0.55 &$-$& 2.98$\pm$0.60 &7.55$^{+1.14}_{-3.70}$ &$-$ &9.82$^{+1.50}_{-4.62}$\\
        Fe$^{+}$/H$^{+}$  $\times$10$^{-8}$(B15) &3.50$\pm$1.10 &$-$& 5.76$\pm$1.20& 8.90$^{+1.34}_{-4.15}$ & $-$ &11.1$^{+1.77}_{-5.42}$\\
       Fe$^{2+}$/H$^{+}$  $\times$10$^{-8}$&5.00$\pm$1.90 &3.18$\pm$1.10& 3.35$\pm$2.10& $-$ &$-$  &$-$ \\
       Fe/H $\times$10$^{-8}$ (\ironii~$\lambda$8617)  &4.40$\pm$1.10 &$-$& 5.77$\pm$1.20&8.90$\pm$4.53 &$-$ &11.1$^{+6.6}_{-5.7}$\\ 
        Fe/H $\times$10$^{-8}$ (sum) &8.32$^{+1.34}_{-4.15}$ &$-$& 9.11$\pm$2.70&$-$ &$-$ &$-$\\
        Fe/H  $\times$10$^{-7}$  (ICF eq2) &1.90$\pm$0.5  &0.97$\pm$0.36&0.64$\pm$0.40&$-$ &$-$ &$-$\\
         Fe/H  $\times$10$^{-7}$ (ICF eq3)  &2.24$\pm$0.30  &2.30$\pm$1.10&0.73$\pm$4.60&$-$&$-$&$-$\\
        
        \hline
        
    \end{tabular}
    \\
\tablefoot{Ionic and total iron abundances derived for K1a, K1b, and K4 in this work, together with the corresponding values reported by \citet{Bouvis_2025} for comparison. }
\end{table}

\section{Comparison with shock and PDR models}

To investigate whether \hmolec~emission originates from shocks or from dense condensations resembling PDRs, we employed the models of \cite{Burton_1992}. In this study, the authors modeled the \hmolec~emission of bright \hmolec~lines, which are commonly observed in both shocked environments and PDRs. These include the pure rotational transitions 0–0 S(0), S(2), and S(3), as well as the rovibrational transitions 1–0 S(1) and 2–1 S(1). 

For the PDR models, two scenarios were considered: (i) neutral clouds pumped by far-UV photons, and (ii) high-density environments, where \hmolec~exhibits characteristics similar to collisional excitation of \hmolec. For the shock models, the code of \cite{Hollenbach_1989} was used to calculate the \hmolec~ spectrum of non-dissociative shocks, employing different initial conditions for J-type non-dissociative shocks and C-type shocks.

The results also examine how the fluxes of the bright \hmolec~lines change as a function of the radiation field (G$_0$) for PDR models, or as a function of shock velocity for C- and J-type shocks. We overlaid our observed values of the 1-0 S(1) and 2-1 S(1) \hmolec~transitions, green and black stars on Fig. \ref{Burton_models}, respectively, onto these model predictions. We found that the model that better reproduces both line intensities is a J-type shock model with a density of 10$^4$\,cm$^{-3}$ and shock velocities below 20 km\,s$^{-1}$. In all other cases, the observed line fluxes do not align with the model predictions.

\begin{figure*}[htbp]
    \centering
    \begin{subfigure}[b]{0.3\textwidth}
        \centering
        \includegraphics[width=\textwidth]{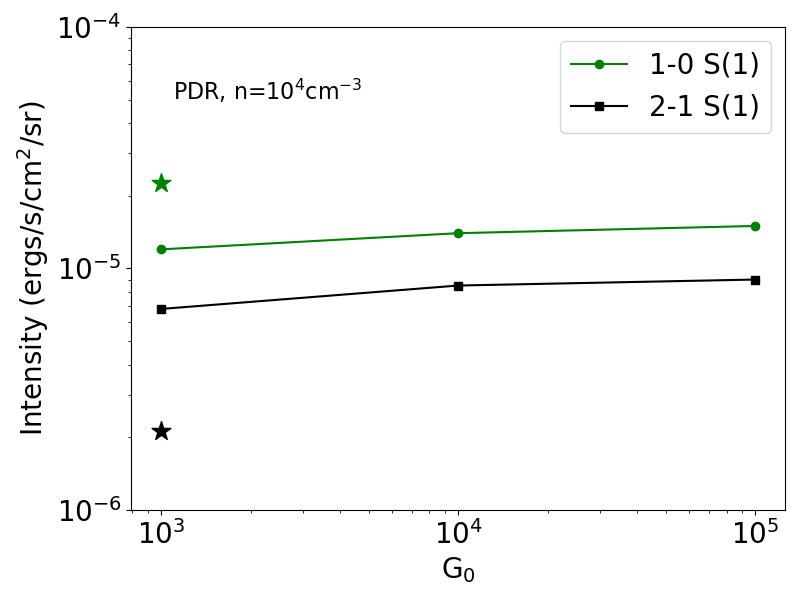}
    \end{subfigure}
    \hfill
    \begin{subfigure}[b]{0.3\textwidth}
        \centering
        \includegraphics[width=\textwidth]{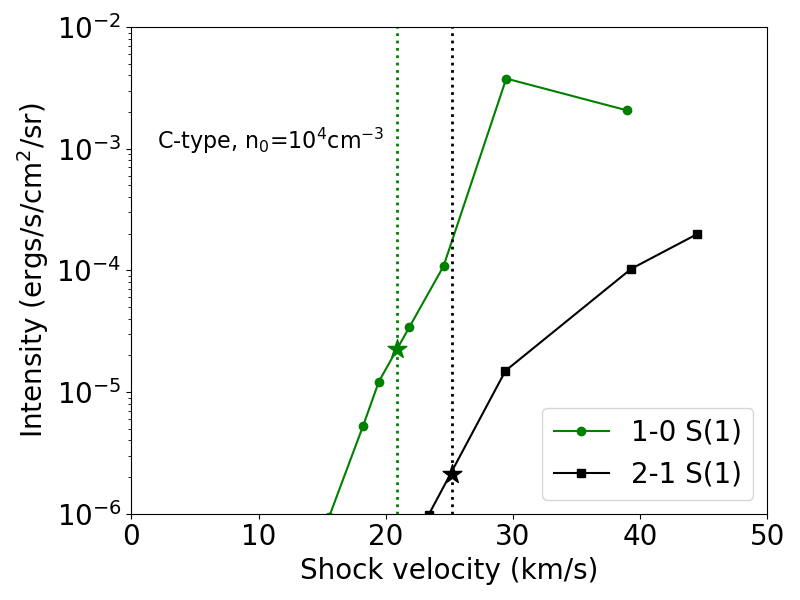}
    \end{subfigure}
    \hfill
    \begin{subfigure}[b]{0.3\textwidth}
        \centering
        \includegraphics[width=\textwidth]{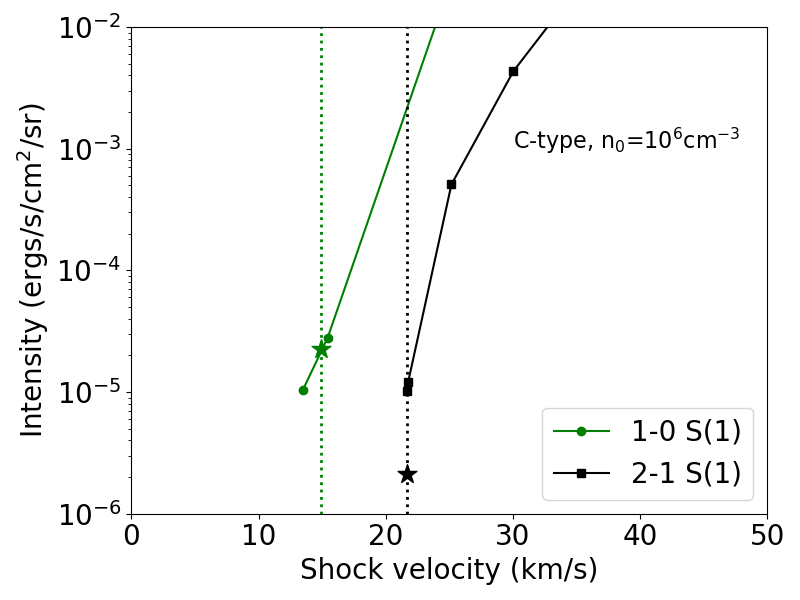}
    \end{subfigure}

    \vskip\baselineskip

    \begin{subfigure}[b]{0.4\textwidth}
        \centering
        \includegraphics[width=\textwidth]{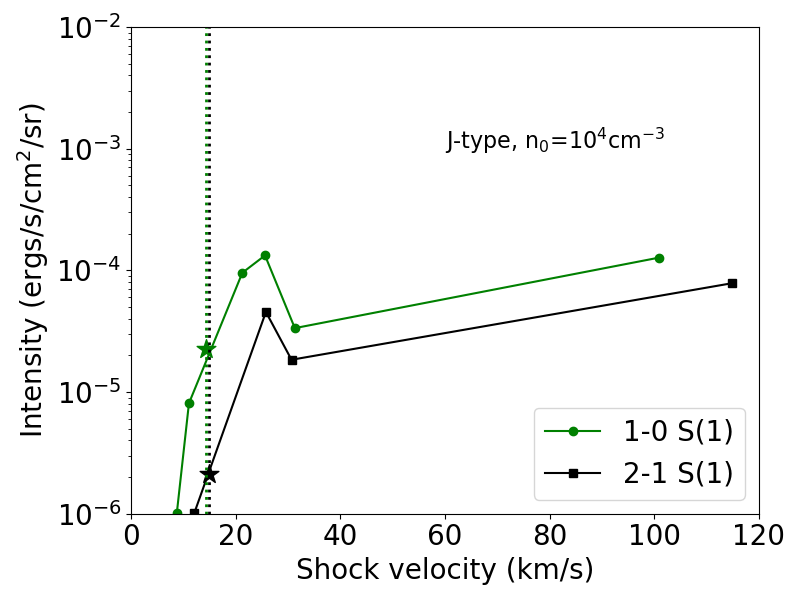}
    \end{subfigure}
    \hfill
    \begin{subfigure}[b]{0.4\textwidth}
        \centering
        \includegraphics[width=\textwidth]{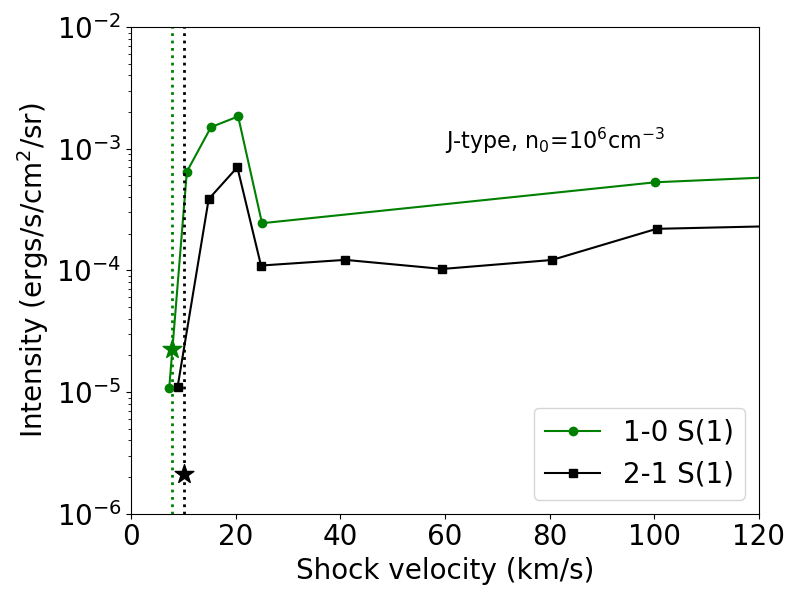}
    \end{subfigure}

    \caption{PDR and shock model predictions for the 1-0 S(1) and 2-1 S(1) \hmolec~lines from \cite{Burton_1992}, for K1a. The top-left panel shows a PDR model; the two middle panels present C-type shock models, considering different pre-shock densities, and the bottom two panels show J-type shock models with two different pre-shock densities.}
    \label{Burton_models}
\end{figure*}

\end{appendix}

\end{document}